\documentclass[longauth]{aa}  
\usepackage{graphicx}
\usepackage{txfonts}
\usepackage[colorlinks=true,linkcolor=blue,citecolor=blue]{hyperref}
\usepackage{float}
\usepackage{subfloat}
\usepackage{subcaption}
\usepackage[flushleft]{threeparttable}

\newcommand{\feii}{\mbox{Fe\,{\sc ii}}}
\newcommand{\znii}{\mbox{Zn\,{\sc ii}}}

\newcommand{\siii}{\mbox{Si\,{\sc ii}}}
\newcommand{\sii}{\mbox{S\,{\sc ii}}}
\newcommand{\mgi}{\mbox{Mg\,{\sc i}}}
\newcommand{\mgii}{\mbox{Mg\,{\sc ii}}}
\newcommand{\hi}{\mbox{H\,{\sc i}}}

\newcommand{\alii}{\mbox{Al\,{\sc ii}}}

\newcommand{\cii}{\mbox{C\,{\sc ii}}}
\newcommand{\civ}{\mbox{C\,{\sc iv}}}
\newcommand{\siiv}{\mbox{Si\,{\sc iv}}}

\newcommand{\oi}{\mbox{O\,{\sc i}}}

\newcommand{\nv}{\mbox{N\,{\sc v}}}

\begin{document}

\title{Dissecting the interstellar medium of a $z=6.3$ galaxy}

\subtitle{X-shooter spectroscopy and \textit{HST} imaging of the afterglow\\and environment of the \textit{Swift} GRB\,210905A\thanks{Based on observations carried out under ESO prog. ID 106.21T6.010 and ID 106.21T6.015 (PI: N. Tanvir) with the X-shooter spectrograph installed at the Cassegrain focus of the Very Large Telescope (VLT), Unit 3 - Melipal, operated by the European Southern Observatory (ESO) on Cerro Paranal, Chile. Partly 
based on observations with the NASA/ESA Hubble Space Telescope, obtained at the Space Telescope Science Institute, which is operated by the Association of Universities for Research in Astronomy (AURA). This observation is primarily associated with proposal GO\,16918, PI: N. Tanvir.\vspace{0.3cm}}}

\author{A. Saccardi
          \inst{1}\fnmsep\thanks{E-mail: andrea.saccardi@obspm.fr},
          S. D. Vergani\inst{1,2,3}, A. De Cia\inst{4}, V. D'Elia\inst{5,6}, K. E. Heintz\inst{7,8,9}, L. Izzo\inst{10}, J. T. Palmerio\inst{1}, P. Petitjean\inst{3}, A. Rossi\inst{11}, A. de Ugarte Postigo\inst{12}, L. Christensen\inst{8,9}, C. Konstantopoulou\inst{4}, A. J. Levan\inst{13}, D. B. Malesani\inst{13,8,9}, P. M{\o}ller\inst{13,14}, T. Ramburuth-Hurt\inst{4}, R. Salvaterra\inst{15}, N. R. Tanvir\inst{16}, C. C. Th\"one\inst{17}, S. Vejlgaard\inst{8,9}, J. P. U. Fynbo\inst{8,9}, D. A. Kann\inst{18}, P. Schady\inst{19}, D. J. Watson\inst{8,9}, K. Wiersema\inst{20}, S. Campana\inst{2}, S. Covino\inst{2}, M. De Pasquale\inst{21}, H. Fausey\inst{22}, D. H. Hartmann\inst{23}, A. J. van der Horst\inst{22,24}, P. Jakobsson\inst{7}, E. Palazzi\inst{11}, G. Pugliese\inst{25,26}, S. Savaglio\inst{27}, R. L. C.  Starling\inst{16}, G. Stratta\inst{11}, T. Zafar\inst{28}
          }

\institute{GEPI, Observatoire de Paris, Université PSL, CNRS, 5 Place Jules Janssen, 92190 Meudon, France
\and
INAF - Osservatorio Astronomico di Brera, via E. Bianchi 46, I23807, Merate (LC), Italy
\and
Institut d’Astrophysique de Paris, UMR 7095, CNRS-SU, 98 bis
boulevard Arago, 75014, Paris, France
\and
Department of Astronomy, University of Geneva, Chemin Pegasi 51, 1290 Versoix, Switzerland
\and
Space Science Data Center (SSDC) - Agenzia Spaziale Italiana
(ASI), I-00133 Roma, Italy
\and
INAF - Osservatorio Astronomico di Roma, via Frascati 33, 00040
Monte Porzio Catone, Italy
\and
Centre for Astrophysics and Cosmology, Science Institute, University of Iceland, Dunhagi 5, 107 Reykjav\'ik, Iceland
\and
Cosmic Dawn Center (DAWN), Denmark
\and
Niels Bohr Institute, University of Copenhagen, Jagtvej 128, DK-2200 Copenhagen N, Denmark
\and
DARK, Niels Bohr Institute, University of Copenhagen, Jagtvej 128,
2200 Copenhagen, Denmark
\and
INAF - Osservatorio di Astrofisica e Scienza dello Spazio, via Piero
Gobetti 93/3, I-40129 Bologna, Italy
\and
Artemis, Observatoire de la C\^ote d’Azur, Université C\^ote d’Azur,
CNRS, 06304 Nice, France
\and
Department of Astrophysics/IMAPP, Radboud University, 6525 AJ Nijmegen, The Netherlands
\and
European Southern Observatory, Karl-Schwarzschildstrasse 2, D-85748 Garching bei München, Germany
\and
INAF IASF-Milano, via Alfonso Corti 12, I-20133 Milano, Italy
\and
School of Physics and Astronomy, University of Leicester, University Road, Leicester LE1 7RH, UK
\and
Astronomical Institute (ASU CAS), Ondřejov, Czech Republic
\and
Instituto de Astrofísica de Andalucía (IAA-CSIC), Glorieta de la
Astronomía s/n, 18008 Granada, Spain
\and
Department of Physics, University of Bath, Bath, BA2 7AY, UK
\and
Physics Department, Lancaster University, Lancaster, LA1 4YB,
UK
\and
Università degli Studi di Messina - Dipartimento MIFT, Polo Papardo, via F. S. D’Alcontres 31, I-98166 Messina, Italy
\and
Department of Physics, the George Washington University, 725 21st Street NW, Washington, DC 20052, USA
\and
Department of Physics and Astronomy, Clemson University, Kinard Lab of Physics, Clemson, SC 29634-0978, USA
\and
Astronomy, Physics and Statistics Institute of Sciences (APSIS), 725 21st Street NW, Washington, DC 20052, USA
\and
Anton Pannekoek Institute for Astronomy, University of Amsterdam, Science Park 904, 1098 XH Amsterdam, The Netherlands
\and
Leiden Observatory, University of Leiden, Niels Bohrweg 2, 2333
CA Leiden, The Netherlands
\and
Physics Department, University of Calabria, 87036 Arcavacata di
Rende (Cs), Italy
\and
Australian Astronomical Optics, Macquarie University, 105 Delhi Road, North Ryde, NSW 2113, Australia
}

\date{}

 
  \abstract
   {The study of the properties of galaxies in the first billion years after the Big Bang is one of the major topics
   of current astrophysics. Optical and near-infrared spectroscopy of the afterglows of long gamma-ray bursts (GRBs) provides a powerful diagnostic tool to 
   probe the interstellar medium (ISM) of their host galaxies and foreground absorbers, even up to the highest redshifts. 

We analyze
the VLT/X-shooter afterglow spectrum of GRB\,210905A, triggered by the {\it Swift Neil Gehrels Observatory}, 
and detect neutral hydrogen, low-ionization, high-ionization, and fine-structure absorption lines from a complex system at $z=6.3118$, which we associate with the GRB host galaxy. We use them to study the ISM properties of the host system, revealing the metallicity, kinematics, and chemical abundance pattern of its gas along the GRB line of sight. We also detect absorption lines from at least two foreground absorbers at $z=5.7390$ and $z=2.8296$. 
  
The total metallicity of the $z\sim6.3$ system is $\text{[M/H]}_{\rm tot}=-1.72\pm0.13$, after correcting for dust depletion and taking $\alpha$-element enhancement into account, as suggested by our analysis. This is consistent with the values found for the other two GRBs at $z\sim6$ with spectroscopic data showing metal absorption lines (GRB\,050904 and GRB\,130606A), and it is at the higher end of the metallicity distribution of quasar damped Lyman-$\alpha$ systems (QSO-DLAs) extrapolated to such a high redshift. In addition, we determine the overall amount of dust and dust-to-metal mass ratio ($DTM$) ([Zn/Fe]$_{\rm fit} = 0.33 \pm 0.09$ and $DTM=0.18\pm0.03$). 
We find indications of nucleosynthesis due to massive stars and, for some of the components of the gas clouds, we find evidence of peculiar nucleosynthesis, with an overabundance of aluminum (as also found for GRB\,130606A).

From the analysis of fine-structure lines, we determine distances of several kiloparsecs for the low-ionization gas clouds closest to the GRB. Those are farther
distances than usually found for GRB host absorption systems, possibly due to the very high number of ionizing photons produced by the GRB that could ionize the line of sight up to several hundreds of parsecs. 

Using the $HST$/F140W image of the GRB field, we show the GRB host galaxy (with a possible afterglow contamination) as well as multiple objects within 2\arcsec{} from the GRB position. We discuss the galaxy structure and kinematics that could explain our observations, also taking into account a tentative detection of Lyman-$\alpha$ emission at $z=6.3449$ ($\sim 1200$\,km\,s$^{-1}$ from the GRB redshift in velocity space), and the observational properties of Lyman-$\alpha$\, emitters at very high redshift.
  
This study shows the amazing potential of GRBs to access detailed information on the properties (metal enrichment, gas kinematic, dust content, nucleosynthesis...) of very high-redshift galaxies, independently of the galaxy luminosity. Deep spectroscopic observations with VLT/MUSE and JWST will offer the unique possibility of combining the information presented in this paper with the properties of the ionized gas, with the goal of better understanding how galaxies in the reionization era form and evolve.
  
  }


   \titlerunning{Dissecting the interstellar medium of a $z=6.3$ galaxy}
   \authorrunning{A. Saccardi et al.}

   \maketitle
%
\section{Introduction}
The identification of galaxies at the highest redshifts remains one of the central goals in contemporary astrophysics. Significant investments of time with premier facilities continue to be required to identify and subsequently characterize these distant galaxies, and these are prime goals of many legacy surveys with the Hubble Space Telescope (HST) and the James Webb Space Telescope (JWST). The challenge in undertaking these observations arises from the combination of extreme luminosity distance, which makes the majority of individual galaxies invisible to ground-based observatories, to HST, and possibly even to JWST. However, such investments are strongly motivated because they provide unique diagnostics on the physical processes that shaped these earliest galaxies. These initial conditions are then vital for exploring the evolution of stars, galaxies, and the intergalactic medium (IGM) across cosmic history.

However, the faintness of these galaxies limits the available diagnostics, even with the largest telescopes. As a result, many galaxies are limited to photometric observations from which the physical properties are then derived. In a smaller number of cases, individual far-UV emission lines such as Lyman-$\alpha$ can also be observed (e.g., \citealt{Inoue2016,Herenz2019}). Stronger constraints have been obtained via stacked spectra of very large numbers of galaxies at $z \sim 3$ \citep{Berry2012}, but beyond this, and for almost all individual high redshift galaxies, few constraints on the chemical or dynamical properties of the galaxies are possible. 

Selection via gamma-ray bursts (GRBs) bypasses many of these concerns. GRBs are instantaneously the most luminous explosions in nature and have been detected out to the highest redshifts \citep{Salvaterra2009,Tanvir2009,Cucchiara2011}. At peak, their optical afterglows can fleetingly be a factor of 10$^8$ or more brighter than the galaxies that host them \citep{Bloom2009,Racusin2008_nature}, providing the ability to obtain detailed spectroscopic observations of the afterglow. Furthermore, this afterglow carries the imprint of material within the host galaxy and along the line of sight. Hence, while most frequently used to determine burst redshifts, GRB afterglows can also directly measure: dynamics within the host galaxy, its chemical content (e.g., \citealt{Sparre2014}), the hydrogen column density \citep{Jakobsson2006,Selsing2019}, the distance of absorbing material from the GRB (e.g., \citealt{Prochaska2006,Vreeswijk2007}), the state of the intergalactic medium around the host \citep{Tanvir2019}, and the presence of any other absorption systems along the line of sight \citep{Vergani2009}. 

In this regard, GRBs offer many of the benefits that quasar sightlines offer, but also further advantages. Long duration GRBs are securely associated with 
the collapse of massive stars \citep{Hjorth2011_gamma,Cano2017}. Their sightlines point directly to star-forming regions, unlike the random sightlines in quasar damped Lyman-alpha absorption (DLA) systems. If we are interested in star formation processes at these early cosmic epochs, then such sightlines are of great value.  Perhaps most notably, GRB afterglows fade revealing their faint host galaxies \citep{Tanvir2012,McGuire2016}, partly belonging to the bulk of the high-redshift galaxy population \citep{Vergani2015,Palmerio2019,Perley2016,Salvaterra2013}, and therefore making it possible to pair information about the galaxy in absorption (e.g., metallicity, dynamics) with information in emission (luminosity, star formation rate). Thanks to high redshift GRBs, we can build pictures of the complete properties of high redshift galaxies in remarkable detail. 

In this paper we study the VLT/X-shooter optical and NIR afterglow spectrum of GRB\,210905A at redshift $z=6.3$. In Sect. \S\ref{data} and \S\ref{data analysis}, we present our dataset and the fitting of the absorption lines. In Sect. \S\ref{results} we present and discuss our results. The conclusions are drawn in Sect. \S\ref{conclusions}. A $\Lambda$CDM cosmological model with $\Omega_{M}=0.308$, $\Omega_{\Lambda}=0.692$, and $H_{0}=67.8$\,km s$^{-1}$ Mpc$^{-1}$ \citep{Planck2016} has been assumed for calculations. All data are in observer frame and 1\,$\sigma$ errors are reported throughout the paper, unless otherwise specified.
Two companion papers (\cite{Rossi2022}; Fausey et al. in prep) present and analyze the GRB and afterglow multiwavelength spectral and temporal properties, and the use of the X-shooter spectrum to infer the neutral fraction of the intergalactic medium, respectively.


\section{VLT/X-shooter observation of GRB 210905A}
\label{data}
GRB\,210905A was discovered by the \textit{Neil Gehrels Swift Observatory} (Swift hereafter, \citealt{Gehrels2004}) Burst Alert Telescope (BAT) on September 5, 2021, at 00:12:41 UT. The X-Ray Telescope (XRT) began observing the field 91.7\,s after the BAT trigger and found a bright,
uncataloged source at the enhanced position of coordinates RA (J2000) = 20h 36m 11.58s Dec. (J2000) = $-$44\degr{} 26$'$ 22.4\arcsec{} with an uncertainty of 2.6\arcsec{} \citep{Beardmore2021}. The {\it Swift} Ultra-Violet and Optical Telescope (UVOT) observed the field 157\,s after the BAT trigger without detecting any sources consistent with the XRT position. Ground-based observations indicated a possible high-redshift origin of the GRB \citep{GCN2021GRB210905AStrausbaugh_2,GCN2021GRB210905AStrausbaugh, GCN2021GRB210905ADavanzo, GCN2021GRB210905Cooke, GCN2021GRB210905Nicuesa}, and provided a precise position of its afterglow with ALMA at RA (J2000) = 20h 36m 11.5685s ($\pm 0.0002$ s)
Dec. (J2000) = $-$44\degr{} 26$'$ 24.840\arcsec{} ($\pm0.002\arcsec$) \citep{Laskar2021}. We refer the reader to \cite{Rossi2022} for an extensive multiwavelength analysis of the GRB and afterglow properties.

After $\sim2.53$ hours (observer frame) from the GRB detection, we observed the location of the afterglow of GRB\,210905A with the ESO 
VLT UT3 equipped with the X-shooter spectrograph \citep{Vernet2011} under good atmospheric conditions. The observing set-up is detailed in Table \ref{tab:1}. We obtained a set of four exposures of 1200\,s in the three different arms of X-shooter, covering the wavelength range 3\,000-21\,000~\AA. Observations were executed using the ABBA nod-on-slit mode, with a nod throw of 6\arcsec{} along the slit. We reduced each single spectrum of the UVB and VIS arm using the STARE mode reduction, with the extraction window at the position of the GRB afterglow trace, and selecting two background windows at both sides of the spectral trace. The NIR arm was extracted using the standard X-shooter NOD mode pipeline \citep{Goldoni2006,Modigliani2010}. Then, for each single exposure, the flux calibrated spectra were corrected for slit-loss in each arm, residual sky features  were subtracted and finally the individual spectra were stacked into a final science spectrum, using the procedures reported in \citet{Selsing2019}. We have also applied a telluric correction to the final stacked VIS spectrum, which was initially estimated from the observations of a telluric star, and finally applied to the GRB afterglow spectrum, after applying a wavelength-dependent correcting factor for the different airmass of the GRB compared to the standard telluric star. Wavelengths
were corrected to the vacuum-heliocentric system.

\begin{table*}
\centering
\caption{Log of the observations.}
 \label{tab:1}
 \begin{tabular}{lccccc}
  \hline
  \hline
  Epoch & Arm & Exp. time & Wav. range & Slit & Resolution $R$\\
  (Hours) &  & $(s)$ & $(nm)$ & $('')$ & $(\lambda/\delta\lambda)$\\ 
  \hline
2.53 & UVB & 4x1200 & 300-560 & 1.0x11 & 5400\\
2.53 & VIS & 4x1200 & 560-1020 & 0.9x11 & 8900\\
2.53 & NIR & 4x1200 & 1020-2100 & 0.9x11JH & 5600\\
  \hline
  \hline
 \end{tabular}
\end{table*}


\section{Data analysis}
\label{data analysis}

\begin{figure*}
   \centering
   \includegraphics[width=0.65\textwidth]{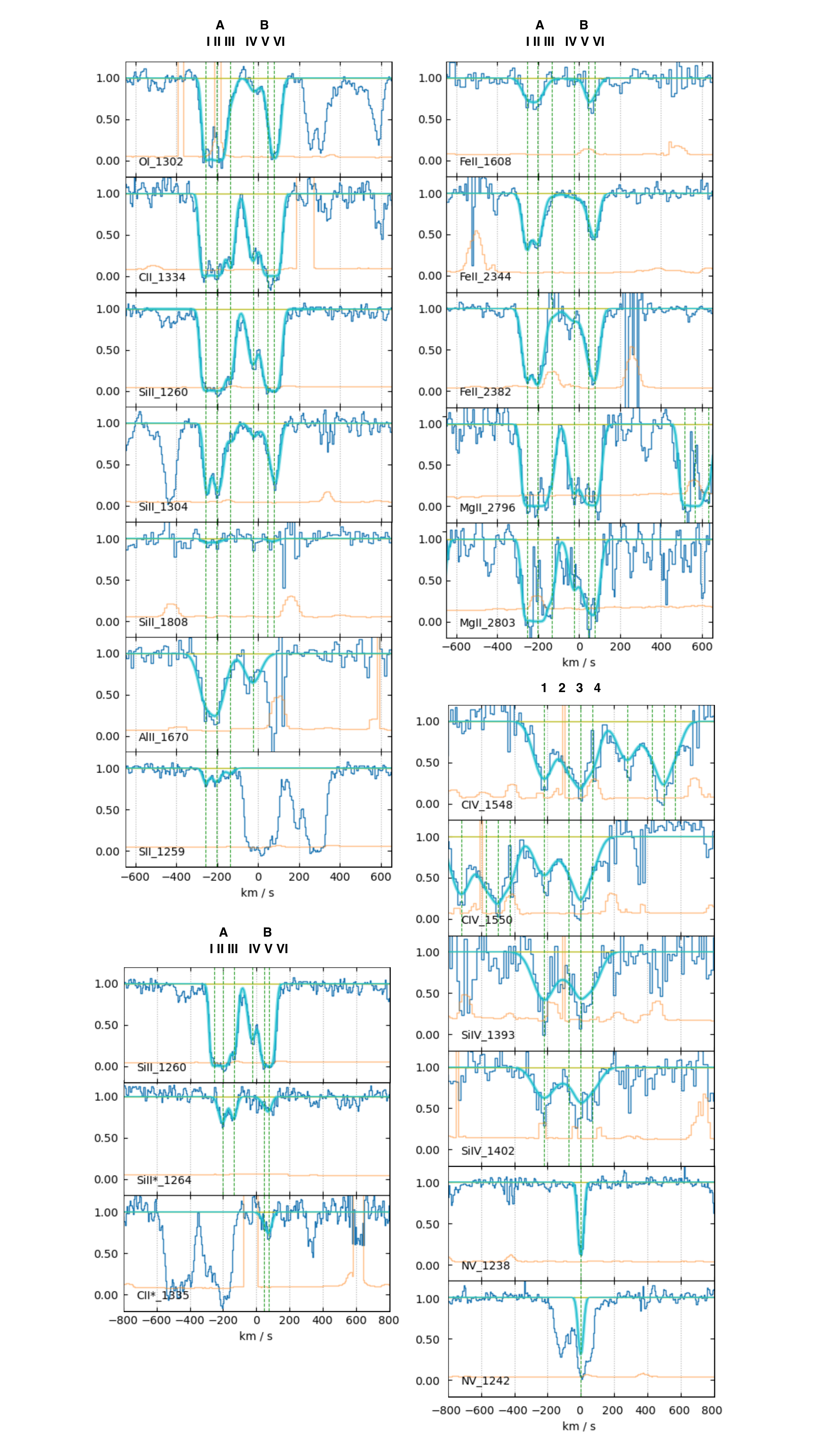}\caption{VLT/X-shooter optical and NIR afterglow spectrum of GRB\,210905A at redshift $z=6.3$. {\it Top left and right}: Low-ionization absorption lines of the GRB host galaxy system. Here and in the following panels, data are in blue, the fit is in cyan, the error spectrum is in orange, and the vertical green dashed lines indicate the center of the components. {\it Bottom left}: Fine-structure absorption lines. The top panel shows the \siii{}$\lambda1260$ absorption lines corresponding to the ground-level, as reference. {\it Bottom right}: High-ionization absorption lines of the GRB host galaxy system. We stress that \siiv{} and \civ{} lines are noisy because they are affected by sky-line residuals and telluric absorptions. The $\lambda1242$\,\AA{} absorption of \nv{} is blended with the \feii{}$\lambda2734$\,\AA{} line of the $z=2.8296$ foreground absorber. All the plots are in the velocity space and 0 has been fixed arbitrarily to $z=6.3168$  (see Sect.\,\ref{nv}), corresponding to the  \nv{}$\lambda1238$\,\AA{} absorption line (and to the stronger high-ion ionization line component). }
         \label{low}
   \end{figure*}

We identify in the spectrum several absorbing systems. The highest redshift one is at $z=6.3$ and spans $\sim360$\,km\,s$^{-1}$. The presence of a damped Lyman-$\alpha$ (DLA) absorption and of fine-structure absorption lines allows us to associate it with the host galaxy of GRB\,210905A (\citealt{GCN2021GRB210905ATanvir} and following sections; see however discussion in Sect.\,\ref{fine}, \ref{nv} and \ref{lya}). A very strong foreground system at $z=2.8296$ (\mgii{} and \feii{} lines) and another at $z=5.7390$ (\cii{}, \feii{}, \civ{}, \siiv{} lines) are also present (see Fig. \ref{intervening}), plus a tentative \civ{} system at $z=5.3092$.

The DLA absorption imprinted from {\sc Hi} in the host-galaxy ISM is studied in detail in Fausey et al. in prep. The DLA column density measurement includes a careful modeling of the red-part of the damping wing due to absorption from a significant neutral {\sc Hi} fraction in the line-of-sight through the IGM from this burst. Fausey et al. in prep measures $\log (N$(H\,{\sc i})/cm$^{-2}$) = $21.10\pm 0.10$ assuming a single velocity component fixed to the systemic redshift $z_{\rm GRB} = 6.3118$. We here focus on the low-, high-ionization and fine-structure metal-line transitions to determine the physical properties of the absorbing neutral gas.

\begin{table*}
\begin{threeparttable}
\caption{Column density of low ionization lines. The velocity shift of the components with respect to the \nv{} line is indicated. The last row reports the Doppler parameter $b$ of each component as resolved by the X-shooter observations.}            
\label{table_N} 
\centering
\begin{tabular}{c c c c c c c}       
\hline\hline
Species & $I$ & $II$ & $III$ & $IV$ & $V$ & $VI$ \\
\hline
Velocity & $-255\,{\rm km~s}^{-1}$ & $-203\,{\rm km~s}^{-1}$ & $-136\,{\rm km~s}^{-1}$ & $-25\,{\rm km~s}^{-1}$ & $+46\,{\rm km~s}^{-1}$ & $+75\,{\rm km~s}^{-1}$ \\
\hline
\cii{}$\lambda1334$ & $>15.79$ & $>15.02$ & $>14.35$ & $>14.30$ & $>14.72$ & $>15.33$ \\

\cii{}{*}$\lambda1335$ & & & & & $13.16\pm0.17$ & $13.43\pm0.09$ \\

\oi{}$\lambda1302$ & $>15.66$ & $>15.68$ & $14.02\pm0.12$ & $13.83\pm0.18$ & $>14.01$ & $>15.02$\\

\mgii{}$\lambda2796,\lambda2803^\mathsection$ & $>14.32$ &  $>15.01$ & $>13.36$ & $>13.30$ & $>13.76$ & $>13.51$ \\

\alii{}$\lambda1670^\ddag$ & $>13.25$ & $>13.37$ & $>12.23$ & $>12.77$ \\

\siii{}$\lambda1260,\lambda1304,\lambda1808$ & $14.33\pm0.04$ & $14.50\pm0.03$ & $13.32\pm0.05$ & $13.15\pm0.03$ & $13.51\pm0.07$ & $14.27\pm0.02$\\

\siii{}{*}$\lambda1264$ & & $12.76\pm0.04$ & $12.49\pm0.08$ & & $12.02\pm0.09$ & $12.44\pm0.07$\\

\sii{}$\lambda1259^\dagger$ & $13.99\pm0.09$ & $14.22\pm0.09$& $13.69\pm0.09$& \\

\feii{}$\lambda1608,\lambda2344,\lambda2382$ & $14.03\pm0.04$ & $13.88\pm0.02$ & $12.55\pm0.09$ & $12.79\pm0.09$ & $13.26\pm0.04$ & $13.68\pm0.02$\\

\hline
\hline
$b\,{\rm (km~s^{-1})}$ & $15.6$ & $27.7$ & $21.7$ & $28.4$ & $29.6$ & $23.2$\\
\hline
\hline
\end{tabular}
\begin{tablenotes}
    \item[$^\mathsection$]\mgii{} lines are particularly uncertain because they are found in a very noisy region at the end of the NIR arm spectrum.
    \item[$^\ddagger$]The $V, VI$ (and partially $IV$) components of \alii{} are strongly affected by a sky line and could not be determined.
    \item[$^\dagger$]The $IV, V, VI$ components of \sii{} are blended with the \siii{}$\lambda1260$\,\AA{} absorption.
\end{tablenotes}
\end{threeparttable}
\end{table*}

\begin{table*}
\caption{Column density of high ionization lines. The velocity shift of the components with respect to the \nv{} line are indicated. The last row reports the Doppler parameter $b$ used of each component as resolved by the X-shooter observations.}            
\label{table_N2} 
\centering
\begin{tabular}{c c c c c c}       
\hline\hline
Species & $1$ & $2$ & $3$ & $4$ \\
\hline
Velocity & $-220\,{\rm km~s}^{-1}$ & $-72\,{\rm km~s}^{-1}$ & $0\,{\rm km~s}^{-1}$ & $+71\,{\rm km~s}^{-1}$ \\
\hline

\civ{}$\lambda1548$,$\lambda1550$ & $>14.36$ & $>14.13$ & $>16.2$ & $>14.08$ \\

\nv{}$\lambda1238$, $\lambda1242$ & & & $>14.25$ \\

\siiv{}$\lambda1393$,$\lambda1402$ & $>13.89$ & $>13.41$ & $>13.82$ & $>13.63$\\
\hline        
\hline
$b\,{\rm (km~s^{-1})}$ & $56.2$ & $46.0$ & $15.5$ & $31.2$\\
\hline
\hline
\end{tabular}
\end{table*}

\subsection{The $z=6.3$ system}
\label{z63}
Fig. \ref{low} shows the absorption lines detected at $z=6.3$. Thanks to the wavelength range coverage of X-shooter, we can detect transitions up to the \mgii{}$\lambda2796\,\text{\AA},\lambda2803$\,\AA{} doublet. Low-ionization lines are detected in two main systems, indicated by $A$ (components $I$, $II$, $III$) and $B$ (slightly weaker and formed by components $IV$, $V$, $VI$) 
having their strongest components at $z=6.3118$ and $z=6.3186$ ($\Delta v = 278$\,km\,s$^{-1}$), respectively (see Fig.\ref{low} and Tab.\,\ref{table_N}). The \siii{}$^*\lambda1264$\,\AA{} fine-structure line is detected for both systems, whereas \cii{}$^*\lambda1335$\,\AA{} is detected for the reddest components while the bluest ones are blended with the \cii{}$\lambda1334$\,\AA{} absorption of the red system\footnote{There is an absorption coincident with \feii{}$^*\lambda2396$\,\AA{} of component $II$, but it is due to the \feii{}$\lambda2600$\,\AA{} absorption of the $z=5.7390$ intervening system.}. High-ionization lines (\civ{}, \siiv{}, \nv{}) are also detected for both systems, with their strongest component shifted from that of low-ionization lines (see also Tab.\,\ref{table_N2}). While we can determine  that \civ{}, \siiv{} span more than 300\,km\,s$^{-1}$ (despite the poor quality of the spectrum at that wavelength range, due to the presence of telluric emission and absorption lines), the \nv{} absorption is much narrower ($\sim100$\,km\,s$^{-1}$) and detected in the $B$ system only. We can analyze only the $\lambda1238$\,\AA{} line of the \nv{} doublet, because the $\lambda1242$\,\AA{} one is blended with the \feii{}$\lambda2734$\,\AA{} line of the $z=2.8296$ foreground absorber.

We fit the systems with the \texttt{Astrocook} code \citep{Cupani2020}, a \texttt{Python} software environment to analyze spectra which includes a set of algorithms to model spectral features in emission and absorption (continuum, spectral lines, complex absorption systems). The column densities we determined are presented in Tables \ref{table_N} and \ref{table_N2}. The errors reported are those obtained by the line fitting, but, considering the intermediate resolution of the X-shooter spectrograph, they can be underestimated.

The fact that we detect fine-structure lines
in both systems (components {\it II, III, V, VI}), and that the overall $N_{Si}/N_{Fe}$ column density ratio of systems $A$ and $B$ are very similar leads us to assume that both systems belong to the ISM of the same galaxy complex. The presence of fine-structure lines, as well as that of the Lyman-$\alpha$ break associated with them, allow us to associate the entire system with the host galaxy complex of the GRB (see however Sect.\,\ref{fine} and \ref{lya}). Component {\it II} has the highest column densities of low-ionization lines, therefore we take its redshift ($z=6.3118$) as that of the GRB host.

\subsection{HST image of the GRB\,210905A field}
\label{HST}
\begin{figure}[h]
\centering
\includegraphics[scale=0.55]{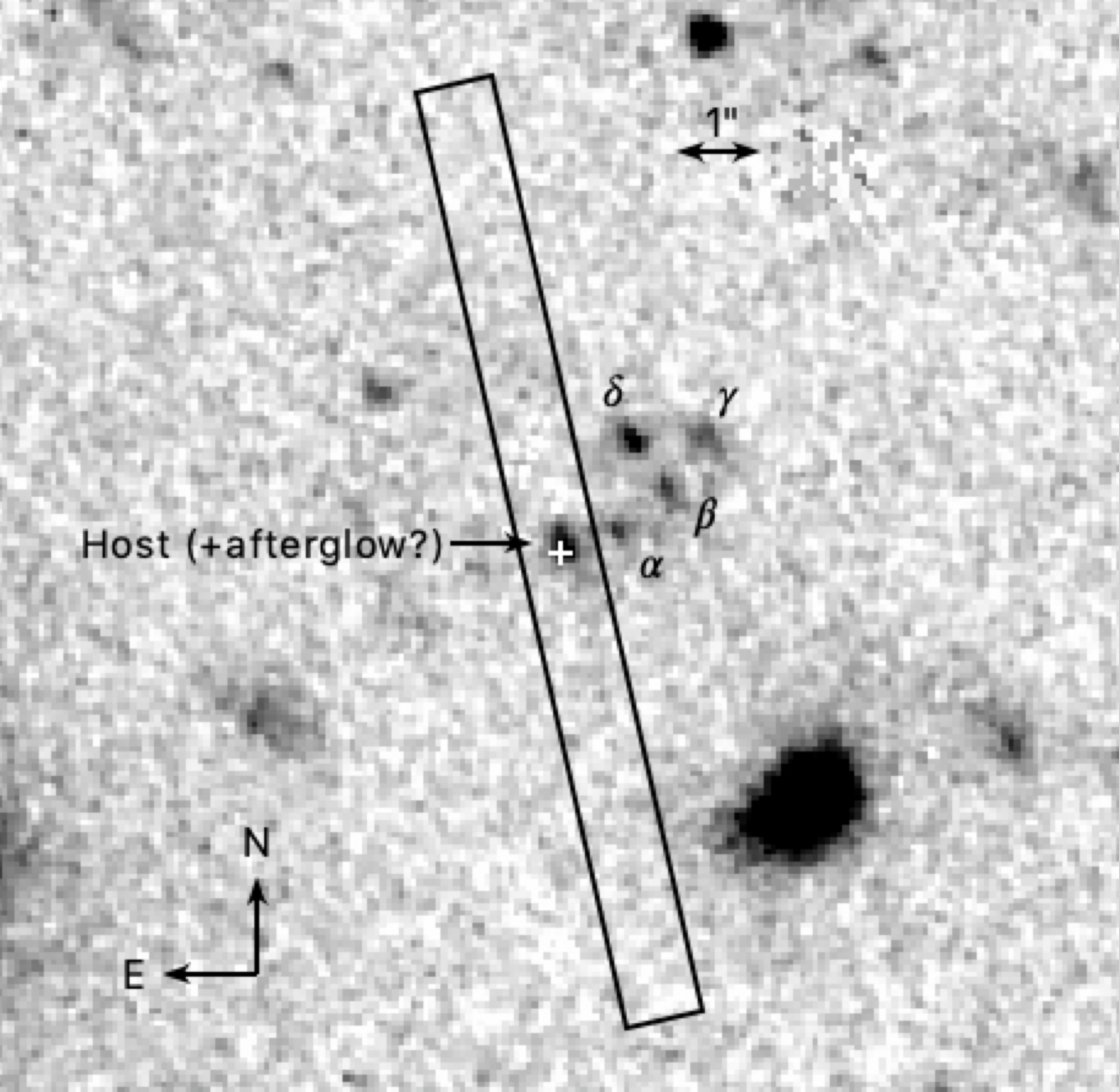}
\caption{Zoom-in (about $12\arcmin\times12\arcmin$) on the deep {\it HST}/WFC3 F140W-band image obtained 250 days after the GRB trigger. The cross indicates the ALMA localization of the afterglow and the box the X-shooter slit position, while the Greek letters label the four objects close to the afterglow position.}
\label{fig:hst_slit}
\end{figure} 

Fig. \ref{fig:hst_slit} shows the HST image of GRB\,210905A field. The afterglow position coincides with an object that we consider to be the GRB host galaxy. Its $F140W$ magnitude is $25.66\pm0.05$ mag (AB), corresponding to SFR$_{\rm UV}\sim20\,M_{\rm \odot}$\,yr$^{-1}$(see \citealt{Rossi2022}). The afterglow contamination is considered as negligible by \citep{Rossi2022}. In the proximity of the GRB, we identified four objects (distinguished by Greek letters), with projected distances of $0.73\arcsec$, $1.43\arcsec$, $1.53\arcsec$, $2.13\arcsec$, for $\alpha$, $\beta$, $\gamma$, $\delta$, respectively.  Assuming that they are at the same redshift as the GRB host ($z=6.3118$), those correspond to 4.14, 8.14, 8.67, and 12.08\,kpc, respectively.  
The $F140W$ magnitudes of the single objects are $26.46\pm0.07$, $26.38\pm0.06$, $26.34\pm0.06$, $25.98\pm0.05$ mag (AB) for $\alpha$, $\beta$, $\gamma$, $\delta$, respectively, considering an aperture of 0.33\arcsec{} radius.

Following the $I$-band detection of a faint source at $\sim1.5\arcsec$ from the GRB position, reported in \cite{Rossi2022}, it is very likely that at least one of these objects is at lower redshift and corresponds to the counterpart of one of the foreground absorbers identified in the afterglow spectrum (see Sect.\,\ref{lya}).

\subsection{Emission lines}
\label{em}
In the X-shooter spectrum we also find a tentative detection of an emission line at $\lambda8929$\,\AA{} (observer frame). The emission line region covers $\sim2.5\arcsec$, and includes the GRB afterglow position (see Fig.\,\ref{figlya}). The emission line coincides with a weak sky-emission line. We verified the persistence of the line candidate in independent combinations of the exposures, and verified that it appears independently of the sky-subtraction
method used in the processing of the X-shooter data.
We extracted the spectrum over the region
$-0.8\arcsec$ to $-3.3\arcsec$ of the 2D spectrum, in order to cover the full spatial extent of the line. 
We measured the line flux directly by integrating the continuum-subtracted flux, as well as by fitting a Gaussian profile.  
In both cases we obtain similar values for the flux $F_{\rm Ly\alpha}=(3.1\pm0.6)\times10^{-18}$\,erg\,s$^{-1}$\,cm$^{-2}$. The error on the flux estimate includes only the error on the spectrum and not on the continuum fitting.

Being the only emission line detected, and considering that the hypothesis of a low-redshift galaxy overlapping with the afterglow position is very unlikely because of the absence of bright objects in our deep FORS and HAWK-I images \citep{Rossi2022}, we tentatively associate it with Lyman-$\alpha$ emission at $z=6.3449$, that is $\sim 1200$\,km\,s$^{-1}$ redward of the DLA. 
We discuss this further in Sect.\,\ref{lya}.
We reobserved the field with X-shooter $\sim205$ days after the GRB trigger in order to try and confirm the detection of the emission line once the afterglow had faded. Using the same position angle (PA=$12.6\degr{}$) as for the afterglow spectrum, we cannot confirm or rule out the detection of the line, since measuring over the same region yields $F_{\rm Ly\alpha}=(3.0\pm1.3)\times10^{-18}$\,erg\,s$^{-1}$\,cm$^{-2}$.


\section{Results and Discussion}
\label{results}

\subsection{Line strength analysis}

Following the method described by \cite{deUgartePostigo2012} we can study the strength of some of the most prominent spectral features of GRB\,210905A in the context of those observed in their sample of 69 GRB afterglow spectra. Our goal is to compare GRB host environments at different redshifts even if the spectral resolution or the signal-to-noise ratio is not comparable. To do this, we measure the equivalent widths of prominent spectral features that are commonly observed in GRB afterglow spectra and compare them with their typical strength distribution. We include in our analysis only the features that fall within our spectral range and that are not significantly affected by telluric absorption. The measurements of the equivalent widths of the $z=6.312$ and $z=6.318$ systems are combined and studied as a single system, for consistency with typically lower resolution spectra of the sample. The results of our measurements are displayed in Table~\ref{tab:EWs}. 

\begin{figure}[h]
   \centering
   \includegraphics[width=\hsize]{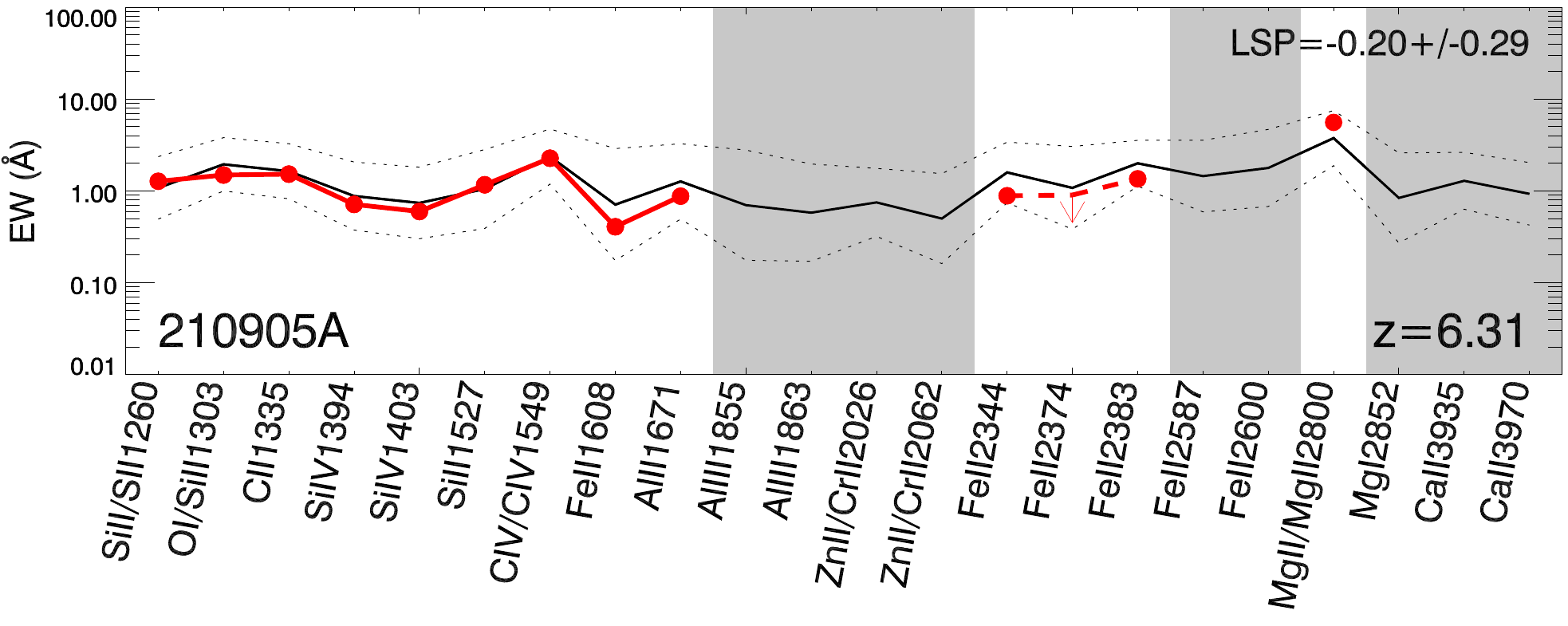}
      \caption{Line strength diagram for the X-shooter spectrum of GRB\,210905A. 
              }
         \label{fig:LSD}
   \end{figure}

Figure~\ref{fig:LSD} shows our measurements, compared with the sample of \cite{deUgartePostigo2012} in a graphical representation called a line strength diagram. In the x-axis we have different spectral features and in the y-axis their rest-frame equivalent widths in logarithmic scale. The thick black line represents the average line strength of the features, whereas the dotted lines are the 1\,$\sigma$ standard deviation in log-normal space. In red we show the rest-frame equivalent widths of the features in our spectrum. Gray regions identify the range where there is no spectral coverage or it is affected by strong telluric features. This diagram allows us to easily identify deviations in the relative strength of the individual features, through which we can identify environments with unusual composition or ionization rate. In our case, we see that the strength distribution is very similar to the typical GRB spectrum, with just some slightly weaker \feii{} and \alii{} and stronger than average \mgii{}. Furthermore, we can also calculate the \textit{line strength parameter} (LSP), which compares the average strength of the lines in our spectrum with those of the sample using a single parameter. We determine a LSP = $-0.20\pm0.29$, which indicates that the spectral features in this spectrum are very close to the average of the sample and just slightly weaker, falling at the 41st percentile of the spectral sample.

\begin{table}[h]
\begin{center}
\caption{List of rest-frame equivalent widths for strong features measured in the spectrum of GRB\,210905A. Each feature may combine several components.
} \label{tab:EWs}
\begin{tabular}{cc} 
\hline
\hline
Feature & EW$_{rest}$ (\AA)\\
\hline
\siii{}/\sii{}$\lambda$1260 &  1.27$\pm$0.02\\
\oi{}/\siii{}$\lambda$1303  &  1.49$\pm$0.04\\
\cii{}$\lambda$1335      &     1.53$\pm$0.01\\
\siiv{}$\lambda$1394     &     0.71$\pm$0.14\\
\siiv{}$\lambda$1403     &     0.60$\pm$0.12\\
\siii{}$\lambda$1527     &     1.17$\pm$0.08\\
\civ{}/\civ{}$\lambda$1549  &  2.27$\pm$0.25\\
\feii{}$\lambda$1608     &     0.41$\pm$0.06\\
\alii{}$\lambda$1671     &     0.88$\pm$0.09\\
\feii{}$\lambda$2344     &     0.89$\pm$0.05\\
\feii{}$\lambda$2374     &     $<0.30$\\
\feii{}$\lambda$2383     &     1.36$\pm$0.06\\
\mgii{}/\mgii{}$\lambda$2800&  5.58$\pm$0.23\\
\hline
\hline
\end{tabular}
\end{center}
\end{table}

\begin{figure}[h]
   \centering
   \includegraphics[width=\hsize]{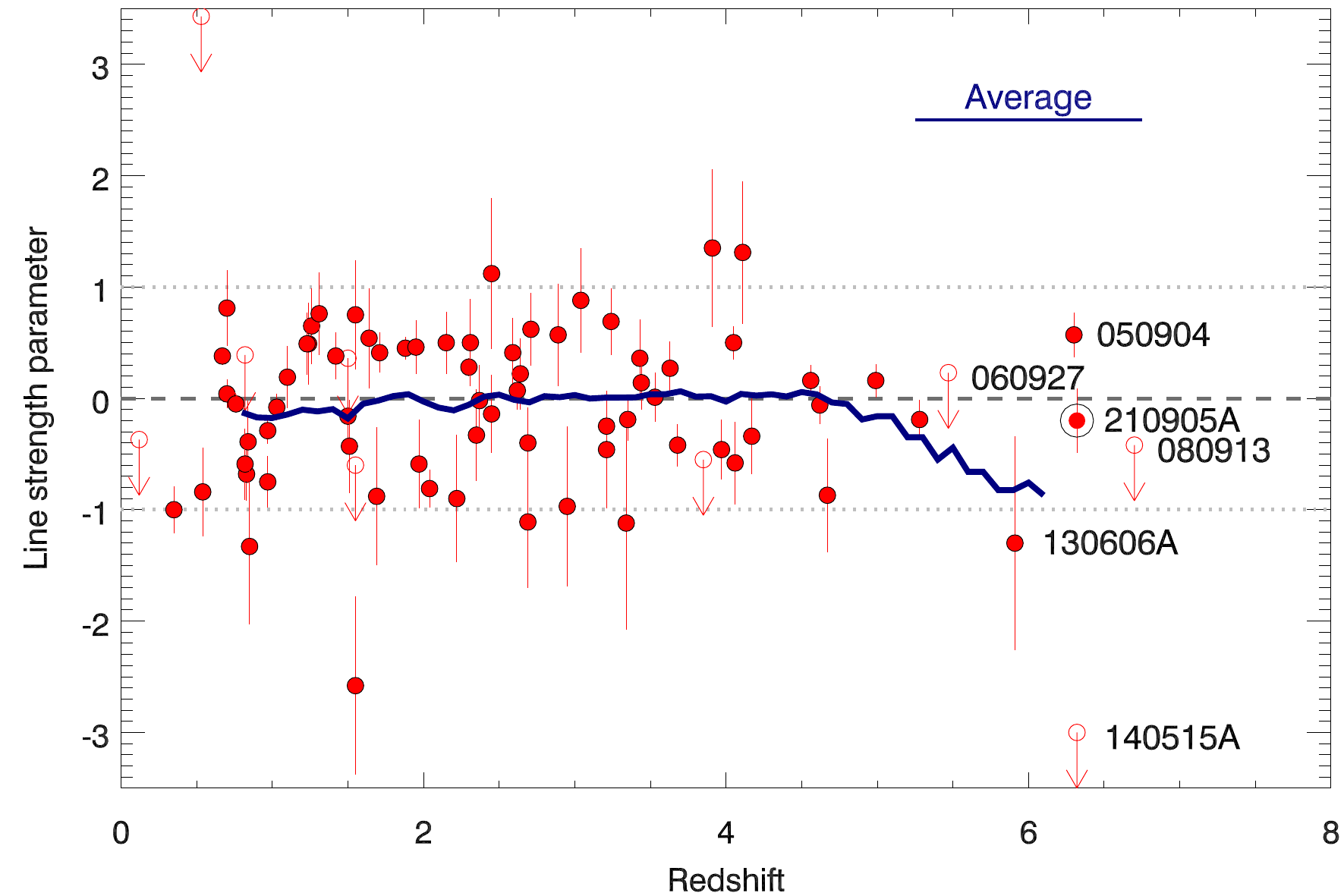}
      \caption{Evolution of the LSP with redshift. The blue line represents a rolling average of the LSP with a window of 1.5 in redshift, as indicated in the plot. The limits below zero have been considered as detections for the calculation of this average, which makes it, strictly speaking, an upper limit.
              }
         \label{fig:LSP_z}
   \end{figure}

The spectroscopic observation of high-redshift GRBs can also tell us about the evolution of the host environments in which GRBs explode. Fig.~\ref{fig:LSP_z} plots the LSP with varying redshift, using the sample of \cite{deUgartePostigo2012} which has been complemented with nine new measurements of GRB spectra above a redshift of 4, including GRB\,210905A. The data were obtained from several sources \citep{Kawai2006,Thone2013,Sparre2014,Jeong2014,Hartoog2015,Melandri2015,Selsing2019,deugartepostigo2020} and are measured using the tools available at GRBSpec\footnote{\url{http://grbspec.eu}\,.} \citep{deUgartePostigo2014,Blazek2020}.

In the paper of \cite{deUgartePostigo2012}, with a very reduced sample of $z>4$ GRBs, the authors already mentioned a possible decline of the LSP at high redshifts, which is in agreement with what we find when adding our complementary sample. The decline begins to be significant at $z>5$, albeit with small number statistics, and could be due both to smaller host galaxies in the early universe and to lower metallicities, as we approach an earlier generations of stars. Further details and analysis will be given in a forthcoming work (de Ugarte Postigo et al. in prep).

\subsection{Fine-structure lines and cloud distances}
\label{fine}

Fine-structure lines are commonly found in GRB afterglow spectra and have been proven to be due to indirect UV pumping excitation produced by the GRB radiation \citep{Vreeswijk2007,Vreeswijk2013,Prochaska2006}. This happens when the afterglow UV photons excite the absorber atoms and ions to a principal quantum number state above the fundamental one, and then, by a spontaneous emission of lower energy photons, the fine-structure lines of the fundamental state are populated.
Under this assumption, we can use the fine-structure lines to infer the distance of the corresponding gas clouds from the GRB \citep{Vreeswijk2007,Vreeswijk2013,DElia2009}.
The details of the photoionization code 
used can be found in \cite{DElia2009}, and references therein. The inputs to the code are the incident flux coming from the GRB afterglow as a function of time and frequency, and the initial column densities of the components analyzed.
In our case, \cii{} fine-structure lines are observed only for components $V$ and $VI$, where the corresponding ground state line can only be constrained with a lower limit. 
No \feii{} fine-structure lines are observed in this GRB, despite being found very commonly in GRB absorption spectra (see e.g., \citealt{DElia2009,DElia2014}). In fact, taking into account the \feii{} column densities measured in the GRB\,210905A spectrum, and assuming for \feii{} the same ground-based and fine-structure line column density ratio as \siii{} and \cii{}, would result in absorption lines too weak to be detected in our spectrum.
Thus, the only usable lines to constrain the distance between the GRB and the absorber are the \siii{} ground state and its fine-structure level. These absorption features are detected for components {$II$, $III$, $V$} and {$VI$}. 

The initial column densities of the four components analyzed were obtained summing the \siii{} and \siii{}{*} values in Tab.\,\ref{table_N} and assuming that, before the GRB afterglow switches on, all the \siii{} is in the ground state. The afterglow flux was taken from the companion paper \cite{Rossi2022}, adopting in particular power law indices $ \alpha=0.695$ (temporal decay at t$<1$ day) and  $ \beta = 0.6$ (spectral slope).

Intriguingly, the distances derived for the GRB\,210905A components are the highest ever recorded for a GRB showing excited transitions. In detail, we found $d_{II} \sim 11$\,kpc, $d_{III} \sim 7$\,kpc, $d_{V} \sim 17$\,kpc, and $d_{VI} \sim 16 $\,kpc.  These values must be considered as rough estimates of the distances, for two reasons. Firstly, they are derived using just two abundances for each component, that are the \siii{} ground state level and its first fine-structure level. Given that the total column density is computed from the data, and that the distance is a free parameter of the calculation, our measurements cannot be strongly constrained. Secondly, given the limited X-shooter spectral resolution, it may happen that the \siii{} ground state column density of the gas involved in the UV pumping process may be blended with other components with similar velocities. However, even artificially assuming lower {\it b} parameters, the column density of the ground state would increase by 0.1-0.2 dex only. The fine-structure lines are far from being saturated and therefore their column density values are robust. 
In addition, these high distance values are in agreement with the nondetection of the \feii{} fine-structure line system, commonly found in many GRB afterglow spectra.
In order not to see these common transitions, the absorber must be several kiloparsecs away from the GRB explosion site. From the \feii$^*${} column density lower limit that we can estimate from our data, we can only establish a lower limit on the distance of 2.2\,kpc.

\begin{figure}[h]
   \centering
   \includegraphics[width=0.48\textwidth]{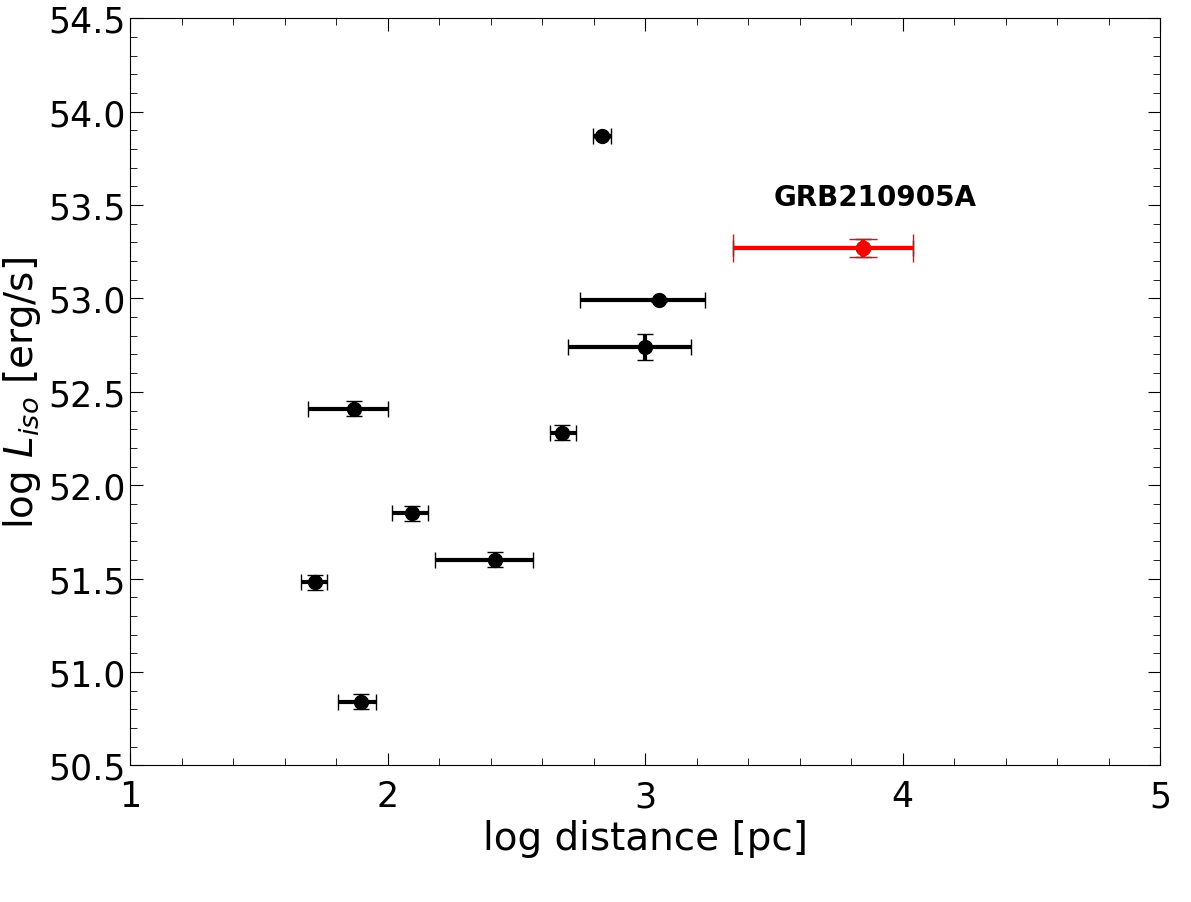}
      \caption{Relation between luminosity ($L_{\rm iso}$) and distance to the closest absorbers exhibiting fine-structure transitions for GRB sight lines,
      where relevant information is available (black points and error bars: GRB\,020813, GRB\,050730, GRB\,060206, GRB\,060418, GRB\,080310, GRB\,080319B, GRB\,080330, GRB\,081008, GRB\,090926A \citealt{Vreeswijk2012,Ghirlanda2018}; red point and error bar: GRB\,210905A). For GRB\,210905A, we used $L_{\rm iso}$ from \cite{Rossi2022} and error bars determined conservatively by the lower limit obtained from iron (2.2\,kpc) and the distance of the second cloud (11\,kpc). The correlation is found independently of that. }
        \label{Liso-d}
   \end{figure}

The distances determined should correspond to the closest low-ionization gas clouds from the GRBs. The large distances reported above are therefore surprising (taking into account the fact that GRBs explode in star-forming regions of galaxies), 
because they would imply the lack of such gas clouds along the GRB line of sight up to several kiloparsecs. This can be accommodated if we suppose that the GRB was able to ionize its host ISM gas up to such large distances, and therefore the GRB afterglow UV photons would not cross any closer low-ionization gas clouds with high enough column density to be able to produce the observed fine-structure transitions. Another possibility is that the GRB happened at the edge of a galaxy disk toward the observer and that the transitions we are observing come from a foreground galaxy. However, the lack of HI absorption detection at higher redshift in the X-shooter spectra (event with low column densities) makes this hypothesis unlikely. In the first hypothesis, we should see in general a relation among the GRB ionizing flux and the distance of the low-ionization gas clouds as determined by the fine-structure line analysis. Indeed, Fig. \ref{Liso-d} shows a clear correlation ($\rho_\text{Spearman}=0.78^{+0.10}_{-0.10}$, see Appendix Fig. \ref{Correlation}) between the distance of the closest low-ionization gas clouds and the GRB luminosity (L$_{\rm iso}$), for GRBs having these quantities determined in the past.

In \cite{Krongold2013}, time dependent photo-ionization models are developed using average GRB parameters to determine the number of ionizing photons, and assuming some preexisting ionization, as expected in young star-forming regions. Following \cite{Rossi2022}, GRB\,210905A is an extremely energetic burst (among the 7\% brightest). Using the burst luminosity and the spectral and temporal parameters published in their paper, we determined a number of ionizing photons $\sim30$ times higher than the average value assumed in \cite{Krongold2013}. Even if a detailed calculation is beyond the scope of our paper, looking at the \cite{Krongold2013} results, it is easy to infer qualitatively that at such a high number of ionizing photons a significant fraction of silicon may be ionized to \siiv{} or higher ionization states to distances of several hundreds of parsecs at the time of our observations (1244\,s rest-frame), under  plausible conditions implying some preionization (up to 100\,pc, see \citealt{Whalen2008,Watson2013}) and electron densities of 1-10$^2$\,cm$^{-3}$. This would explain the absence of low-ionization gas clouds closer to the GRB and the L$_{\rm iso}$-distance correlation. This scenario is also in agreement with the large $N{_{\rm H}}$ value determined from the X-ray data \citep{Rossi2022} compared to the \hi{} column density ($7.7^{+3.6}_{-3.2}\times 10^{22}$\,cm$^{-2}$ and $1.26^{+0.3}_{-0.3}\times 10^{21}$\,cm$^{-2}$, respectively). Such discrepancy is often found in GRBs (see i.e., \citealt{Watson2007,Schady2011}). However, considering the high redshift, large $N{_{\rm H}}$ can also be due to the amount of metals present along the line of sight \citep{Behar2011,Campana2015,Dalton2021}.

On the face of it, a size of $\sim15$\,kpc is large compared to typical observed sizes of even bright galaxies at $z\sim6$ \citep[e.g.,][]{Shibuya2015,Yang2022}. On the other hand, such sizes are based on the UV brightest regions, whereas gas is likely to be distributed on larger scales, as also found in hydrodynamic simulations of massive reionization era haloes \cite[e.g.,][]{Rosdahl2018}.
It is also possible that the GRB host is part of a larger complex of galaxies (see also the HST image of the field and Sect.\,\ref{lya} discussion). In this case, components $V$ and $VI$ must be infalling on the GRB host so as to explain their being redshifted compared to the components closest to the GRB ($II$ and $III$).
Ultimately, the answer may come only from deep photometric imaging, as discussed in Sect.\,\ref{lya}.

Finally, we notice that despite showing high-column density of low-ionization gas, component {\it I} does not have any associated fine-structure absorption. Therefore we associate this component with ISM at even larger distances than the others. In Sects.\,\ref{metal}, \ref{DTM} and \ref{sectnucleo}, we determine global properties of the whole absorbing gas complex, as well as those obtained by a component-by-component analysis.

\subsection{\nv}
\label{nv}

Another interesting feature in the X-shooter spectrum is the presence of a strong, narrow \nv{} absorption line. Such \nv{} absorption in GRB afterglow spectra has been associated with gas very close to the GRB site ($\sim$tens of parsecs; \citealt{Prochaska2008,Fox2008,Heintz2018}).
The \nv{} column density value we determined is typical of those of GRB afterglows ($\log N(\nv)_{\rm med}/\text{cm}^{-2} = 14.50$). The \nv{} narrow component is coincident with the strongest component of the other high-ionization lines. At the corresponding velocity, the low-ionization line fit does not require any components and there are no fine-structure lines. We can interpret this \nv{} cloud as a high-ionization cloud between the GRB and the low-ionization clouds. It is not possible to determine if it is associated with the GRB circumburst environment, but the high number of ionizing photons  and the consideration discussed in Sect.\,\ref{fine} may place it at $\sim100$\,pc at least. 
It is also possible that the \nv{} absorption is due to a high-ionization cloud present between systems A and B, and it is completely unrelated to the GRB environment and ionization. However, in this case, the fact that its profile and column density are typical of other GRB spectra would be a coincidence.

\subsection{Metallicity and dust depletion}
\label{metal}
We use the value $\log (N$(H\,{\sc i})/cm$^{-2}) = 21.10\pm 0.10$ determined by Fausey et al. in prep and the metal column densities measured in this work to obtain the metallicity of the host along the line of sight to the GRB. The values are reported in Table\,\ref{metallicity}. They are comparable with those found for GRB\,050904 ([Si/H] $=-1.6\pm0.3$; \citealt{Thone2013}) at $z=6.3$ and for GRB\,130606A ([Si/H] $=-1.3\pm0.08$, [Fe/H] $=-2.09\pm0.08$; \citealt{Hartoog2015}) at $z=5.9$
(the only two other very high-redshift GRBs for which spectroscopic data are available to determine metallicity), and with those typically found for GRB host galaxies \citep[e.g.,][]{Sparre2014,Wiseman2017,Bolmer2019}.

\begin{table}[h]
\caption{Metal abundances. For each element (first column), the total column density (second column), the ratio over iron (third column) and the metallicity are reported.}      
\label{metallicity} 
\centering
\begin{tabular}{c c c c}       
\hline\hline
$X$ & $\log(N/{\rm cm}{^{-2}})$ & [$X$/Fe] & [X/H] \\
\hline
C & $>16.02$ & $>0.66$ & $>-1.5$\\
O & $>16.03$ & $>0.41$ & $>-1.8$\\
Mg & $>15.13$ & $>0.61$ & $>-1.6$\\
Al & $>13.69$ & $>0.31$ & $>-1.8$\\
Si & $14.90\pm0.02$ & $0.46\pm0.04$ & $-1.71\pm0.11$\\
S & $>14.50$ & $>0.45$ & $>-1.7$\\
Fe & $14.43\pm0.02$ & & $-2.17\pm0.11$\\
\hline
\hline
\end{tabular}
\end{table}

\begin{figure}
\centering
\includegraphics[width=0.5\textwidth]{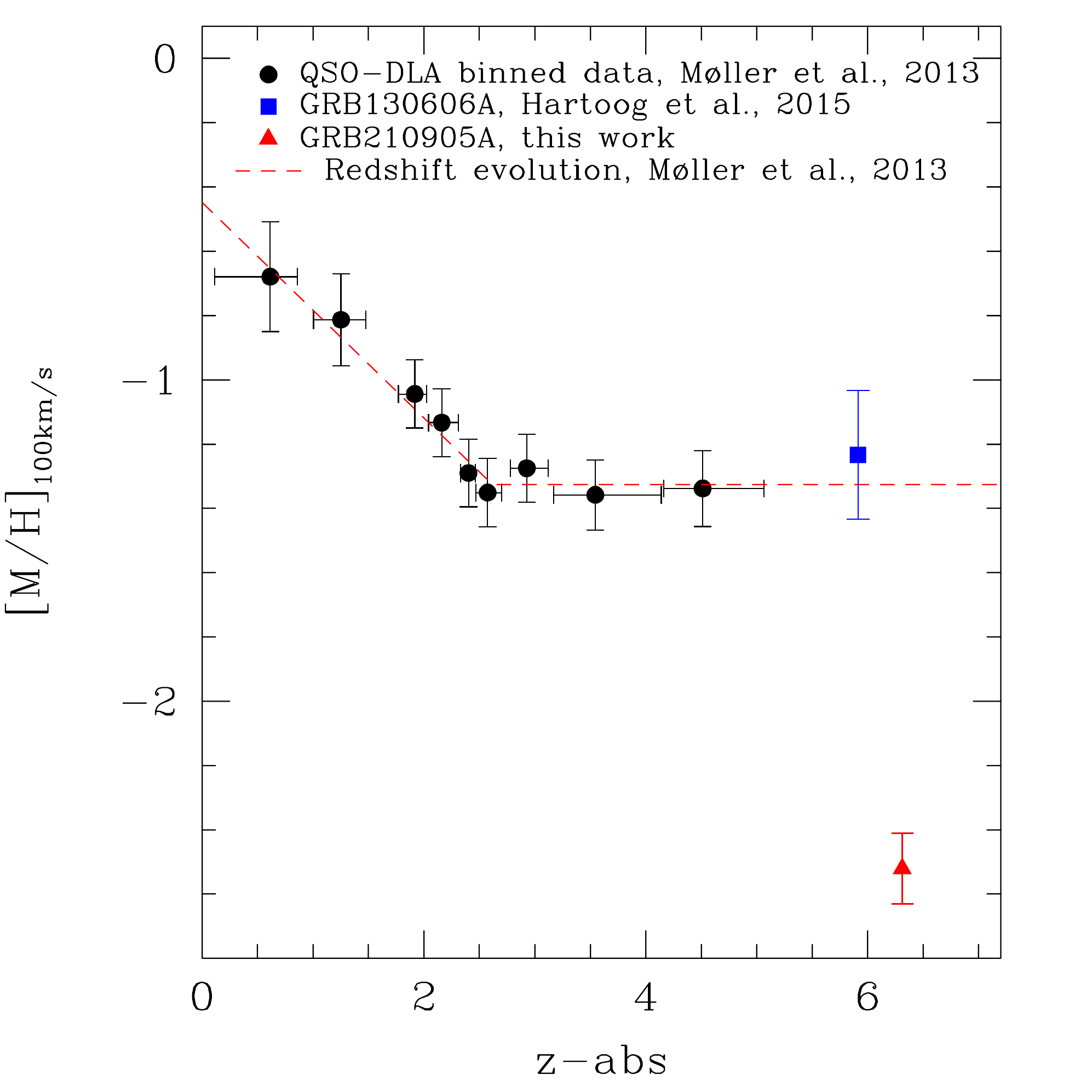}
\caption{QSO-DLAs binned data from \cite{Moller2013} with the high-redshift measurement of GRB\,130606A \citep{Hartoog2015} and GRB\,210905A included. $\text{[M/H]}_{\rm 100\,km~s^{-1}}$ is the metallicity expected for DLAs with $\Delta V_{90}=100$\,km\,s$^{-1}$ (zero-point normalization).
The data of \cite{Moller2013} (filled
circles) show a break in the evolution of the mass-metallicity relation at $z\sim2.6$ (dashed line), which is not confirmed by our measurement of GRB\,210905A (filled triangle). We assumed for this point $\Delta V_{90}=359$\,km\,s$^{-1}$ and $\text{[M/H]} = -1.71\pm0.11$, as explained in the text.}
\label{metallicity_redshift}
\end{figure}

We determine the $\Delta V_{90}$ as defined by \cite{Ledoux2006} to place the GRB\,210905A values in the velocity-metallicity relation \citep{Ledoux2006,Prochaska2008, Arabsalmani2015,Moller2013}. From \siii{}$\lambda1304$\,\AA{} and \feii{}$\lambda2344$\,\AA{} we determined $\Delta V_{90}=358$\,km\,s$^{-1}$ and $\Delta V_{90}=367$\,km\,s$^{-1}$, respectively. Those values are at the high end of the $\Delta V_{90}$ found for QSO- and GRB-DLAs, but other systems were found in the past with similar or larger values at lower redshift.
We compare the GRB\,210905A values with the $\Delta V_{90}$-metallicity relation \citep{Arabsalmani2015} and to the so-called mass-metallicity relation for QSO-DLAs presented by \cite{Moller2013}. This last work studies also the redshift evolution of this relation and it is of great interest to enlarge the sample to constrain it at very high redshift.
To perform the comparison, we use the \feii{} $\Delta V_{90}$ and $\text{[M/H]}=-1.71\pm0.11$, taken from silicon since the metallicity of iron is greatly influenced by dust depletion. We applied a correction for the spectral resolution of our data, but this is negligible in our case ($\Delta V_{90}=359$\,km\,s$^{-1}$ instead of $\Delta V_{90}=367$\,km\,s$^{-1}$). 
GRB\,210905A is at the lower end of the dispersion of the $\Delta V_{90}$ - metallicity relation \citep{Arabsalmani2015}, but still consistent with it, considering the errors on the metallicity measurement. 
The GRB\,210905A galaxy complex lies 1.2 dex below the \cite{Moller2013} relation (see Fig. \ref{metallicity_redshift}), which is at $3.0\,\sigma$ from the relation, taking into account the [M/H] measurement uncertainty. This can indicate that the scaling relation is not in place at such high redshift or that clearly there are still too few points to constrain its evolution at $z>5$ (but see also Sect.\,\ref{lya}).

Following the relation reported by \cite{Neeleman2013}, which is less steep and without a redshift break with respect to \cite{Moller2013}, we find a better agreement for GRB\,210905A. This work indeed predicts a metallicity, $\text{[M/H]} = -2.03\pm0.82$, which is more consistent with our measurement despite the wide dispersion.

  \begin{figure*}
   \centering
   \includegraphics[width=0.45\textwidth]{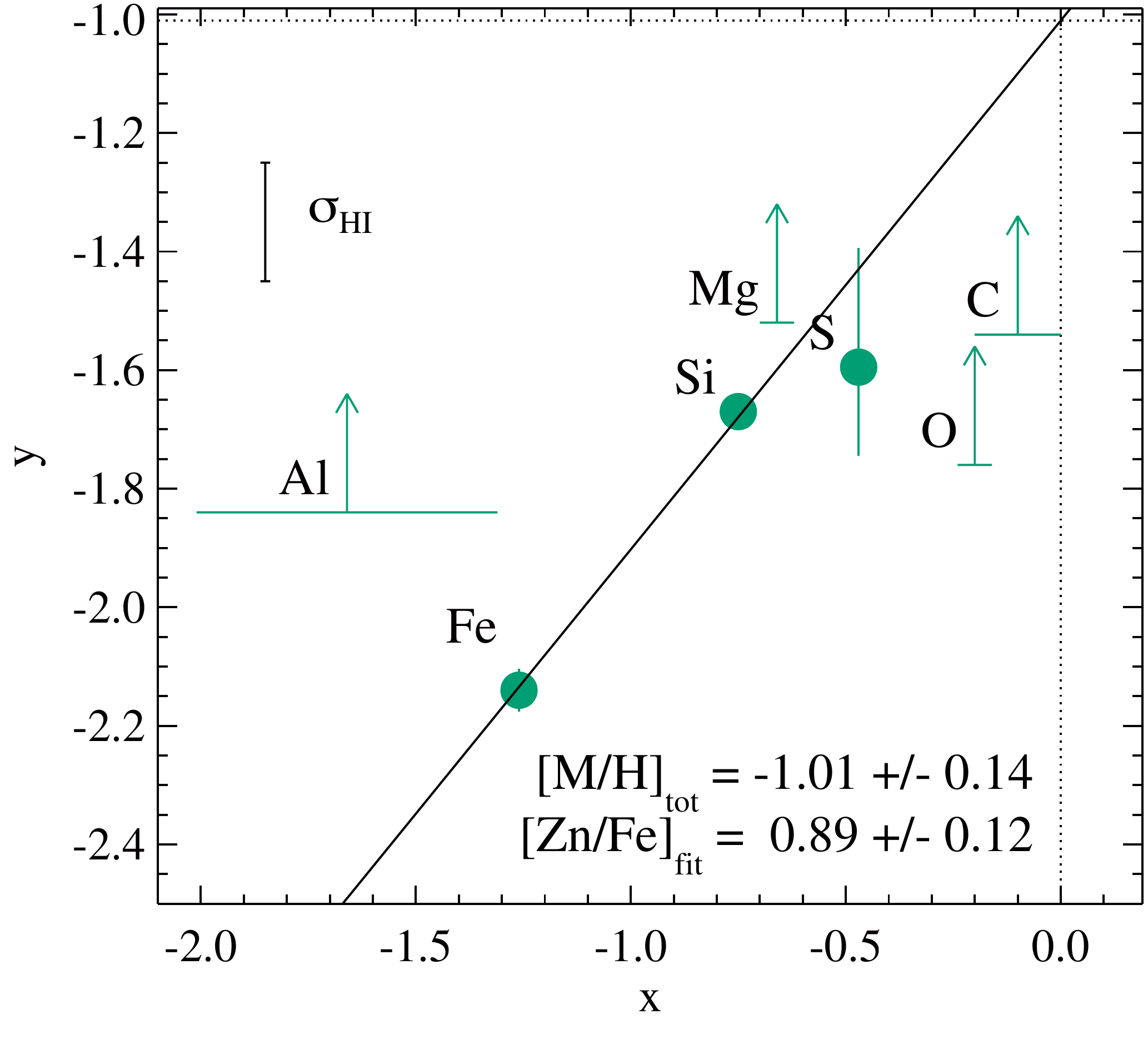}
   \includegraphics[width=0.45\textwidth]{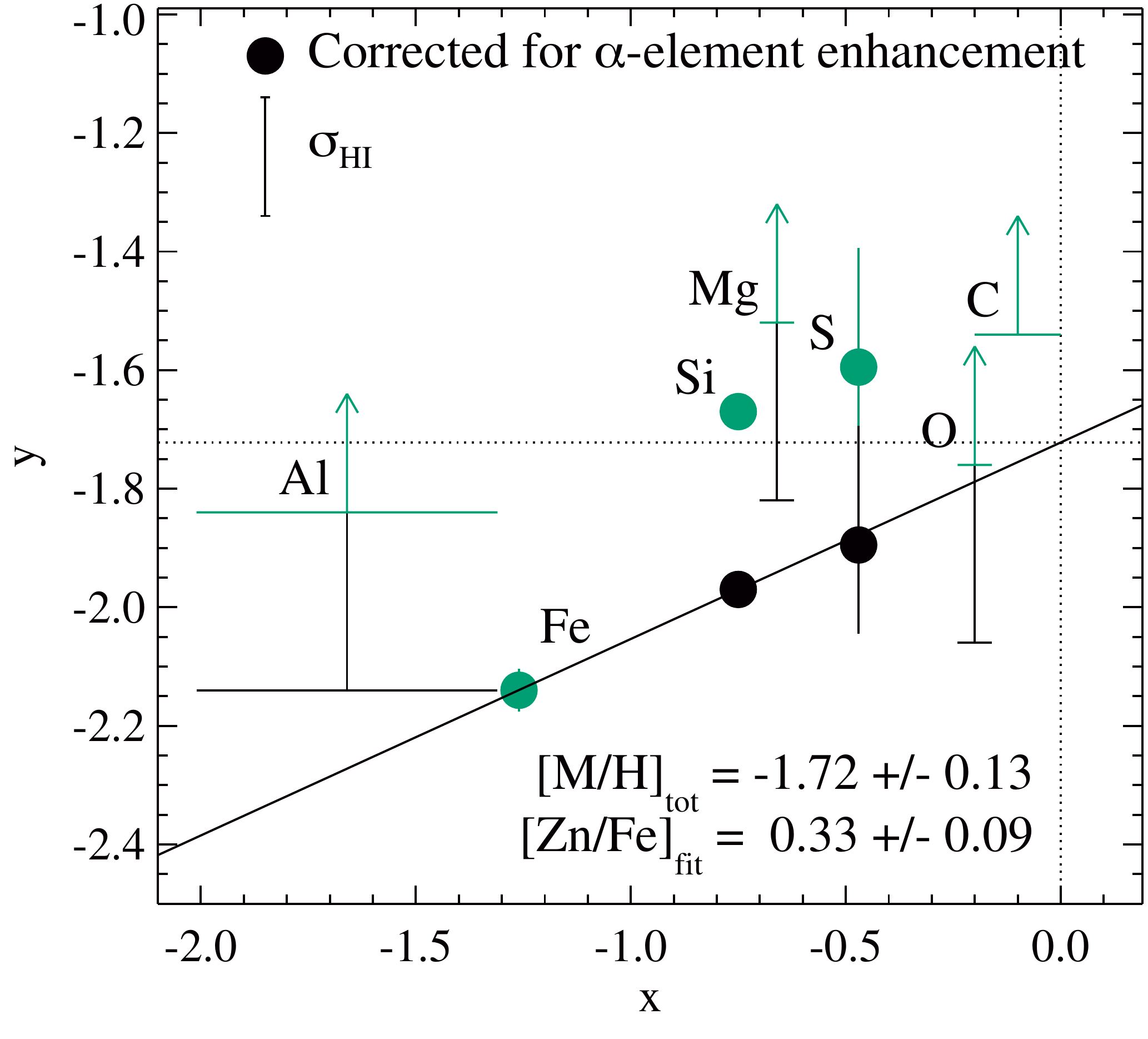}
      \caption{Abundance pattern observed in the host of GRB\,210905A from the total absorption-line profile.
      The linear fit to the data (solid line) determines the total metallicity $\text{[M/H]}_{\rm tot}$ (intercept) and overall strength of depletion [Zn/Fe]$_{\rm fit}$ (slope), as labeled. See Sect.\,\ref{metal} for a full description. The right panel represents the fit obtained after correcting the relevant elements for $\alpha$-element enhancement of 0.3 dex (black points and limits). 
      }
         \label{depletion tot}
   \end{figure*}

We analyze the abundances and relative abundances of different metals with the aim of characterizing the chemical enrichment in the GRB host galaxy. First, we use the measurements of the column density of \hi{}, \alii{}, \feii{}, \siii{}, \mgii{}, \sii{}, \oi{}, and \cii{} for the whole line profile, listed in Table \ref{metallicity}, assuming \cite{Asplund2009} solar abundances. 
Concerning sulfur, in Table \ref{metallicity} we report only the measurements of the components $I$, $II$, and $III$ since the $IV, V, VI$ components of \sii{}$\lambda1259$\,\AA{} are blended with the \siii{}$\lambda1260$\,\AA{} absorption.
To calculate the total sulfur column density for this analysis, we considered the similar ratio of the column densities of different elements of system A and B (see Sect.\,\ref{z63}) and we estimate the sulfur column density of system B by applying this ratio to the sulfur column density of system A. To be very conservative, we fixed as the lowest possible value of the total sulfur column density that of system A, and as the uppermost value twice the sulfur column density of system A and its 1\,$\sigma$ error.
We do not detect \znii{}, and we can only determine an upper limit for its column density of $\log N( \znii)/\text{cm}^{-2} < 12.5$, too poorly constrained to be used in the following analysis.  In Fig. \ref{depletion tot} we present the observed abundance patterns along the total line of sight of the $z=6.3$ galaxy. In this formalism, $x$ represents how refractory an element is (i.e., how easily it is normally incorporated into dust grains), and $y$ is closely related to the abundances of different metals, as defined and tabulated in \cite{DeCia2021}, except for carbon and aluminum, which are measured in \cite{Konstantopoulou2022}. If there is no dust depletion in a system, the observed abundances are expected to line up horizontally. If there is dust depletion, the abundances are expected to line up in a linear relation, the steeper the higher amount of dust. The slope of this linear relation is the parameter [Zn/Fe]$_{\rm fit}$, which represents how strong the overall dust depletion is, based on the observations of all metals. The $y$-intercept at $x=0$ is the total metallicity (in the gas + dust phase) of the system. In addition, any deviations from this straight line could be due to nucleosynthesis, for example $\alpha$-element enhancements, or other peculiar abundances. 

\begin{table*}
\centering
\caption{Properties derived from the total metal abundances and component by component. The total metallicity ($\text{[M/H]}_{\rm tot}$), dust depletion [Zn/Fe]$_{\rm fit}$ and dust-to-metal ratio ($DTM$) are reported for the analysis performed taking $\alpha$-element enhancement into account.}
\label{table_results_alpha}
\begin{tabular}{c c c c c c c c}       
\hline\hline
With $\alpha$-element corr.  & $I$ & $II$ & $III$ & $IV$ & $V$ & $VI$ & Tot\\
\hline
\rule{0pt}{3ex}
[M/H]$_{\rm tot}$ & & & & & & & $-1.72\pm0.13$\\[2ex]

[Zn/Fe]$_{\rm fit}$&
$0.00^{+0.11}_{-0.00}$&
$0.53^{+0.09}_{-0.09}$&
$0.57^{+0.14}_{-0.14}$&
$0.00^{+0.17}_{-0.00}$&
$0.00^{+0.21}_{-0.00}$&
$0.53^{+0.10}_{-0.10}$&
$0.33\pm0.09$\\[2ex]

$DTM$ & 
$0.00^{+0.04}_{-0.00}$& 
$0.26^{+0.03}_{-0.03}$&
$0.27^{+0.04}_{-0.04}$&
$0.00^{+0.06}_{-0.00}$&
$0.00^{+0.07}_{-0.00}$&
$0.26^{+0.03}_{-0.03}$&
$0.18\pm0.03$\\[1ex]

\hline\hline
\end{tabular}
\end{table*}

For GRB\,210905A we observe an abundance pattern that, despite the large uncertainties, is roughly described with a linear relation\footnote{We fit a linear relation to the constrained data in Fig. \ref{depletion tot}, and not including constraints from the limits. We obtain a best $\chi^2$ lower than one, because of the large uncertainties in the data. However we note that the y-axis uncertainties are dominated by the uncertainty in the column density of \hi{}, and is the same for all the data points. 
The component-by-component analysis (see below) confirms our results, but without the large \hi{} uncertainties.}, as shown in Fig. \ref{depletion tot}, indicating significant dust depletion, [Zn/Fe]$_{\rm fit}=0.89\pm0.12$ and a dust-corrected metallicity of $[\rm M/\mbox{H}]_{\rm tot}=-1.01\pm0.14$. Such a value is placed roughly in the upper envelope of the fit of the total metallicity evolution of the population of GRB hosts and QSO-DLAs at $2<z<4$ \citep{DeCia2018}, but is higher than the values determined at $z>4$.
The depletion factor found, which can be used as an indication of the overall chemical enrichment of the system, is at the high-end of the values of the population of GRB hosts and QSO-DLAs at $2<z<4$, at similar metallicity \citep{DeCia2016}, and higher than the values found for GRB hosts at $z>4$ (though the statistics are limited; \citealt{Bolmer2019}).

We also perform a component-by-component analysis of the depletion patterns (see Table \ref{table_N} and Fig. \ref{depletion comp}), with the caveat that some components may not be associated with neutral gas.
The individual components of gas toward GRB\,210905A show diversity in their strength of depletion, and therefore their chemical enrichment. Components $II$, $III$, and $VI$ have very high depletion levels
([Zn/Fe]$_{\rm fit}\gtrsim1$) that are difficult to fit with dust production (and destruction) models within such a relatively short amount of time (1\,Gyr).

\begin{figure*}
\centering
\includegraphics[width=0.8\textwidth]{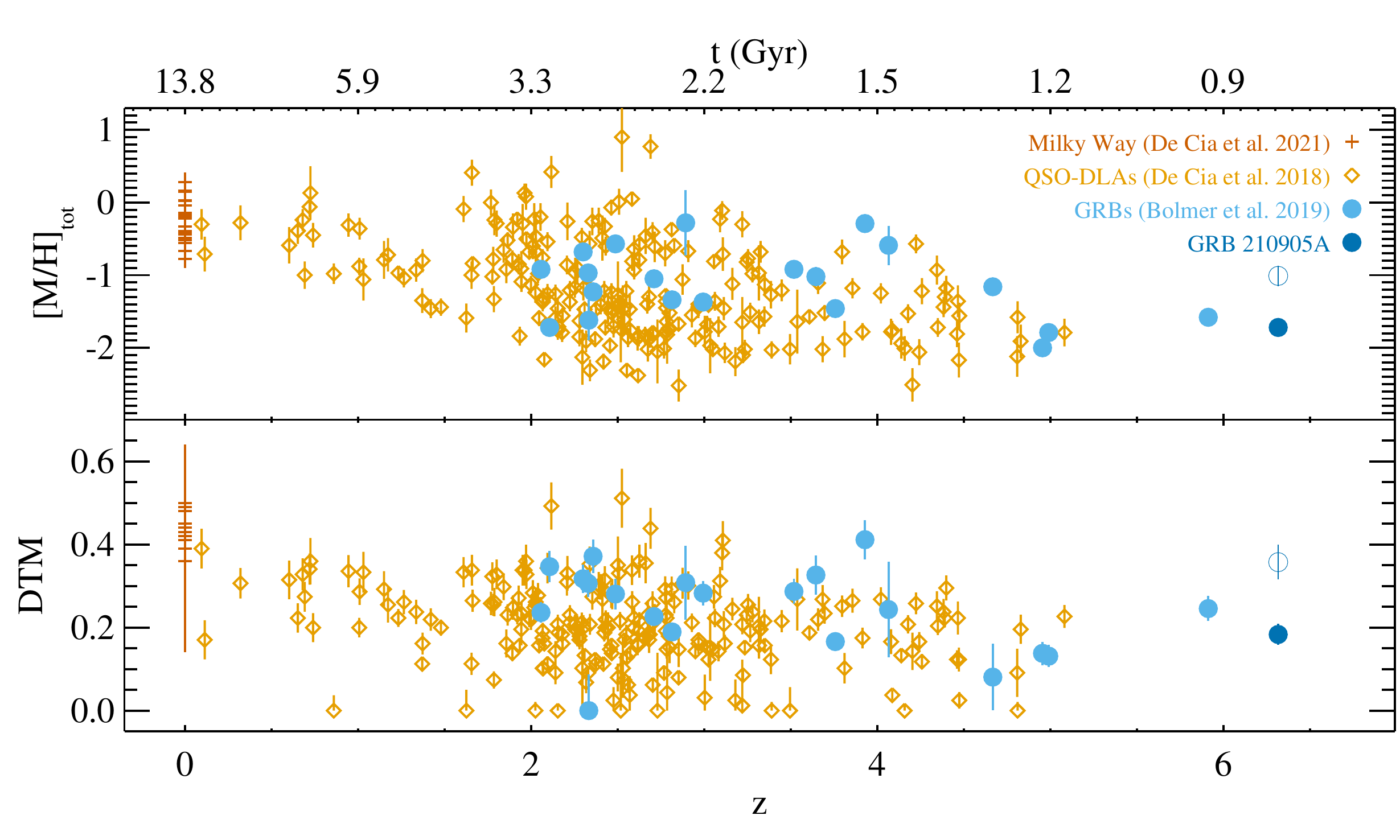}
\caption{Evolution of [M/H]$_{\rm tot}$ (top panel) and $DTM$ (bottom panel) over cosmic time of the MW and of QSO- and GRB-DLAs. GRB\,210905A values are shown in dark blue dots (the empty circles refer to the values determined not considering $\alpha$-element enhancement).}
\label{MH_DTM_z}
\end{figure*}

In the analysis above, we did not take the effects of potential $\alpha$-element enhancement into account. With the current data, it is not possible to characterize and disentangle metallicity, dust depletion, and nucleosynthesis simultaneously. However, we may expect $\alpha$-elements to be enhanced in the host galaxy of GRB 210905A. At $z\sim6$ the universe is about 1\,Gyr old, and therefore we expect that there was no time for SNe Type Ia to have exploded in the host galaxy,
thus we could expect an enhancement of about a factor of two with respect to iron \citep{Tinsley1979,McWilliams1997}. We therefore repeat the analysis assuming an $\alpha$-element enhancement in the host galaxy of 0.3~dex, as such amount of enhancement is measured in stellar abundances at low metallicity (e.g., \citealt{Tinsley1979,McWilliams1997}), and have been observed in galaxies at $z\sim3$ (e.g., \citealt{Steidel2016,Cullen2021}) and QSO-DLAs \citep{Dessauges-Zavadsky2006,Cooke2011,Becker2012,Ledoux2006,DeCia2016}.
This assumption would significantly lower the amount of dust that we derive, as well as the ISM metallicity.
The right panel of Fig.\,\ref{depletion tot} shows that, after assuming a decrease of 0.3 dex in the amount of oxygen, sulfur, silicon, magnesium, and, conservatively, also of aluminum, due to $\alpha$-element enhancement, we derive a lower metallicity [M/H]$_{\rm tot} = -1.72 \pm 0.13$ and a lower overall amount of dust depletion [Zn/Fe]$_{\rm fit} = 0.33 \pm 0.09$. This is consistent with the average values observed in DLAs. 
The assumption of $\alpha$-element enhancement results in better linear fits (in terms of $\chi^2$).
In light of all the aforementioned considerations, and of the reasonable dust depletion value for a ~1 Gyr old galaxy, in the following, we consider $\alpha$-element enhancement as the favored scenario for our analysis and the results we report are obtained in this context (unless otherwise mentioned). Table \ref{table_results_alpha} reports the results for the global values and for the component-by component analysis.

In Fig. \ref{MH_DTM_z} (upper panel) we compare the metallicity that we derive for GRB\,210905A with other GRB- and QSO DLAs \citep{Bolmer2019,DeCia2018}.  GRB\,210905A (as well as GRB\,130606A at $z\sim6$) is above the extrapolation at higher $z$ of the evolution of metallicity of QSO-DLAs with cosmic time ($\text{[M/H]}_{\rm tot} = -2.3$ at $z=6.3$; \citealt{DeCia2018}) but within its scatter. $z\sim6$ GRB host values are not consistent with the drop of metallicity suggested by \cite{Rafelski2014}, and may indicate a lack of quick evolution of chemical enrichment (at least for GRB host galaxies) at very high redshift. Of course the statistics do not allow any conclusion yet, but it is clear that GRBs are key tools to investigate the chemical enrichment at very high redshift.
It is meaningful to compare GRB hosts to QSO-DLAs, because they both correspond to galaxies selected independently of their luminosity that may belong to the same population (\citealt{Prochaska2007b,Fynbo2008a}; Krogager et al. in prep). However, GRB afterglows probe the central star-forming regions of their host galaxies, whereas QSO-DLAs probe on average gas at larger impact parameter from the galaxy centers \citep{Fynbo2008a}.

\subsection{Dust-to-metal mass ratio and dust extinction}
\label{DTM}

We derive the dust-to-metal mass ratio ($DTM$), that is the ratio between the mass of dust and the total mass of the metals, as follows: 
\begin{equation}
    DTM = \frac{\sum_{X}   (1-10^{\delta_X}) \times 10^{\mathrm{[X/H]}_\odot}  \times W_X}{\sum_{X}(10^{\mathrm{[X/H]}_\odot}\times W_X)} 
\end{equation}
where the metal depletions $\delta_X$ are derived from the [Zn/Fe]$_{\rm fit}$ following \citet{DeCia2016}, $\mbox{[X/H]}_\odot$ are the solar abundances $X_\odot$ - 12, $W_X$ are the atomic weights of the metals considered. In practice, we include all the elements with an elemental abundance above 3 (in the scale where H has an abundance of 12, \citealt{Asplund2009}), in the same way as \cite{Konstantopoulou2022}.
In absolute values (i.e., to be compared with the 0.45 value of the Milky Way), we find $DTM=0.18\pm0.03$ ($0.36\pm0.04$ for the case without taking into account $\alpha$-element enhancement) for the total line profile. This means that, given the metal content in this galaxy, about 18\% (36\%) of its maximum potential amount of dust is already in place at $z\sim6$.

\begin{figure*}
\centering
\raisebox{2cm}{\includegraphics[width=0.3\textwidth]{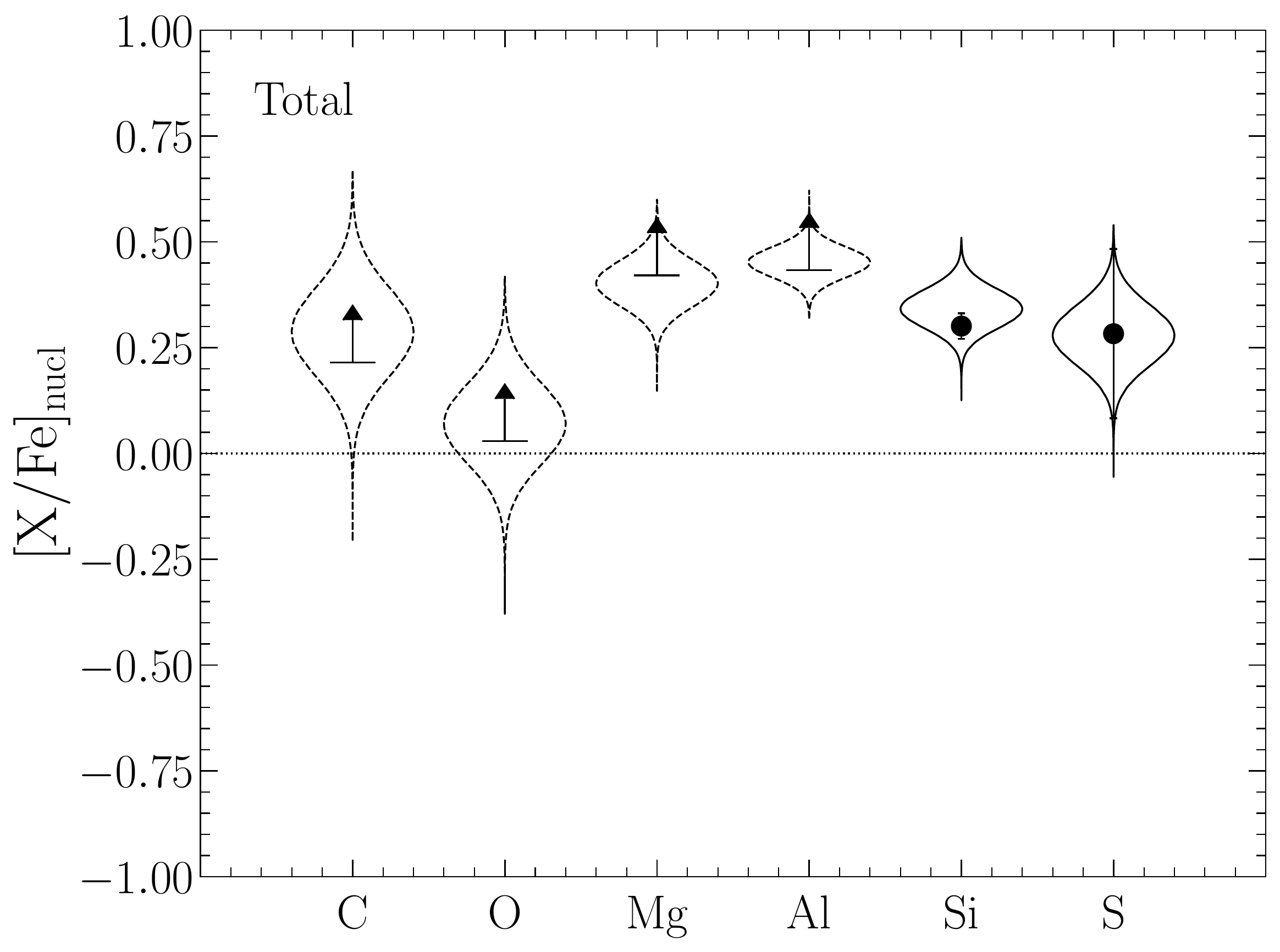}}
\hspace{0.5cm}
\includegraphics[width=0.65\textwidth]{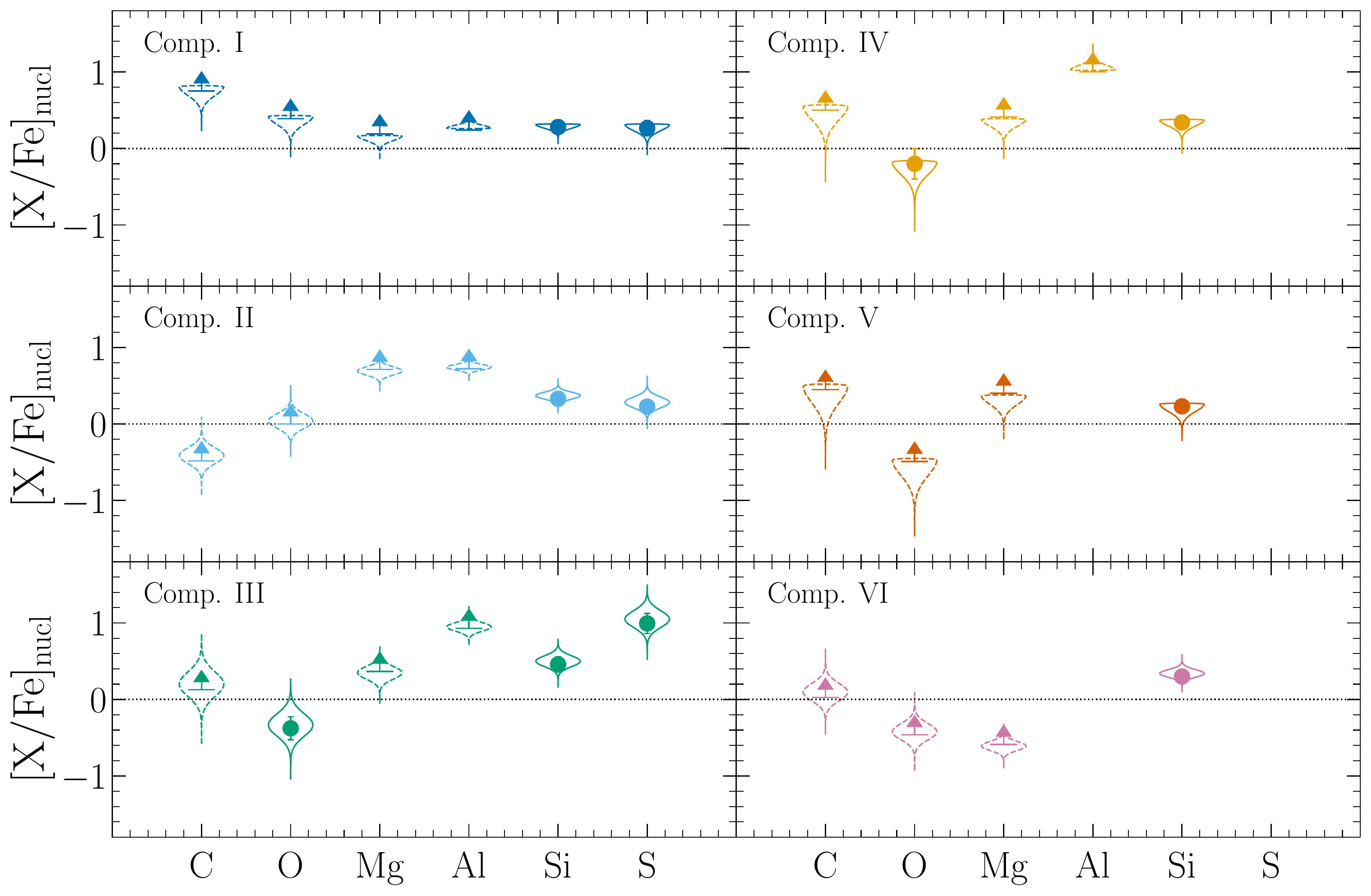}
\caption{Abundances
of different elements with respect to iron after correcting for dust depletion, for
the total (left panel) and component-by-component (right panel) analysis. Limits are indicated by arrows and the error bars represent the uncertainty propagated from column density measurements; the violin plots represents the uncertainty caused by dust depletion estimated using one million Monte Carlo realizations of the [Zn/Fe]$_{\rm fit}$. The impact of dust depletion on the nuclear abundances is correlated between elements; i.e., a higher dust depletion correction lowers all [$X$/Fe] values except for [Al/Fe] which it raises, and vice-versa.}
\label{multicomp_alpha}
\end{figure*}

This may be seen as a rather high value (considering that the Universe is about 1~Gyr old at this redshift), giving constraints to the potential sources of dust production. However, the uncertainties are significant and include negligible $DTM$ values. Fig.\,\ref{MH_DTM_z}, bottom panel, shows the comparison of the $DTM$ of GRB\,210905A with the ones of other GRBs and QSO-DLAs.
Breaking this down in a component-by-component analysis, we obtain values consistent with 0 for components $I$, $IV$ and $V$, whereas the other components have higher values with $DTM\sim0.20-0.30$.

Following \cite{DeCia2016}, we estimate the dust extinction from the global $DTM$ to be null. This is consistent with the low amount of extinction ($A_{\rm V,SED}<0.03$~mag) needed to reproduce the afterglow spectral energy distribution (SED; \citealt{Rossi2022}).

Cosmic dust dimming (CDD($z$)) should be considered at redshifts higher than about $z>3$ \citep{zafar19}. The amount of neutral hydrogen per unit redshift decreases \citep{zafar13} while metallicity increases with cosmic time \citep{rafelski12}. These two effects cause total Fe (gas $+$ dust) per unit redshift to be constant throughout cosmic time \citep{zafar19}. Following the \cite{zafar19} method to estimate CDD($z$) for the sightline of GRB\,210905A, we find that there is a slight increase in extinction by intervening absorbers, that is $A_V=0.012$\,mag ($z=2.8296$) and $A_V=0.023$\,mag ($z=5.739$), resulting in cumulative extinction (dimming) of $A_V=0.03\pm0.02$\,mag at the redshift of GRB\,210905A. This is consistent with \cite{DeCia2016} suggesting negligible dust for this GRB and probably dust dimming induced by the effect of cosmic distance.

Table \ref{table_results_alpha} reports the global and the component by component $DTM$ and dust extinction values, while for the case that does not take into account $\alpha$-element corrections, the values are reported in Table \ref{table_results_noalpha}.

\subsection{Nucleosynthesis}
\label{sectnucleo}

The deviations from the linear fits of Figs. \ref{depletion tot} and \ref{depletion comp alpha} are likely due to the effects of nucleosynthesis, or peculiar abundances in the host ISM. Figure \ref{multicomp_alpha} shows the over- and underabundance of different elements with respect to iron due to nucleosynthesis, [$X$/Fe]$_{\rm nucl}$, that is after correcting for the dust depletion estimate, for the total and component-by-component analysis (the case without taking into account $\alpha$-element enhancement is shown in Fig. \ref{multicomp_noalpha}). 
The corresponding values and errors are reported in Tables \ref{table_abb} and \ref{table_abb_depl}. 

In general, the global observed pattern can be explained by nucleosynthesis due to core-collapse SNe and massive (S-)AGB stars (see e.g., Fig.\,6 of \citealt{Masseron2020} and references therein). 
Both in the global and component-by-component analysis there is no evidence of the typical {\it odd-even} pattern predicted for pair-instability supernovae nucleosynthesis (see e.g., \citealt{Salvadori2019} and references therein). 

Components $II$, $III$ and $IV$ exhibit aluminum overabundance. It is the second time that aluminum overabundance has been observed for a GRB at $z\sim6$, the other being GRB\,130606A \citep{Hartoog2015}, the only other high-$z$ case where a detailed analysis of the chemical abundances and relative abundances was possible.

In the SAGA Database \citep{Suda2008}, the stars having more similar [Al/Fe] values to those we determined are O-deficient stars with peculiar properties, the so-called Na-O anticorrelation \citep{Fulbright2007,Alves-Brito2010}, typical of some stars found in globular clusters and some dwarf galaxies, such as the
Sagittarius dwarf spheroidal galaxy (\citealt{Fulbright2007}, and references therein).
The peculiar pattern shown by these giant stars may be due to dredging-up material that has undergone nucleosynthetic
processing in H-burning shells, or to primordial contamination by massive stars. The best candidates for such a process are massive AGB stars and fast rotating (very) massive stars (see e.g., \citealt{Prantzos2007} and references therein). This last possibility is particularly intriguing as it could correspond to long GRB progenitor stars. 
Due to the errors affecting our measurements, and to the very limited elements for which we can determine [$X$/Fe]$_{\rm nucl}$, our aim here is only to show some potential qualitative similarity with the nucleosynthetic patterns of all the aforementioned stars. The development and analysis of models  matching the data is beyond the scope of this paper.
We stress also that, at least for components $III$ and $IV$, oxygen underabundance can be due to a high ionization of the gas (see Sect.\,\ref{fine}). Indeed, among the components for which we could constrain the distance from the GRB, component $III$ is the closest to the GRB (component $IV$ is too faint to show fine-structure lines, and therefore to be able to determine its distance from the GRB; see Sect.\,\ref{fine}).
\mgii{} lines (having a close ionization potential value) are saturated, and \mgi{} is at the very end of the spectrum, in a region that is too noisy to infer any detection and column density value. It is also interesting to note that oxygen is not underabundant in component $I$, that is the farthest from the GRB (see Sect.\,\ref{fine}).

\subsection{The [\cii]-to-\hi\ relative abundance}
\label{ciitohi}

The high quality afterglow spectrum of this GRB, along with the accurate measurement of, in particular, \cii$^*\lambda1335.7${} allows us to measure the relative abundance of [\cii]-to-\hi{} in the ISM of the highest redshift absorbing galaxy to date. This transition can be used to measure the cooling rate and the SFR per unit area \cite[e.g.,][]{Wolfe_SFRmedium2003,Wolfe2008}, and provide unique predictions for the \hi{} gas mass of high-$z$ [\cii]-emitting galaxies \citep{Heintz21}. Since the blue-most components of \cii$^*\lambda1335.7${} are blended with the stronger, red components from \cii$\lambda1334${} (see Fig. \ref{low} in Sect. \ref{data analysis}), we assume that the total column density is equal to $2.38$ times
that inferred from the red-most components (to match the relative ratio observed for \siii$^*$), $\log N_\text{\cii$^*$}/\text{cm}^{-2}~\text{(tot)} = 2.38\times N_\text{\cii$^*$}/\text{cm}^{-2}~\text{(obs)} = 13.98\pm 0.08$. We can then derive the [\cii] spontaneous emission rate per H atom, $l_c$, following \cite{Wolfe_SFRmedium2003}.
\begin{equation}
    l_c = \frac{N_{\rm C\textsc{ii}^*} h \nu_{ul} A_{ul}}{N_{\rm H\textsc{i}}} = (2.14\pm 0.41) \times 10^{-27} ~ {\rm erg\, s^{-1}\, H^{-1}} 
\end{equation}
and the equivalent [\cii]-to-\hi\ conversion factor, $\beta_\text{\cii}$, converted to typical emission-derived units of $M_\odot / L_\odot$, following \cite{Heintz21}
\begin{equation}
    \log \beta_\text{[\cii]} = \log (N_{\rm HI} / N_{\rm C\textsc{ii}^*}) - 3.934 = 3.22\pm 0.08 ~ {[M_\odot / L_\odot]}
\end{equation}
where $h \nu_{ul}$ and $A_{ul}$ are the energy and the Einstein coefficient for spontaneous photon decay of the $^2 P_{3/2}$ $\rightarrow$ $^ 2P_{1/2}$ transition of C$^+$, respectively. The value of $l_c$ is consistent with typical QSO-DLA \citep{Wolfe_SFRmedium2003} and extremely strong DLA measurements \citep[ES-DLA, often representative of GRB sightlines][]{Guimaraes12,Noterdaeme14,Telikova2022}, and may suggest that it originates from the cold neutral medium \cite[CNM, see e.g.,][]{Balashev2022}. The measurement of the absorption-based [\cii]-to-\hi\ conversion factor, $\beta_{\rm [CII]}$, is also perfectly consistent with the empirical relation between $\beta_{\rm [CII]}$ and metallicity derived by \cite{Heintz21}, at the observed GRB host metallicity. This suggests that there is a fundamental connection between [\cii] and \hi\ in the star-forming ISM of galaxies from $z\approx 0$ and out to $z\gtrsim 6$ as revealed by this burst.

We note that if UV pumping from the GRB (see Sect.\,\ref{fine}) is contributing to the \cii$^*$ column density, the \cii$^*$ value used for this analysis should be considered as an upper limit. In this case, the fact that the \cii$^*$ abundances match what is expected from \cii$^*$ collisionally excited by star formation at the respective metallicity \citep{Heintz21} would be a coincidence.

\subsection{Galaxy structure and kinematics}
\label{lya}

  \begin{figure*}
   \centering
   \includegraphics[width=1\textwidth]{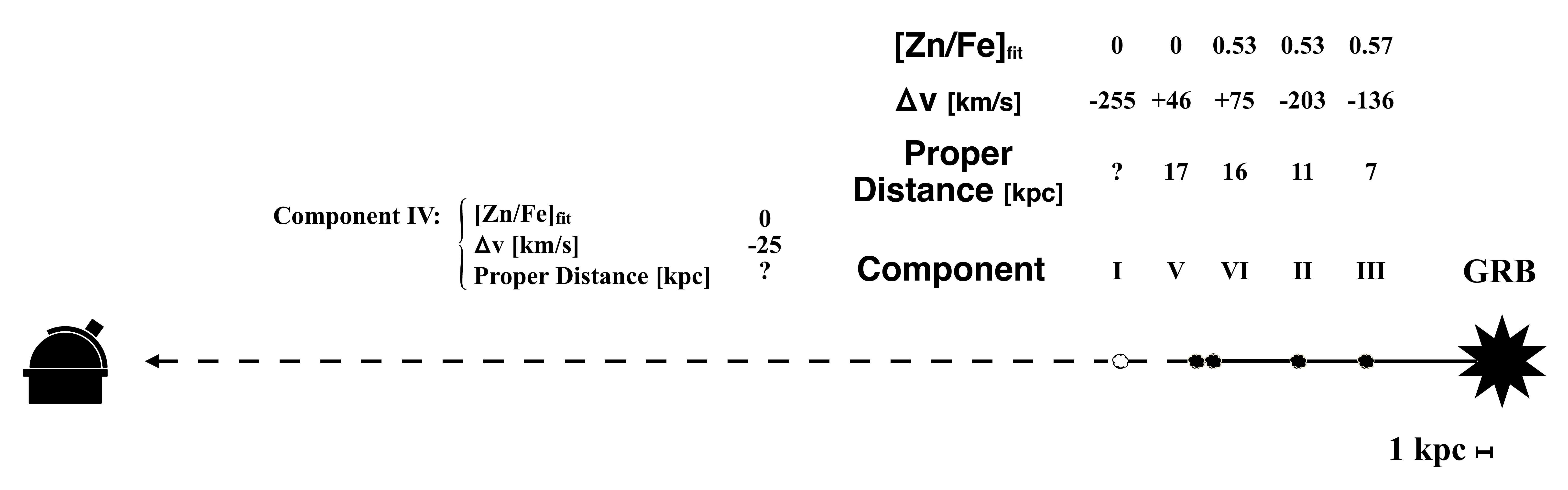}
      \caption{Schematic summary of part of the information obtained by our observations.}
        \label{sketch}
   \end{figure*}

On the basis of the results presented in the previous Sections, we can try to build a schematic view of the structure and kinematics of the galaxy complex. Part of the information obtained in previous sections is illustrated in the sketch presented in Fig.\,\ref{sketch}.

We detected absorption lines from gas belonging to different gas clouds at large distances (several kiloparsecs) from the GRB, which very likely was able to ionize its line of sight at least up to several hundreds of parsecs. A high ionization gas is detected all over these systems (as testified by the \civ{} and \siiv{} absorption lines), and a cloud at a higher ionization state is present (showing a strong narrow \nv{} absorption), likely between the GRB and the low-ionization gas clouds, at distances of at least 100\,pc. 
Furthermore, globally, systems $A$ and $B$ are chemically similar, and there are also strong similarities between components of systems $A$ and $B$, with components $II$, $III$ and $VI$ having similar significant dust depletion, whereas for $I$, $IV$ and $V$ it is consistent with zero.
We note also that the proper distances corresponding to the redshift of the two systems (represented for example by those of components $II$ and $VI$) would correspond to proper distances about forty times larger than those determined from the fine-structure line analysis, hence implying that the $\Delta v$ of the two systems is due to gas motion.

To the information retrieved from the absorbing gas signatures in the afterglow spectrum, we can add the information obtained through the HST image of the field of GRB\,210905A (see Sect.\,\ref{HST}).
Considering the small impact parameter of the objects nearby the GRB afterglow (also in function of redshift), their circumgalactic gas should be detected in the afterglow spectrum \citep{Tumlinson2017}. Therefore, even if we do not know the redshift of the objects in the proximity of GRB\,210905A, they should be similar to those of the absorbing systems identified in the spectrum, that are the host galaxy system or the two absorbers at $z=2.8296$ and $z=5.7390$.
It is possible that one or some of these objects are at the same redshift as the GRB and are interacting with or forming a galaxy group. In this case, it is plausible that the absorbing gas includes gas from the GRB host galaxy and from the potentially interacting circumgalactic gas of the objects or of the entire galaxy complex. A rough comparison between the low-ionization gas distances determined from the fine-structure lines and the measured projected distances from the HST image is not at odds with this scenario. Of course, a redshift estimate of the different objects would be needed to confirm and define it. This scenario has also the advantage of accommodating the fact that the components $V$ and $VI$ are farther than components $III$ and $II$, but red-shifted. The presence of inflowing (and outflowing) gas of interacting systems can easily explain that. We also note the tentative presence of features that can be interpreted as tidal tails, connecting the southern objects.

We do not favor a scenario where the $z=6.3$ absorbing gas complex detected in the spectrum does not include the GRBs host galaxy, because in this case we should have seen some HI absorption at higher redshift than the DLA, with a Lyman break.
If we would consider system $A$ as the GRB host and $B$ as a foreground unrelated galaxy, i) the chemical similarities would be surprising, ii) system $B$ must be in-falling on system $A$ with velocities of about 250\,km\,s$^{-1}$, iii) it would be difficult to accommodate component $I$ in this scenario. The opposite scenario (system $B$ as the GRB host and $A$ as a foreground galaxy) would be at odds with the distances determined from the fine-structure lines (see Sect.\,\ref{fine}).

Galaxy mergers and interactions are often considered as triggers of the star formation episode producing GRBs, and evidence for galaxy mergers has been found in long GRB host galaxy studies (e.g., \citealt{Chary2002,Arabsalmani2019,Savaglio2012,Savaglio2015,Chen2012,Wiseman2017,Penprase2007}; see however \citealt{Lyman2017}). Indeed, our observations show some similarities with GRB\,090323 afterglow spectrum \citep{Savaglio2012}, even if in that case the systems had a higher velocity separation ($\sim600$\,km\,s$^{-1}$), and the FORS2 spectral resolution did not allow a component-by-component analysis. The merger or galaxy group scenario would also explain the discrepancy with the mass-metallicity relation derived from the gas velocity (see Sect.\,\ref{metal}).

{\it HST} and ALMA observations have shown that clumpy structures extending over several kiloparsecs can be a common morphology for some very high-redshift galaxies \citep{Matthee2017,Carniani2018a,Carniani2018b}. Indeed, the galaxy we are probing has some similarities with B14-65666 \citep{Hashimoto2019} at $z=7.15$, a starburst galaxy with very high star-formation rate ($\sim200\,M_\odot$\,yr$^{-1}$) and stellar mass of $\sim8\times 10^8\,M_\odot$. The galaxy is formed of two clumps with similar properties separated from each other by a few kiloparsecs and kinematically separated by $\sim 200$\,km\,s$^{-1}$. If object $\alpha$ in the HST image of GRB\,210905A (Fig.\,\ref{fig:hst_slit}) is associated with the GRB host galaxy complex, our case can be similar to the B14-65666 case.

In Sect.\,\ref{em}, we reported a tentative detection of Lyman-$\alpha$ emission at $z=6.3449$, that is $\sim 1200$\,km\,s$^{-1}$ from the GRB host absorption lines. The line is detected over $\sim2.5$\arcsec, corresponding at such a redshift to $\sim14$\,kpc. Its luminosity ($L_{\rm Ly\alpha}\sim 1.4\times10^{42}$ erg\,s$^{-1}$), would be in agreement with typical values found for GRB host galaxies showing Lyman-$\alpha$ emission (see \citealt{Vielfaure2021} and references therein).
Due to its resonant nature, the Lyman-$\alpha$ emission interpretation is not trivial as it depends on many characteristics of the emitting galaxy (e.g., gas kinematics, outflows, \hi{} absorption, dust content, \ldots{} \citealt{Verhamme2006,Verhamme2008,Gronke2015}) and its line peak is usually found shifted from the redshift of the host galaxy systemic emission. This applies also to GRB host galaxies with detected Lyman-$\alpha$ emission (see \citealt{Jakobsson2005,Milvang-Jensen2012,Vielfaure2021} and references therein).
Lyman-$\alpha$ halos are usually found around galaxies, and high-redshift galaxies with $>10$\,kpc halos have been previously detected \citep{Leclercq2017,Leclercq2020}. A similar consideration can be applied to the centroid/peak difference between the Lyman-$\alpha$ and the UV galaxy emission, supposing that the UV light is traced by the afterglow position. Indeed \cite{Claeyssens2022} report a centroid/peak difference of several kiloparsecs. However, the velocity shift is extremely high if the Lyman-$\alpha$ emission comes from the same galaxy producing the absorption lines (even supposing a shift between the absorption line and the systemic redshift of the galaxy). Nevertheless, such a high shift might have been already observed, at least at very high redshift. Indeed, again, our case can be compared with B14-65666 \citep{Hashimoto2019}, where spectroscopic observations revealed Lyman-$\alpha$ emission at $\sim 800$\,km\,s$^{-1}$ from the [CII] and [OIII] emission lines of the star-forming clumps. A shift of a few 100\,km\,s$^{-1}$ is often found between GRB afterglow absorption and emission lines (see e.g., \citealt{Friis2015,Vielfaure2020}) and more in general in many Lyman-$\alpha$ emitters (e.g., \citealt{Orlitova2018}). Taking into account this additional velocity shift, would result in velocities  $> 1000$\,km\,s$^{-1}$ as in our case. \cite{Hashimoto2019} explain this large shift as due to large \hi{} content or to large outflows velocities, and demonstrate a positive correlation between high Lyman-$\alpha$ emission velocity shift and galaxy UV luminosity.

The high velocity issue would be solved under the hypothesis that Lyman-$\alpha$ emission comes from the GRB host galaxy and that the absorption lines are due to the gas of a galaxy close to the GRB along the GRB line of sight. However, as discussed above, in this case we should have detected at least some absorption features at a redshift similar to the Lyman-$\alpha$ emission one.

MUSE observations would allow the confirmation of the detection of the Lyman-$\alpha$ emission line, and the possibility to investigate its origin and the association of the emission and absorption lines with the GRB host and the galaxy complex identified in the $HST$ image.

More in general, further deep {\it HST}/F606W observations will be fundamental to assess which of the objects identified in the proximity of the GRB host are at low redshift. Furthermore, new {\it HST}/F140W observation are planned for Spring 2023 so as to assess the afterglow contribution to the current image and to allow a robust determination of the magnitude and SFR$_{\rm {UV}}$ of the host galaxy. 

Most importantly, GRB\,210905A would be an amazing target for $JWST$ observations, allowing the extremely rare possibility for a galaxy at $z=6.3$ to combine emission line spectroscopy with the detailed information retrieved from the afterglow spectrum absorption lines and with the $HST$ photometry.

\section{Conclusions}
\label{conclusions}

We investigated the properties of the gas along the line of sight of GRB\,210905A, thanks to the VLT/X-shooter spectroscopic observations of its afterglow.
We detected neutral hydrogen, low--ionization, high-ionization and fine-structure metal lines, likely from the GRB host galaxy complex at redshift $z=6.3118$. We detected also absorption lines from, at least, two intervening systems (at $z=5.7390$ and $z=2.8296$). 

The $z=6.3$ complex spans $\sim 360\,$km\,s$^{-1}$ and is composed of two major systems ($A$ at $z=6.3118$ and $B$ at $z=6.3186$) separated by $\sim 300$\,km\,s$^{-1}$, and formed by six components. 
We studied the abundance patterns and determined the galaxy metallicity, the overall strength of dust depletion and the $DTM$ of the complex. We found evidence of $\alpha$-element overabundance, a dust-corrected metallicity of $[\rm M/\mbox{H}]_{\rm tot}=-1.72\pm0.13$, and overall strength of the dust depletion [Zn/Fe]$_{\rm fit}=0.33\pm0.09$, indicating a moderate dust depletion. The metallicity and $DTM$ values are lower than then average metallicity and $DTM$ of GRB hosts at $2<z<4$, and than the extrapolation at high-redshift of 
the metallicity distribution of QSO-DLAs. However the values are comparable to those of QSO- and GRB-DLAs at $z>4.5$.
Indeed, the metallicities of the other two GRBs at $z\sim6$ with suitable afterglow spectroscopy available (GRB\,050904 at $z\sim6.3$ and GRB\,130606A at $z\sim5.9$) are consistent with that of the GRB\,210905A complex. 
The overall dust depletion corresponds to average values.

We performed also a component-by-component analysis of the gas and found that both $A$ and $B$ systems have a component with significant dust depletion and dust to metal ratio, whereas the others show no dust content.
We found evidence for $\alpha$-element enhancement, with nucleosyntethic patterns  due to massive stars. 
Interestingly, similarly to GRB\,130606A, at least in three components we find evidence of an aluminum overabundance, together with an oxygen underabundance in two components, which are reminiscent of those found for some globular cluster stars and could be due to very massive young rotating stars. 
While further conclusions are beyond the scope of this paper, we show that GRB afterglows can be effectively used to investigate the role of massive stars in the chemical enrichment of early Universe.

From the modeling of the fine-structure lines, under the assumption that they are due to UV pumping by the GRB radiation, we estimated the distance of the absorbing gas clouds and determined distances of several kiloparsecs, larger than those usually found for GRB sight-lines. This may be due to the high ionization present in the star-forming environment of very high redshift galaxies, and is correlated with the high number of ionizing photons produced by the GRBs, which would have ionized gas up to several hundreds of parsecs. Indeed, taking into account the GRBs with available suitable information, we found a correlation between the GRB luminosity and the distance of the closer low-ionization gas clouds, as determined from the fine-structure lines identified in the GRB afterglow spectra.

We also obtained $HST$/F140W observations of the GRB field, which allowed us to detect the GRB host galaxy as well as a complex of four objects in the proximity of the GRB afterglow position. One or more of them could be part of the same galaxy group and/or complex of the GRB host. Taking into account the results obtained through the GRB afterglow spectroscopy, as well as from the $HST$ observations, and the properties of other very high-redshift galaxies, we discussed the structure and kinematics of the galaxy complex, including also the tentative detection of Lyman-$\alpha$ emission at $z=6.3449$ ($\sim 1200$\,km\,s$^{-1}$ from the host redshift in velocity space). We discarded the possibility that the absorbing gas does not belong to the GRB host complex but to a nearby galaxy, and we suggested that a scenario of clumpy and merging galaxies could explain our observations.

Further {\it HST}/F140W observations are planned at more than one year from the GRB trigger, so that any afterglow contribution can be firmly excluded and it will be possible to confirm the host properties such as the GRB SFR$_{\rm UV}$ and the morphology. 
{\it HST}/F606W observations would allow us to identify which of the objects detected in the GRB proximity are at low redshift and therefore cannot be associated with the GRB host galaxy complex. Furthermore, VLT/MUSE observations would make it possible to confirm or discard the Lyman-$\alpha$ emission tentative detection, and, if confirmed, we would be able to analyze its position and extension, compared to the GRB afterglow position and the other objects in the field. These observations, combined with {\it JWST} data, to cover the host galaxy rest-frame optical emission, will offer the unique opportunity to combine the properties of the neutral and warm gas identified in this work with those of the continuum and ionized gas that would be determined through the galaxy photometry and the emission lines spectroscopy. 

This study shows the great potential of GRBs as tools to investigate the gas of very high-redshift galaxies.
The metallicities obtained so far from the three GRB afterglows at $z\sim
6$, for which these measurements have been possible, consistently show that high-z GRB hosts are
already enriched to a metallicity of $>2$\% solar, 1 Gyr after the Big Bang. Such metallicity enrichment
is indeed expected if GRBs select those galaxies populating the faint-end of the galaxy luminosity
function at these redshifts \citep{Salvaterra2012,Salvaterra2013}, which are believed to provide the major contribution to cosmic reionization.
It would be very difficult, if not impossible, to obtain the information presented in this work for field galaxies observed even with the best telescopes of the future ({\it JWST}, ELT,...). To increase the number of these kind of studies, it is fundamental to develop space satellites with energy range coverage and sensitivity suitable to detect the very high-redshift GRB population (\citealt{Ghirlanda2015}; e.g., {\it THESEUS, Gamow}; \citealt{Amati2018,Tanvir2021,White2021}, respectively). Onboard NIR afterglow detection and preselection of very high-redshift GRBs would be a significant added value to optimize ground-based spectroscopic observations with sensitive medium-high resolution spectrographs.  

\begin{acknowledgements}
This work was supported by CNES. AS and SDV acknowledge support from DIM-ACAV+. DAK acknowledges support from Spanish National Research Project RTI2018-098104-J-I00 (GRBPhot). AJL and DBM are supported by the European Research Council (ERC) under the European Union’s Horizon 2020 research and innovation programme (grant agreement No.~725246). The Cosmic Dawn Center is funded by the Danish National Research Foundation under grant No{.} 140.
We thank J. Blaizot, P. Bonifacio, E. Caffau, C. Charbonnel, N. Prantzos, S. Salvadori for fruitful discussions.
\end{acknowledgements}


\bibliographystyle{aa} 
\bibliography{susy20.bib} 

\begin{thebibliography}{132}
\expandafter\ifx\csname natexlab\endcsname\relax\def\natexlab#1{#1}\fi

\bibitem[{{Alves-Brito} {et~al.}(2010){Alves-Brito}, {Mel{\'e}ndez}, {Asplund},
  {Ram{\'\i}rez}, \& {Yong}}]{Alves-Brito2010}
{Alves-Brito}, A., {Mel{\'e}ndez}, J., {Asplund}, M., {Ram{\'\i}rez}, I., \&
  {Yong}, D. 2010, \aap, 513, A35

\bibitem[{{Amati} {et~al.}(2018){Amati}, {O'Brien}, {G{\"o}tz}, {Bozzo},
  {Tenzer}, {Frontera}, {Ghirlanda}, {Labanti}, {Osborne}, {Stratta}, {Tanvir},
  {Willingale}, {Attina}, {Campana}, {Castro-Tirado}, {Contini}, {Fuschino},
  {Gomboc}, {Hudec}, {Orleanski}, {Renotte}, {Rodic}, {Bagoly}, {Blain},
  {Callanan}, {Covino}, {Ferrara}, {Le Floch}, {Marisaldi}, {Mereghetti},
  {Rosati}, {Vacchi}, {D'Avanzo}, {Giommi}, {Piranomonte}, {Piro}, {Reglero},
  {Rossi}, {Santangelo}, {Salvaterra}, {Tagliaferri}, {Vergani}, {Vinciguerra},
  {Briggs}, {Campolongo}, {Ciolfi}, {Connaughton}, {Cordier}, {Morelli},
  {Orland ini}, {Adami}, {Argan}, {Atteia}, {Auricchio}, {Balazs}, {Baldazzi},
  {Basa}, {Basak}, {Bellutti}, {Bernardini}, {Bertuccio}, {Braga}, {Branchesi},
  {Brandt}, {Brocato}, {Budtz-Jorgensen}, {Bulgarelli}, {Burderi}, {Camp},
  {Capozziello}, {Caruana}, {Casella}, {Cenko}, {Chardonnet}, {Ciardi},
  {Colafrancesco}, {Dainotti}, {D'Elia}, {De Martino}, {De Pasquale}, {Del
  Monte}, {Della Valle}, {Drago}, {Evangelista}, {Feroci}, {Finelli},
  {Fiorini}, {Fynbo}, {Gal-Yam}, {Gendre}, {Ghisellini}, {Grado}, {Guidorzi},
  {Hafizi}, {Hanlon}, {Hjorth}, {Izzo}, {Kiss}, {Kumar}, {Kuvvetli}, {Lavagna},
  {Li}, {Longo}, {Lyutikov}, {Maio}, {Maiorano}, {Malcovati}, {Malesani},
  {Margutti}, {Martin-Carrillo}, {Masetti}, {McBreen}, {Mignani}, {Morgante},
  {Mundell}, {Nargaard-Nielsen}, {Nicastro}, {Palazzi}, {Paltani}, {Panessa},
  {Pareschi}, {Pe'er}, {Penacchioni}, {Pian}, {Piedipalumbo}, {Piran}, {Rauw},
  {Razzano}, {Read}, {Rezzolla}, {Romano}, {Ruffini}, {Savaglio}, {Sguera},
  {Schady}, {Skidmore}, {Song}, {Stanway}, {Starling}, {Topinka}, {Troja}, {van
  Putten}, {Vanzella}, {Vercellone}, {Wilson-Hodge}, {Yonetoku}, {Zampa},
  {Zampa}, {Zhang}, {Zhang}, {Zhang}, {Zhang}, {Antonelli}, {Bianco}, {Boci},
  {Boer}, {Botticella}, {Boulade}, {Butler}, {Campana}, {Capitanio}, {Celotti},
  {Chen}, {Colpi}, {Comastri}, {Cuby}, {Dadina}, {De Luca}, {Dong}, {Ettori},
  {Gandhi}, {Geza}, {Greiner}, {Guiriec}, {Harms}, {Hernanz}, {Hornstrup},
  {Hutchinson}, {Israel}, {Jonker}, {Kaneko}, {Kawai}, {Wiersema}, {Korpela},
  {Lebrun}, {Lu}, {MacFadyen}, {Malaguti}, {Maraschi}, {Meland ri}, {Modjaz},
  {Morris}, {Omodei}, {Paizis}, {P{\'a}ta}, {Petrosian}, {Rachevski}, {Rhoads},
  {Ryde}, {Sabau-Graziati}, {Shigehiro}, {Sims}, {Soomin}, {Sz{\'e}csi},
  {Urata}, {Uslenghi}, {Valenziano}, {Vianello}, {Vojtech}, {Watson}, \&
  {Zicha}}]{Amati2018}
{Amati}, L., {O'Brien}, P., {G{\"o}tz}, D., {et~al.} 2018, Advances in Space
  Research, 62, 191

\bibitem[{{Arabsalmani} {et~al.}(2015){Arabsalmani}, {M{\o}ller}, {Fynbo},
  {Christensen}, {Freudling}, {Savaglio}, \& {Zafar}}]{Arabsalmani2015}
{Arabsalmani}, M., {M{\o}ller}, P., {Fynbo}, J. P.~U., {et~al.} 2015, \mnras,
  446, 990

\bibitem[{{Arabsalmani} {et~al.}(2019){Arabsalmani}, {Roychowdhury},
  {Starkenburg}, {Christensen}, {Le Floc'h}, {Kanekar}, {Bournaud}, {Zwaan},
  {Fynbo}, {M{\o}ller}, \& {Pian}}]{Arabsalmani2019}
{Arabsalmani}, M., {Roychowdhury}, S., {Starkenburg}, T.~K., {et~al.} 2019,
  \mnras, 485, 5411

\bibitem[{{Asplund} {et~al.}(2009){Asplund}, {Grevesse}, {Sauval}, \&
  {Scott}}]{Asplund2009}
{Asplund}, M., {Grevesse}, N., {Sauval}, A.~J., \& {Scott}, P. 2009, \araa, 47,
  481

\bibitem[{{Balashev} {et~al.}(2022){Balashev}, {Telikova}, \&
  {Noterdaeme}}]{Balashev2022}
{Balashev}, S.~A., {Telikova}, K.~N., \& {Noterdaeme}, P. 2022, \mnras, 509,
  L26

\bibitem[{{Beardmore} {et~al.}(2021){Beardmore}, {Evans}, {Goad}, {Osborne}, \&
  {Swift-XRT Team.}}]{Beardmore2021}
{Beardmore}, A.~P., {Evans}, P.~A., {Goad}, M.~R., {Osborne}, J.~P., \&
  {Swift-XRT Team.} 2021, GRB Coordinates Network, 30768, 1

\bibitem[{{Becker} {et~al.}(2012){Becker}, {Sargent}, {Rauch}, \&
  {Carswell}}]{Becker2012}
{Becker}, G.~D., {Sargent}, W. L.~W., {Rauch}, M., \& {Carswell}, R.~F. 2012,
  \apj, 744, 91

\bibitem[{{Behar} {et~al.}(2011){Behar}, {Dado}, {Dar}, \& {Laor}}]{Behar2011}
{Behar}, E., {Dado}, S., {Dar}, A., \& {Laor}, A. 2011, \apj, 734, 26

\bibitem[{{Berry} {et~al.}(2012){Berry}, {Gawiser}, {Guaita}, {Padilla},
  {Treister}, {Blanc}, {Ciardullo}, {Francke}, \& {Gronwall}}]{Berry2012}
{Berry}, M., {Gawiser}, E., {Guaita}, L., {et~al.} 2012, \apj, 749, 4

\bibitem[{{Bla{\v{z}}ek} {et~al.}(2020){Bla{\v{z}}ek}, {de Ugarte Postigo},
  {Kann}, {Th{\"o}ne}, {Ag{\"u}{\'\i} Fern{\'a}ndez}, \& {Izzo}}]{Blazek2020}
{Bla{\v{z}}ek}, M., {de Ugarte Postigo}, A., {Kann}, D.~A., {et~al.} 2020, in
  Society of Photo-Optical Instrumentation Engineers (SPIE) Conference Series,
  Vol. 11452, Society of Photo-Optical Instrumentation Engineers (SPIE)
  Conference Series, 1145218

\bibitem[{{Bloom} {et~al.}(2009){Bloom}, {Perley}, {Li}, {Butler}, {Miller},
  {Kocevski}, {Kann}, {Foley}, {Chen}, {Filippenko}, {Starr}, {Macomber},
  {Prochaska}, {Chornock}, {Poznanski}, {Klose}, {Skrutskie}, {Lopez}, {Hall},
  {Glazebrook}, \& {Blake}}]{Bloom2009}
{Bloom}, J.~S., {Perley}, D.~A., {Li}, W., {et~al.} 2009, \apj, 691, 723

\bibitem[{{Bolmer} {et~al.}(2019){Bolmer}, {Ledoux}, {Wiseman}, {De Cia},
  {Selsing}, {Schady}, {Greiner}, {Savaglio}, {Burgess}, {D'Elia}, {Fynbo},
  {Goldoni}, {Hartmann}, {Heintz}, {Jakobsson}, {Japelj}, {Kaper}, {Tanvir},
  {Vreeswijk}, \& {Zafar}}]{Bolmer2019}
{Bolmer}, J., {Ledoux}, C., {Wiseman}, P., {et~al.} 2019, \aap, 623, A43

\bibitem[{{Campana} {et~al.}(2015){Campana}, {Salvaterra}, {Ferrara}, \&
  {Pallottini}}]{Campana2015}
{Campana}, S., {Salvaterra}, R., {Ferrara}, A., \& {Pallottini}, A. 2015, \aap,
  575, A43

\bibitem[{{Cano} {et~al.}(2017){Cano}, {Wang}, {Dai}, \& {Wu}}]{Cano2017}
{Cano}, Z., {Wang}, S.-Q., {Dai}, Z.-G., \& {Wu}, X.-F. 2017, Advances in
  Astronomy, 2017, 8929054

\bibitem[{{Carniani} {et~al.}(2018{\natexlab{a}}){Carniani}, {Maiolino},
  {Amorin}, {Pentericci}, {Pallottini}, {Ferrara}, {Willott}, {Smit},
  {Matthee}, {Sobral}, {Santini}, {Castellano}, {De Barros}, {Fontana},
  {Grazian}, \& {Guaita}}]{Carniani2018a}
{Carniani}, S., {Maiolino}, R., {Amorin}, R., {et~al.} 2018{\natexlab{a}},
  \mnras, 478, 1170

\bibitem[{{Carniani} {et~al.}(2018{\natexlab{b}}){Carniani}, {Maiolino},
  {Smit}, \& {Amor{\'\i}n}}]{Carniani2018b}
{Carniani}, S., {Maiolino}, R., {Smit}, R., \& {Amor{\'\i}n}, R.
  2018{\natexlab{b}}, \apjl, 854, L7

\bibitem[{{Chary} {et~al.}(2002){Chary}, {Becklin}, \& {Armus}}]{Chary2002}
{Chary}, R., {Becklin}, E.~E., \& {Armus}, L. 2002, \apj, 566, 229

\bibitem[{{Chen}(2012)}]{Chen2012}
{Chen}, H.-W. 2012, \mnras, 419, 3039

\bibitem[{{Claeyssens} {et~al.}(2022){Claeyssens}, {Richard}, {Blaizot},
  {Garel}, {Kusakabe}, {Bacon}, {Bauer}, {Guaita}, {Jeanneau}, {Lagattuta},
  {Leclercq}, {Maseda}, {Matthee}, {Nanayakkara}, {Pello}, {Thai}, {Tuan-Anh},
  {Verhamme}, {Vitte}, \& {Wisotzki}}]{Claeyssens2022}
{Claeyssens}, A., {Richard}, J., {Blaizot}, J., {et~al.} 2022, arXiv e-prints,
  arXiv:2201.04674

\bibitem[{{Cooke} {et~al.}(2021){Cooke}, {Webb}, {Dobie}, {Zhang}, {Hegarty},
  {Moller}, {Goode}, {Strausbaugh}, {Abbott}, {Gupta}, \&
  {Suhr}}]{GCN2021GRB210905Cooke}
{Cooke}, J., {Webb}, S., {Dobie}, D., {et~al.} 2021, GRB Coordinates Network,
  30773, 1

\bibitem[{{Cooke} {et~al.}(2011){Cooke}, {Pettini}, {Steidel}, {Rudie}, \&
  {Nissen}}]{Cooke2011}
{Cooke}, R., {Pettini}, M., {Steidel}, C.~C., {Rudie}, G.~C., \& {Nissen},
  P.~E. 2011, \mnras, 417, 1534

\bibitem[{{Cucchiara} {et~al.}(2011){Cucchiara}, {Levan}, {Fox}, {Tanvir},
  {Ukwatta}, {Berger}, {Kr{\"u}hler}, {K{\"u}pc{\"u} Yolda{\c s}}, {Wu},
  {Toma}, {Greiner}, {Olivares}, {Rowlinson}, {Amati}, {Sakamoto}, {Roth},
  {Stephens}, {Fritz}, {Fynbo}, {Hjorth}, {Malesani}, {Jakobsson}, {Wiersema},
  {O'Brien}, {Soderberg}, {Foley}, {Fruchter}, {Rhoads}, {Rutledge}, {Schmidt},
  {Dopita}, {Podsiadlowski}, {Willingale}, {Wolf}, {Kulkarni}, \&
  {D'Avanzo}}]{Cucchiara2011}
{Cucchiara}, A., {Levan}, A.~J., {Fox}, D.~B., {et~al.} 2011, \apj, 736, 7

\bibitem[{{Cullen} {et~al.}(2021){Cullen}, {Shapley}, {McLure}, {Dunlop},
  {Sanders}, {Topping}, {Reddy}, {Amor{\'\i}n}, {Begley}, {Bolzonella},
  {Calabr{\`o}}, {Carnall}, {Castellano}, {Cimatti}, {Cirasuolo}, {Cresci},
  {Fontana}, {Fontanot}, {Garilli}, {Guaita}, {Hamadouche}, {Hathi},
  {Mannucci}, {McLeod}, {Pentericci}, {Saxena}, {Talia}, \&
  {Zamorani}}]{Cullen2021}
{Cullen}, F., {Shapley}, A.~E., {McLure}, R.~J., {et~al.} 2021, \mnras, 505,
  903

\bibitem[{{Cupani} {et~al.}(2020){Cupani}, {D'Odorico}, {Cristiani}, {Russo},
  {Calderone}, \& {Taffoni}}]{Cupani2020}
{Cupani}, G., {D'Odorico}, V., {Cristiani}, S., {et~al.} 2020, in Society of
  Photo-Optical Instrumentation Engineers (SPIE) Conference Series, Vol. 11452,
  Society of Photo-Optical Instrumentation Engineers (SPIE) Conference Series,
  114521U

\bibitem[{{Dalton} {et~al.}(2021){Dalton}, {Morris}, \&
  {Fumagalli}}]{Dalton2021}
{Dalton}, T., {Morris}, S.~L., \& {Fumagalli}, M. 2021, \mnras, 502, 5981

\bibitem[{{D'Avanzo} {et~al.}(2021){D'Avanzo}, {Melandri}, {Covino}, {Fugazza},
  \& {REM Team}}]{GCN2021GRB210905ADavanzo}
{D'Avanzo}, P., {Melandri}, A., {Covino}, S., {Fugazza}, D., \& {REM Team}.
  2021, GRB Coordinates Network, 30772, 1

\bibitem[{{De Cia} {et~al.}(2021){De Cia}, {Jenkins}, {Fox}, {Ledoux},
  {Ramburth-Hurt}, {Konstantopoulou}, {Petitjean}, \& {Krogager}}]{DeCia2021}
{De Cia}, A., {Jenkins}, E.~B., {Fox}, A.~J., {et~al.} 2021, \nat, 597, 206

\bibitem[{{De Cia} {et~al.}(2016){De Cia}, {Ledoux}, {Mattsson}, {Petitjean},
  {Srianand}, {Gavignaud}, \& {Jenkins}}]{DeCia2016}
{De Cia}, A., {Ledoux}, C., {Mattsson}, L., {et~al.} 2016, \aap, 596, A97

\bibitem[{{De Cia} {et~al.}(2018){De Cia}, {Ledoux}, {Petitjean}, \&
  {Savaglio}}]{DeCia2018}
{De Cia}, A., {Ledoux}, C., {Petitjean}, P., \& {Savaglio}, S. 2018, \aap, 611,
  A76

\bibitem[{{de Ugarte Postigo} {et~al.}(2014){de Ugarte Postigo}, {Blazek},
  {Janout}, {Sprimont}, {Th{\"o}ne}, {Gorosabel}, \&
  {S{\'a}nchez-Ram{\'\i}rez}}]{deUgartePostigo2014}
{de Ugarte Postigo}, A., {Blazek}, M., {Janout}, P., {et~al.} 2014, in Society
  of Photo-Optical Instrumentation Engineers (SPIE) Conference Series, Vol.
  9152, Software and Cyberinfrastructure for Astronomy III, ed. G.~{Chiozzi} \&
  N.~M. {Radziwill}, 91520B

\bibitem[{{de Ugarte Postigo} {et~al.}(2012){de Ugarte Postigo}, {Fynbo},
  {Th{\"o}ne}, {Christensen}, {Gorosabel}, {Milvang-Jensen}, {Schulze},
  {Jakobsson}, {Wiersema}, {S{\'a}nchez-Ram{\'\i}rez}, {Leloudas}, {Zafar},
  {Malesani}, \& {Hjorth}}]{deUgartePostigo2012}
{de Ugarte Postigo}, A., {Fynbo}, J.~P.~U., {Th{\"o}ne}, C.~C., {et~al.} 2012,
  \aap, 548, A11

\bibitem[{{de Ugarte Postigo} {et~al.}(2020){de Ugarte Postigo}, {Kann},
  {Blazek}, {Agui Fernandez}, {Thoene}, {Gomez Velarde}, \& {Perez
  Romero}}]{deugartepostigo2020}
{de Ugarte Postigo}, A., {Kann}, D.~A., {Blazek}, M., {et~al.} 2020, GRB
  Coordinates Network, 28650, 1

\bibitem[{{D'Elia} {et~al.}(2009){D'Elia}, {Fiore}, {Perna}, {Krongold},
  {Vergani}, {Campana}, {Covino}, {D'Avanzo}, {Fugazza}, {Goldoni}, {Guidorzi},
  {Meurs}, {Norci}, {Piranomonte}, {Tagliaferri}, \& {Ward}}]{DElia2009}
{D'Elia}, V., {Fiore}, F., {Perna}, R., {et~al.} 2009, \aap, 503, 437

\bibitem[{{D'Elia} {et~al.}(2014){D'Elia}, {Fynbo}, {Goldoni}, {Covino}, {de
  Ugarte Postigo}, {Ledoux}, {Calura}, {Gorosabel}, {Malesani}, {Matteucci},
  {S{\'a}nchez-Ram{\'{\i}}rez}, {Savaglio}, {Castro-Tirado}, {Hartoog},
  {Kaper}, {Mu{\~n}oz-Darias}, {Pian}, {Piranomonte}, {Tagliaferri}, {Tanvir},
  {Vergani}, {Watson}, \& {Xu}}]{DElia2014}
{D'Elia}, V., {Fynbo}, J.~P.~U., {Goldoni}, P., {et~al.} 2014, \aap, 564, A38

\bibitem[{{Dessauges-Zavadsky} {et~al.}(2006){Dessauges-Zavadsky}, {Prochaska},
  {D'Odorico}, {Calura}, \& {Matteucci}}]{Dessauges-Zavadsky2006}
{Dessauges-Zavadsky}, M., {Prochaska}, J.~X., {D'Odorico}, S., {Calura}, F., \&
  {Matteucci}, F. 2006, \aap, 445, 93

\bibitem[{{Fox} {et~al.}(2008){Fox}, {Ledoux}, {Vreeswijk}, {Smette}, \&
  {Jaunsen}}]{Fox2008}
{Fox}, A.~J., {Ledoux}, C., {Vreeswijk}, P.~M., {Smette}, A., \& {Jaunsen},
  A.~O. 2008, \aap, 491, 189

\bibitem[{{Friis} {et~al.}(2015){Friis}, {De Cia}, {Kr{\"u}hler}, {Fynbo},
  {Ledoux}, {Vreeswijk}, {Watson}, {Malesani}, {Gorosabel}, {Starling},
  {Jakobsson}, {Varela}, {Wiersema}, {Drachmann}, {Trotter}, {Th{\"o}ne}, {de
  Ugarte Postigo}, {D'Elia}, {Elliott}, {Maturi}, {Goldoni}, {Greiner},
  {Haislip}, {Kaper}, {Knust}, {LaCluyze}, {Milvang-Jensen}, {Reichart},
  {Schulze}, {Sudilovsky}, {Tanvir}, \& {Vergani}}]{Friis2015}
{Friis}, M., {De Cia}, A., {Kr{\"u}hler}, T., {et~al.} 2015, \mnras, 451, 167

\bibitem[{{Fulbright} {et~al.}(2007){Fulbright}, {McWilliam}, \&
  {Rich}}]{Fulbright2007}
{Fulbright}, J.~P., {McWilliam}, A., \& {Rich}, R.~M. 2007, \apj, 661, 1152

\bibitem[{{Fynbo} {et~al.}(2008){Fynbo}, {Prochaska}, {Sommer-Larsen},
  {Dessauges-Zavadsky}, \& {M{\o}ller}}]{Fynbo2008a}
{Fynbo}, J.~P.~U., {Prochaska}, J.~X., {Sommer-Larsen}, J.,
  {Dessauges-Zavadsky}, M., \& {M{\o}ller}, P. 2008, \apj, 683, 321

\bibitem[{{Gehrels} {et~al.}(2004){Gehrels}, {Chincarini}, {Giommi}, {Mason},
  {Nousek}, {Wells}, {White}, {Barthelmy}, {Burrows}, {Cominsky}, {Hurley},
  {Marshall}, {M{\'e}sz{\'a}ros}, {Roming}, {Angelini}, {Barbier}, {Belloni},
  {Campana}, {Caraveo}, {Chester}, {Citterio}, {Cline}, {Cropper}, {Cummings},
  {Dean}, {Feigelson}, {Fenimore}, {Frail}, {Fruchter}, {Garmire}, {Gendreau},
  {Ghisellini}, {Greiner}, {Hill}, {Hunsberger}, {Krimm}, {Kulkarni}, {Kumar},
  {Lebrun}, {Lloyd-Ronning}, {Markwardt}, {Mattson}, {Mushotzky}, {Norris},
  {Osborne}, {Paczynski}, {Palmer}, {Park}, {Parsons}, {Paul}, {Rees},
  {Reynolds}, {Rhoads}, {Sasseen}, {Schaefer}, {Short}, {Smale}, {Smith},
  {Stella}, {Tagliaferri}, {Takahashi}, {Tashiro}, {Townsley}, {Tueller},
  {Turner}, {Vietri}, {Voges}, {Ward}, {Willingale}, {Zerbi}, \&
  {Zhang}}]{Gehrels2004}
{Gehrels}, N., {Chincarini}, G., {Giommi}, P., {et~al.} 2004, \apj, 611, 1005

\bibitem[{{Ghirlanda} {et~al.}(2018){Ghirlanda}, {Nappo}, {Ghisellini},
  {Melandri}, {Marcarini}, {Nava}, {Salafia}, {Campana}, \&
  {Salvaterra}}]{Ghirlanda2018}
{Ghirlanda}, G., {Nappo}, F., {Ghisellini}, G., {et~al.} 2018, \aap, 609, A112

\bibitem[{{Ghirlanda} {et~al.}(2015){Ghirlanda}, {Salvaterra}, {Ghisellini},
  {Mereghetti}, {Tagliaferri}, {Campana}, {Osborne}, {O'Brien}, {Tanvir},
  {Willingale}, {Amati}, {Basa}, {Bernardini}, {Burlon}, {Covino}, {D'Avanzo},
  {Frontera}, {G{\"o}tz}, {Melandri}, {Nava}, {Piro}, \&
  {Vergani}}]{Ghirlanda2015}
{Ghirlanda}, G., {Salvaterra}, R., {Ghisellini}, G., {et~al.} 2015, \mnras,
  448, 2514

\bibitem[{{Goldoni} {et~al.}(2006){Goldoni}, {Royer}, {Fran{\c c}ois},
  {Horrobin}, {Blanc}, {Vernet}, {Modigliani}, \& {Larsen}}]{Goldoni2006}
{Goldoni}, P., {Royer}, F., {Fran{\c c}ois}, P., {et~al.} 2006, in Society of
  Photo-Optical Instrumentation Engineers (SPIE) Conference Series, Vol. 6269

\bibitem[{{Gronke} {et~al.}(2015){Gronke}, {Bull}, \& {Dijkstra}}]{Gronke2015}
{Gronke}, M., {Bull}, P., \& {Dijkstra}, M. 2015, \apj, 812, 123

\bibitem[{{Guimar{\~a}es} {et~al.}(2012){Guimar{\~a}es}, {Noterdaeme},
  {Petitjean}, {Ledoux}, {Srianand}, {L{\'o}pez}, \& {Rahmani}}]{Guimaraes12}
{Guimar{\~a}es}, R., {Noterdaeme}, P., {Petitjean}, P., {et~al.} 2012, \aj,
  143, 147

\bibitem[{{Hartoog} {et~al.}(2015){Hartoog}, {Malesani}, {Fynbo}, {Goto},
  {Kr{\"u}hler}, {Vreeswijk}, {De Cia}, {Xu}, {M{\o}ller}, {Covino}, {D'Elia},
  {Flores}, {Goldoni}, {Hjorth}, {Jakobsson}, {Krogager}, {Kaper}, {Ledoux},
  {Levan}, {Milvang-Jensen}, {Sollerman}, {Sparre}, {Tagliaferri}, {Tanvir},
  {de Ugarte Postigo}, {Vergani}, {Wiersema}, {Datson}, {Salinas}, {Mikkelsen},
  \& {Aghanim}}]{Hartoog2015}
{Hartoog}, O.~E., {Malesani}, D., {Fynbo}, J.~P.~U., {et~al.} 2015, \aap, 580,
  A139

\bibitem[{{Hashimoto} {et~al.}(2019){Hashimoto}, {Inoue}, {Mawatari}, {Tamura},
  {Matsuo}, {Furusawa}, {Harikane}, {Shibuya}, {Knudsen}, {Kohno}, {Ono},
  {Zackrisson}, {Okamoto}, {Kashikawa}, {Oesch}, {Ouchi}, {Ota}, {Shimizu},
  {Taniguchi}, {Umehata}, \& {Watson}}]{Hashimoto2019}
{Hashimoto}, T., {Inoue}, A.~K., {Mawatari}, K., {et~al.} 2019, \pasj, 71, 71

\bibitem[{{Heintz} {et~al.}(2018){Heintz}, {Watson}, {Jakobsson}, {Fynbo},
  {Bolmer}, {Arabsalmani}, {Cano}, {Covino}, {D'Elia}, {Gomboc}, {Japelj},
  {Kaper}, {Krogager}, {Pugliese}, {S{\'a}nchez-Ram{\'\i}rez}, {Selsing},
  {Sparre}, {Tanvir}, {Th{\"o}ne}, {de Ugarte Postigo}, \&
  {Vergani}}]{Heintz2018}
{Heintz}, K.~E., {Watson}, D., {Jakobsson}, P., {et~al.} 2018, \mnras, 479,
  3456

\bibitem[{{Heintz} {et~al.}(2021){Heintz}, {Watson}, {Oesch}, {Narayanan}, \&
  {Madden}}]{Heintz21}
{Heintz}, K.~E., {Watson}, D., {Oesch}, P.~A., {Narayanan}, D., \& {Madden},
  S.~C. 2021, \apj, 922, 147

\bibitem[{{Herenz} {et~al.}(2019){Herenz}, {Wisotzki}, {Saust}, {Kerutt},
  {Urrutia}, {Diener}, {Schmidt}, {Marino}, {de la Vieuville}, {Boogaard},
  {Schaye}, {Guiderdoni}, {Richard}, \& {Bacon}}]{Herenz2019}
{Herenz}, E.~C., {Wisotzki}, L., {Saust}, R., {et~al.} 2019, \aap, 621, A107

\bibitem[{{Hjorth} \& {Bloom}(2012)}]{Hjorth2011_gamma}
{Hjorth}, J. \& {Bloom}, J.~S. 2012, in Chapter 9 in ``Gamma-Ray Bursts'',
  169--190

\bibitem[{{Inoue} {et~al.}(2016){Inoue}, {Tamura}, {Matsuo}, {Mawatari},
  {Shimizu}, {Shibuya}, {Ota}, {Yoshida}, {Zackrisson}, {Kashikawa}, {Kohno},
  {Umehata}, {Hatsukade}, {Iye}, {Matsuda}, {Okamoto}, \&
  {Yamaguchi}}]{Inoue2016}
{Inoue}, A.~K., {Tamura}, Y., {Matsuo}, H., {et~al.} 2016, Science, 352, 1559

\bibitem[{{Jakobsson} {et~al.}(2005){Jakobsson}, {Fynbo}, {Paraficz},
  {Telting}, {Jensen}, {Hjorth}, \& {Castro Cer{\'o}n}}]{Jakobsson2005}
{Jakobsson}, P., {Fynbo}, J.~P.~U., {Paraficz}, D., {et~al.} 2005, GRB
  Coordinates Network, 4017

\bibitem[{{Jakobsson} {et~al.}(2006){Jakobsson}, {Levan}, {Fynbo}, {Priddey},
  {Hjorth}, {Tanvir}, {Watson}, {Jensen}, {Sollerman}, {Natarajan},
  {Gorosabel}, {Castro Cer{\'o}n}, {Pedersen}, {Pursimo}, {{\'A}rnad{\'o}ttir},
  {Castro-Tirado}, {Davis}, {Deeg}, {Fiuza}, {Mykolaitis}, \&
  {Sousa}}]{Jakobsson2006}
{Jakobsson}, P., {Levan}, A., {Fynbo}, J.~P.~U., {et~al.} 2006, \aap, 447, 897

\bibitem[{{Jeong} {et~al.}(2014){Jeong}, {Sanchez-Ramirez}, {Gorosabel}, \&
  {Castro-Tirado}}]{Jeong2014}
{Jeong}, S., {Sanchez-Ramirez}, R., {Gorosabel}, J., \& {Castro-Tirado}, A.~J.
  2014, GRB Coordinates Network, 15936, 1

\bibitem[{{Kawai} {et~al.}(2006){Kawai}, {Kosugi}, {Aoki}, {Yamada}, {Totani},
  {Ohta}, {Iye}, {Hattori}, {Aoki}, {Furusawa}, {Hurley}, {Kawabata},
  {Kobayashi}, {Komiyama}, {Mizumoto}, {Nomoto}, {Noumaru}, {Ogasawara},
  {Sato}, {Sekiguchi}, {Shirasaki}, {Suzuki}, {Takata}, {Tamagawa}, {Terada},
  {Watanabe}, {Yatsu}, \& {Yoshida}}]{Kawai2006}
{Kawai}, N., {Kosugi}, G., {Aoki}, K., {et~al.} 2006, \nat, 440, 184

\bibitem[{{Konstantopoulou} {et~al.}(2022){Konstantopoulou}, {De Cia},
  {Krogager}, {Ledoux}, {Noterdaeme}, {Fynbo}, {Heintz}, {Watson}, {Andersen},
  {Ramburuth-Hurt}, \& {Jermann}}]{Konstantopoulou2022}
{Konstantopoulou}, C., {De Cia}, A., {Krogager}, J.-K., {et~al.} 2022, \aap,
  666, A12

\bibitem[{{Krongold} \& {Prochaska}(2013)}]{Krongold2013}
{Krongold}, Y. \& {Prochaska}, J.~X. 2013, \apj, 774, 115

\bibitem[{{Laskar} {et~al.}(2021){Laskar}, {Alexander}, {Margutti}, {Berger},
  {Fong}, {Chornock}, {Mundell}, \& {Schady}}]{Laskar2021}
{Laskar}, T., {Alexander}, K.~D., {Margutti}, R., {et~al.} 2021, GRB
  Coordinates Network, 30783, 1

\bibitem[{{Leclercq} {et~al.}(2020){Leclercq}, {Bacon}, {Verhamme}, {Garel},
  {Blaizot}, {Brinchmann}, {Cantalupo}, {Claeyssens}, {Conseil}, {Contini},
  {Hashimoto}, {Herenz}, {Kusakabe}, {Marino}, {Maseda}, {Matthee}, {Mitchell},
  {Pezzulli}, {Richard}, {Schmidt}, \& {Wisotzki}}]{Leclercq2020}
{Leclercq}, F., {Bacon}, R., {Verhamme}, A., {et~al.} 2020, \aap, 635, A82

\bibitem[{{Leclercq} {et~al.}(2017){Leclercq}, {Bacon}, {Wisotzki}, {Mitchell},
  {Garel}, {Verhamme}, {Blaizot}, {Hashimoto}, {Herenz}, {Conseil},
  {Cantalupo}, {Inami}, {Contini}, {Richard}, {Maseda}, {Schaye}, {Marino},
  {Akhlaghi}, {Brinchmann}, \& {Carollo}}]{Leclercq2017}
{Leclercq}, F., {Bacon}, R., {Wisotzki}, L., {et~al.} 2017, \aap, 608, A8

\bibitem[{{Ledoux} {et~al.}(2006){Ledoux}, {Vreeswijk}, {Smette}, {Jaunsen}, \&
  {Kaufer}}]{Ledoux2006}
{Ledoux}, C., {Vreeswijk}, P., {Smette}, A., {Jaunsen}, A., \& {Kaufer}, A.
  2006, GRB Coordinates Network, 5237

\bibitem[{{Lyman} {et~al.}(2017){Lyman}, {Levan}, {Tanvir}, {Fynbo}, {McGuire},
  {Perley}, {Angus}, {Bloom}, {Conselice}, {Fruchter}, {Hjorth}, {Jakobsson},
  \& {Starling}}]{Lyman2017}
{Lyman}, J.~D., {Levan}, A.~J., {Tanvir}, N.~R., {et~al.} 2017, \mnras, 467,
  1795

\bibitem[{{Masseron} {et~al.}(2020){Masseron}, {Garc{\'\i}a-Hern{\'a}ndez},
  {Santove{\~n}a}, {Manchado}, {Zamora}, {Manteiga}, \&
  {Dafonte}}]{Masseron2020}
{Masseron}, T., {Garc{\'\i}a-Hern{\'a}ndez}, D.~A., {Santove{\~n}a}, R.,
  {et~al.} 2020, Nature Communications, 11, 3759

\bibitem[{{Matthee} {et~al.}(2017){Matthee}, {Sobral}, {Darvish}, {Santos},
  {Mobasher}, {Paulino-Afonso}, {R{\"o}ttgering}, \& {Alegre}}]{Matthee2017}
{Matthee}, J., {Sobral}, D., {Darvish}, B., {et~al.} 2017, \mnras, 472, 772

\bibitem[{{McGuire} {et~al.}(2016){McGuire}, {Tanvir}, {Levan}, {Trenti},
  {Stanway}, {Shull}, {Wiersema}, {Perley}, {Starling}, {Bremer}, {Stocke},
  {Hjorth}, {Rhoads}, {Curtis-Lake}, {Schulze}, {Levesque}, {Robertson},
  {Fynbo}, {Ellis}, \& {Fruchter}}]{McGuire2016}
{McGuire}, J.~T.~W., {Tanvir}, N.~R., {Levan}, A.~J., {et~al.} 2016, \apj, 825,
  135

\bibitem[{{McWilliam}(1997)}]{McWilliams1997}
{McWilliam}, A. 1997, \araa, 35, 503

\bibitem[{{Melandri} {et~al.}(2015){Melandri}, {Bernardini}, {D'Avanzo},
  {S{\'a}nchez-Ram{\'\i}rez}, {Nappo}, {Nava}, {Japelj}, {de Ugarte Postigo},
  {Oates}, {Campana}, {Covino}, {D'Elia}, {Ghirlanda}, {Gafton}, {Ghisellini},
  {Gnedin}, {Goldoni}, {Gorosabel}, {Libbrecht}, {Malesani}, {Salvaterra},
  {Th{\"o}ne}, {Vergani}, {Xu}, \& {Tagliaferri}}]{Melandri2015}
{Melandri}, A., {Bernardini}, M.~G., {D'Avanzo}, P., {et~al.} 2015, \aap, 581,
  A86

\bibitem[{{Milvang-Jensen} {et~al.}(2012){Milvang-Jensen}, {Fynbo}, {Malesani},
  {Hjorth}, {Jakobsson}, \& {M{\o}ller}}]{Milvang-Jensen2012}
{Milvang-Jensen}, B., {Fynbo}, J. P.~U., {Malesani}, D., {et~al.} 2012, \apj,
  756, 25

\bibitem[{{Modigliani} {et~al.}(2010){Modigliani}, {Goldoni}, {Royer},
  {Haigron}, {Guglielmi}, {Fran{\c c}ois}, {Horrobin}, {Bristow}, {Vernet},
  {Moehler}, {Kerber}, {Ballester}, {Mason}, \& {Christensen}}]{Modigliani2010}
{Modigliani}, A., {Goldoni}, P., {Royer}, F., {et~al.} 2010, in \procspie, Vol.
  7737, Observatory Operations: Strategies, Processes, and Systems III, 773728

\bibitem[{{M{\o}ller} {et~al.}(2013){M{\o}ller}, {Fynbo}, {Ledoux}, \&
  {Nilsson}}]{Moller2013}
{M{\o}ller}, P., {Fynbo}, J.~P.~U., {Ledoux}, C., \& {Nilsson}, K.~K. 2013,
  \mnras, 430, 2680

\bibitem[{{Neeleman} {et~al.}(2013){Neeleman}, {Wolfe}, {Prochaska}, \&
  {Rafelski}}]{Neeleman2013}
{Neeleman}, M., {Wolfe}, A.~M., {Prochaska}, J.~X., \& {Rafelski}, M. 2013,
  \apj, 769, 54

\bibitem[{{Nicuesa Guelbenzu} {et~al.}(2021){Nicuesa Guelbenzu}, {Klose}, \&
  {Rau}}]{GCN2021GRB210905Nicuesa}
{Nicuesa Guelbenzu}, A., {Klose}, S., \& {Rau}, A. 2021, GRB Coordinates
  Network, 30781, 1

\bibitem[{{Noterdaeme} {et~al.}(2014){Noterdaeme}, {Petitjean}, {P{\^a}ris},
  {Cai}, {Finley}, {Ge}, {Pieri}, \& {York}}]{Noterdaeme14}
{Noterdaeme}, P., {Petitjean}, P., {P{\^a}ris}, I., {et~al.} 2014, \aap, 566,
  A24

\bibitem[{{Orlitov{\'a}} {et~al.}(2018){Orlitov{\'a}}, {Verhamme}, {Henry},
  {Scarlata}, {Jaskot}, {Oey}, \& {Schaerer}}]{Orlitova2018}
{Orlitov{\'a}}, I., {Verhamme}, A., {Henry}, A., {et~al.} 2018, \aap, 616, A60

\bibitem[{{Palmerio} {et~al.}(2019){Palmerio}, {Vergani}, {Salvaterra}, {Sand
  ers}, {Japelj}, {Vidal-Garc{\'\i}a}, {D'Avanzo}, {Corre}, {Perley},
  {Shapley}, {Boissier}, {Greiner}, {Le Floc'h}, \& {Wiseman}}]{Palmerio2019}
{Palmerio}, J.~T., {Vergani}, S.~D., {Salvaterra}, R., {et~al.} 2019, \aap,
  623, A26

\bibitem[{{Penprase} {et~al.}(2007){Penprase}, {Beeler}, \& {Toro
  Martinez}}]{Penprase2007}
{Penprase}, B.~E., {Beeler}, D.~J., \& {Toro Martinez}, I. 2007, in American
  Astronomical Society Meeting Abstracts, Vol. 210, American Astronomical
  Society Meeting Abstracts \#210, 44.04

\bibitem[{{Perley} {et~al.}(2016){Perley}, {Tanvir}, {Hjorth}, {Laskar},
  {Berger}, {Chary}, {de Ugarte Postigo}, {Fynbo}, {Kr{\"u}hler}, {Levan},
  {Micha{\l}owski}, \& {Schulze}}]{Perley2016}
{Perley}, D.~A., {Tanvir}, N.~R., {Hjorth}, J., {et~al.} 2016, \apj, 817, 8

\bibitem[{{Planck Collaboration} {et~al.}(2016){Planck Collaboration}, {Ade},
  {Aghanim}, {Arnaud}, {Ashdown}, {Aumont}, {Baccigalupi}, {Banday},
  {Barreiro}, {Bartlett}, {Bartolo}, {Battaner}, {Battye}, {Benabed},
  {Beno{\^\i}t}, {Benoit-L{\'e}vy}, {Bernard}, {Bersanelli}, {Bielewicz},
  {Bock}, {Bonaldi}, {Bonavera}, {Bond}, {Borrill}, {Bouchet}, {Boulanger},
  {Bucher}, {Burigana}, {Butler}, {Calabrese}, {Cardoso}, {Catalano},
  {Challinor}, {Chamballu}, {Chary}, {Chiang}, {Chluba}, {Christensen},
  {Church}, {Clements}, {Colombi}, {Colombo}, {Combet}, {Coulais}, {Crill},
  {Curto}, {Cuttaia}, {Danese}, {Davies}, {Davis}, {de Bernardis}, {de Rosa},
  {de Zotti}, {Delabrouille}, {D{\'e}sert}, {Di Valentino}, {Dickinson},
  {Diego}, {Dolag}, {Dole}, {Donzelli}, {Dor{\'e}}, {Douspis}, {Ducout},
  {Dunkley}, {Dupac}, {Efstathiou}, {Elsner}, {En{\ss}lin}, {Eriksen},
  {Farhang}, {Fergusson}, {Finelli}, {Forni}, {Frailis}, {Fraisse},
  {Franceschi}, {Frejsel}, {Galeotta}, {Galli}, {Ganga}, {Gauthier}, {Gerbino},
  {Ghosh}, {Giard}, {Giraud-H{\'e}raud}, {Giusarma}, {Gjerl{\o}w},
  {Gonz{\'a}lez-Nuevo}, {G{\'o}rski}, {Gratton}, {Gregorio}, {Gruppuso},
  {Gudmundsson}, {Hamann}, {Hansen}, {Hanson}, {Harrison}, {Helou},
  {Henrot-Versill{\'e}}, {Hern{\'a}ndez-Monteagudo}, {Herranz}, {Hildebrandt},
  {Hivon}, {Hobson}, {Holmes}, {Hornstrup}, {Hovest}, {Huang}, {Huffenberger},
  {Hurier}, {Jaffe}, {Jaffe}, {Jones}, {Juvela}, {Keih{\"a}nen}, {Keskitalo},
  {Kisner}, {Kneissl}, {Knoche}, {Knox}, {Kunz}, {Kurki-Suonio}, {Lagache},
  {L{\"a}hteenm{\"a}ki}, {Lamarre}, {Lasenby}, {Lattanzi}, {Lawrence}, {Leahy},
  {Leonardi}, {Lesgourgues}, {Levrier}, {Lewis}, {Liguori}, {Lilje},
  {Linden-V{\o}rnle}, {L{\'o}pez-Caniego}, {Lubin}, {Mac{\'\i}as-P{\'e}rez},
  {Maggio}, {Maino}, {Mandolesi}, {Mangilli}, {Marchini}, {Maris}, {Martin},
  {Martinelli}, {Mart{\'\i}nez-Gonz{\'a}lez}, {Masi}, {Matarrese}, {McGehee},
  {Meinhold}, {Melchiorri}, {Melin}, {Mendes}, {Mennella}, {Migliaccio},
  {Millea}, {Mitra}, {Miville-Desch{\^e}nes}, {Moneti}, {Montier}, {Morgante},
  {Mortlock}, {Moss}, {Munshi}, {Murphy}, {Naselsky}, {Nati}, {Natoli},
  {Netterfield}, {N{\o}rgaard-Nielsen}, {Noviello}, {Novikov}, {Novikov},
  {Oxborrow}, {Paci}, {Pagano}, {Pajot}, {Paladini}, {Paoletti}, {Partridge},
  {Pasian}, {Patanchon}, {Pearson}, {Perdereau}, {Perotto}, {Perrotta},
  {Pettorino}, {Piacentini}, {Piat}, {Pierpaoli}, {Pietrobon}, {Plaszczynski},
  {Pointecouteau}, {Polenta}, {Popa}, {Pratt}, {Pr{\'e}zeau}, {Prunet},
  {Puget}, {Rachen}, {Reach}, {Rebolo}, {Reinecke}, {Remazeilles}, {Renault},
  {Renzi}, {Ristorcelli}, {Rocha}, {Rosset}, {Rossetti}, {Roudier},
  {Rouill{\'e} d'Orfeuil}, {Rowan-Robinson}, {Rubi{\~n}o-Mart{\'\i}n},
  {Rusholme}, {Said}, {Salvatelli}, {Salvati}, {Sandri}, {Santos},
  {Savelainen}, {Savini}, {Scott}, {Seiffert}, {Serra}, {Shellard}, {Spencer},
  {Spinelli}, {Stolyarov}, {Stompor}, {Sudiwala}, {Sunyaev}, {Sutton},
  {Suur-Uski}, {Sygnet}, {Tauber}, {Terenzi}, {Toffolatti}, {Tomasi},
  {Tristram}, {Trombetti}, {Tucci}, {Tuovinen}, {T{\"u}rler}, {Umana},
  {Valenziano}, {Valiviita}, {Van Tent}, {Vielva}, {Villa}, {Wade}, {Wandelt},
  {Wehus}, {White}, {White}, {Wilkinson}, {Yvon}, {Zacchei}, \&
  {Zonca}}]{Planck2016}
{Planck Collaboration}, {Ade}, P.~A.~R., {Aghanim}, N., {et~al.} 2016, \aap,
  594, A13

\bibitem[{{Prantzos} {et~al.}(2007){Prantzos}, {Charbonnel}, \&
  {Iliadis}}]{Prantzos2007}
{Prantzos}, N., {Charbonnel}, C., \& {Iliadis}, C. 2007, \aap, 470, 179

\bibitem[{{Prochaska} {et~al.}(2006){Prochaska}, {Chen}, \&
  {Bloom}}]{Prochaska2006}
{Prochaska}, J.~X., {Chen}, H.-W., \& {Bloom}, J.~S. 2006, \apj, 648, 95

\bibitem[{{Prochaska} {et~al.}(2007){Prochaska}, {Chen}, {Dessauges-Zavadsky},
  \& {Bloom}}]{Prochaska2007b}
{Prochaska}, J.~X., {Chen}, H.-W., {Dessauges-Zavadsky}, M., \& {Bloom}, J.~S.
  2007, \apj, 666, 267

\bibitem[{{Prochaska} {et~al.}(2008){Prochaska}, {Dessauges-Zavadsky},
  {Ramirez-Ruiz}, \& {Chen}}]{Prochaska2008}
{Prochaska}, J.~X., {Dessauges-Zavadsky}, M., {Ramirez-Ruiz}, E., \& {Chen},
  H.-W. 2008, \apj, 685, 344

\bibitem[{{Racusin} {et~al.}(2008){Racusin}, {Karpov}, {Sokolowski}, {Granot},
  {Wu}, {Pal'Shin}, {Covino}, {van der Horst}, {Oates}, {Schady}, {Smith},
  {Cummings}, {Starling}, {Piotrowski}, {Zhang}, {Evans}, {Holland}, {Malek},
  {Page}, {Vetere}, {Margutti}, {Guidorzi}, {Kamble}, {Curran}, {Beardmore},
  {Kouveliotou}, {Mankiewicz}, {Melandri}, {O'Brien}, {Page}, {Piran},
  {Tanvir}, {Wrochna}, {Aptekar}, {Barthelmy}, {Bartolini}, {Beskin}, {Bondar},
  {Bremer}, {Campana}, {Castro-Tirado}, {Cucchiara}, {Cwiok}, {D'Avanzo},
  {D'Elia}, {Della Valle}, {de Ugarte Postigo}, {Dominik}, {Falcone}, {Fiore},
  {Fox}, {Frederiks}, {Fruchter}, {Fugazza}, {Garrett}, {Gehrels},
  {Golenetskii}, {Gomboc}, {Gorosabel}, {Greco}, {Guarnieri}, {Immler},
  {Jelinek}, {Kasprowicz}, {La Parola}, {Levan}, {Mangano}, {Mazets},
  {Molinari}, {Moretti}, {Nawrocki}, {Oleynik}, {Osborne}, {Pagani}, {Pandey},
  {Paragi}, {Perri}, {Piccioni}, {Ramirez-Ruiz}, {Roming}, {Steele}, {Strom},
  {Testa}, {Tosti}, {Ulanov}, {Wiersema}, {Wijers}, {Winters}, {Zarnecki},
  {Zerbi}, {M{\'e}sz{\'a}ros}, {Chincarini}, \& {Burrows}}]{Racusin2008_nature}
{Racusin}, J.~L., {Karpov}, S.~V., {Sokolowski}, M., {et~al.} 2008, \nat, 455,
  183

\bibitem[{{Rafelski} {et~al.}(2014){Rafelski}, {Neeleman}, {Fumagalli},
  {Wolfe}, \& {Prochaska}}]{Rafelski2014}
{Rafelski}, M., {Neeleman}, M., {Fumagalli}, M., {Wolfe}, A.~M., \&
  {Prochaska}, J.~X. 2014, \apjl, 782, L29

\bibitem[{{Rafelski} {et~al.}(2012){Rafelski}, {Wolfe}, {Prochaska},
  {Neeleman}, \& {Mendez}}]{rafelski12}
{Rafelski}, M., {Wolfe}, A.~M., {Prochaska}, J.~X., {Neeleman}, M., \&
  {Mendez}, A.~J. 2012, \apj, 755, 89

\bibitem[{{Rosdahl} {et~al.}(2018){Rosdahl}, {Katz}, {Blaizot}, {Kimm},
  {Michel-Dansac}, {Garel}, {Haehnelt}, {Ocvirk}, \& {Teyssier}}]{Rosdahl2018}
{Rosdahl}, J., {Katz}, H., {Blaizot}, J., {et~al.} 2018, \mnras, 479, 994

\bibitem[{{Rossi} {et~al.}(2022){Rossi}, {Frederiks}, {Kann}, {De Pasquale},
  {Pian}, {Lamb}, {D'Avanzo}, {Izzo}, {Levan}, {Malesani}, {Melandri}, {Nicuesa
  Guelbenzu}, {Schulze}, {Strausbaugh}, {Tanvir}, {Amati}, {Campana},
  {Cucchiara}, {Ghirlanda}, {Della Valle}, {Klose}, {Salvaterra}, {Starling},
  {Stratta}, {Tsvetkova}, {Vergani}, {D'A{\`\i}}, {Burgarella}, {Covino},
  {D'Elia}, {de Ugarte Postigo}, {Fausey}, {Fynbo}, {Frontera}, {Guidorzi},
  {Heintz}, {Masetti}, {Maiorano}, {Mundell}, {Oates}, {Page}, {Palazzi},
  {Palmerio}, {Pugliese}, {Rau}, {Saccardi}, {Sbarufatti}, {Svinkin},
  {Tagliaferri}, {van der Horst}, {Watson}, {Ulanov}, {Wiersema}, {Xu}, \&
  {Zhang}}]{Rossi2022}
{Rossi}, A., {Frederiks}, D.~D., {Kann}, D.~A., {et~al.} 2022, \aap, 665, A125

\bibitem[{{Salvadori} {et~al.}(2019){Salvadori}, {Bonifacio}, {Caffau},
  {Korotin}, {Andreevsky}, {Spite}, \& {Sk{\'u}lad{\'o}ttir}}]{Salvadori2019}
{Salvadori}, S., {Bonifacio}, P., {Caffau}, E., {et~al.} 2019, \mnras, 487,
  4261

\bibitem[{{Salvaterra} {et~al.}(2012){Salvaterra}, {Campana}, {Vergani},
  {Covino}, {D'Avanzo}, {Fugazza}, {Ghirlanda}, {Ghisellini}, {Melandri},
  {Nava}, {Sbarufatti}, {Flores}, {Piranomonte}, \&
  {Tagliaferri}}]{Salvaterra2012}
{Salvaterra}, R., {Campana}, S., {Vergani}, S.~D., {et~al.} 2012, \apj, 749, 68

\bibitem[{{Salvaterra} {et~al.}(2009){Salvaterra}, {Della Valle}, {Campana},
  {Chincarini}, {Covino}, {D'Avanzo}, {Fern{\'a}ndez-Soto}, {Guidorzi},
  {Mannucci}, {Margutti}, {Th{\"o}ne}, {Antonelli}, {Barthelmy}, {de Pasquale},
  {D'Elia}, {Fiore}, {Fugazza}, {Hunt}, {Maiorano}, {Marinoni}, {Marshall},
  {Molinari}, {Nousek}, {Pian}, {Racusin}, {Stella}, {Amati}, {Andreuzzi},
  {Cusumano}, {Fenimore}, {Ferrero}, {Giommi}, {Guetta}, {Holland}, {Hurley},
  {Israel}, {Mao}, {Markwardt}, {Masetti}, {Pagani}, {Palazzi}, {Palmer},
  {Piranomonte}, {Tagliaferri}, \& {Testa}}]{Salvaterra2009}
{Salvaterra}, R., {Della Valle}, M., {Campana}, S., {et~al.} 2009, \nat, 461,
  1258

\bibitem[{{Salvaterra} {et~al.}(2013){Salvaterra}, {Maio}, {Ciardi}, \&
  {Campisi}}]{Salvaterra2013}
{Salvaterra}, R., {Maio}, U., {Ciardi}, B., \& {Campisi}, M.~A. 2013, \mnras,
  429, 2718

\bibitem[{{Savaglio}(2015)}]{Savaglio2015}
{Savaglio}, S. 2015, Journal of High Energy Astrophysics, 7, 95

\bibitem[{{Savaglio} {et~al.}(2012){Savaglio}, {Rau}, {Greiner}, {Kr{\"u}hler},
  {McBreen}, {Hartmann}, {Updike}, {Filgas}, {Klose}, {Afonso}, {Clemens},
  {K{\"u}pc{\"u} Yolda{\c{s}}}, {Olivares E.}, {Sudilovsky}, \&
  {Szokoly}}]{Savaglio2012}
{Savaglio}, S., {Rau}, A., {Greiner}, J., {et~al.} 2012, \mnras, 420, 627

\bibitem[{{Schady} {et~al.}(2011){Schady}, {Savaglio}, {Kr{\"u}hler},
  {Greiner}, \& {Rau}}]{Schady2011}
{Schady}, P., {Savaglio}, S., {Kr{\"u}hler}, T., {Greiner}, J., \& {Rau}, A.
  2011, \aap, 525, A113

\bibitem[{{Selsing} {et~al.}(2019){Selsing}, {Malesani}, {Goldoni}, {Fynbo},
  {Kr{\"u}hler}, {Antonelli}, {Arabsalmani}, {Bolmer}, {Cano}, {Christensen},
  {Covino}, {D'Avanzo}, {D'Elia}, {De Cia}, {de Ugarte Postigo}, {Flores},
  {Friis}, {Gomboc}, {Greiner}, {Groot}, {Hammer}, {Hartoog}, {Heintz},
  {Hjorth}, {Jakobsson}, {Japelj}, {Kann}, {Kaper}, {Ledoux}, {Leloudas},
  {Levan}, {Maiorano}, {Melandri}, {Milvang-Jensen}, {Palazzi}, {Palmerio},
  {Perley}, {Pian}, {Piranomonte}, {Pugliese}, {S{\'a}nchez-Ram{\'\i}rez},
  {Savaglio}, {Schady}, {Schulze}, {Sollerman}, {Sparre}, {Tagliaferri},
  {Tanvir}, {Th{\"o}ne}, {Vergani}, {Vreeswijk}, {Watson}, {Wiersema},
  {Wijers}, {Xu}, \& {Zafar}}]{Selsing2019}
{Selsing}, J., {Malesani}, D., {Goldoni}, P., {et~al.} 2019, \aap, 623, A92

\bibitem[{{Shibuya} {et~al.}(2015){Shibuya}, {Ouchi}, \&
  {Harikane}}]{Shibuya2015}
{Shibuya}, T., {Ouchi}, M., \& {Harikane}, Y. 2015, \apjs, 219, 15

\bibitem[{{Sparre} {et~al.}(2014){Sparre}, {Hartoog}, {Kr{\"u}hler}, {Fynbo},
  {Watson}, {Wiersema}, {D'Elia}, {Zafar}, {Afonso}, {Covino}, {de Ugarte
  Postigo}, {Flores}, {Goldoni}, {Greiner}, {Hjorth}, {Jakobsson}, {Kaper},
  {Klose}, {Levan}, {Malesani}, {Milvang-Jensen}, {Nardini}, {Piranomonte},
  {Sollerman}, {S{\'a}nchez-Ram{\'{\i}}rez}, {Schulze}, {Tanvir}, {Vergani}, \&
  {Wijers}}]{Sparre2014}
{Sparre}, M., {Hartoog}, O.~E., {Kr{\"u}hler}, T., {et~al.} 2014, \apj, 785,
  150

\bibitem[{{Steidel} {et~al.}(2016){Steidel}, {Strom}, {Pettini}, {Rudie},
  {Reddy}, \& {Trainor}}]{Steidel2016}
{Steidel}, C.~C., {Strom}, A.~L., {Pettini}, M., {et~al.} 2016, \apj, 826, 159

\bibitem[{{Strausbaugh} \&
  {Cucchiara}(2021{\natexlab{a}})}]{GCN2021GRB210905AStrausbaugh_2}
{Strausbaugh}, R. \& {Cucchiara}, A. 2021{\natexlab{a}}, GRB Coordinates
  Network, 30770, 1

\bibitem[{{Strausbaugh} \&
  {Cucchiara}(2021{\natexlab{b}})}]{GCN2021GRB210905AStrausbaugh}
{Strausbaugh}, R. \& {Cucchiara}, A. 2021{\natexlab{b}}, GRB Coordinates
  Network, 30769, 1

\bibitem[{{Suda} {et~al.}(2008){Suda}, {Katsuta}, {Yamada}, {Suwa}, {Ishizuka},
  {Komiya}, {Sorai}, {Aikawa}, \& {Fujimoto}}]{Suda2008}
{Suda}, T., {Katsuta}, Y., {Yamada}, S., {et~al.} 2008, \pasj, 60, 1159

\bibitem[{{Tanvir} {et~al.}(2021{\natexlab{a}}){Tanvir}, {Rossi}, {Xu}, {Zhu},
  {Izzo}, {Kann}, {Levan}, \& {Stargate
  Collaboration}}]{GCN2021GRB210905ATanvir}
{Tanvir}, N., {Rossi}, A., {Xu}, D., {et~al.} 2021{\natexlab{a}}, GRB
  Coordinates Network, 30771, 1

\bibitem[{{Tanvir} {et~al.}(2009){Tanvir}, {Fox}, {Levan}, {Berger},
  {Wiersema}, {Fynbo}, {Cucchiara}, {Kr{\"u}hler}, {Gehrels}, {Bloom},
  {Greiner}, {Evans}, {Rol}, {Olivares}, {Hjorth}, {Jakobsson}, {Farihi},
  {Willingale}, {Starling}, {Cenko}, {Perley}, {Maund}, {Duke}, {Wijers},
  {Adamson}, {Allan}, {Bremer}, {Burrows}, {Castro-Tirado}, {Cavanagh}, {de
  Ugarte Postigo}, {Dopita}, {Fatkhullin}, {Fruchter}, {Foley}, {Gorosabel},
  {Kennea}, {Kerr}, {Klose}, {Krimm}, {Komarova}, {Kulkarni}, {Moskvitin},
  {Mundell}, {Naylor}, {Page}, {Penprase}, {Perri}, {Podsiadlowski}, {Roth},
  {Rutledge}, {Sakamoto}, {Schady}, {Schmidt}, {Soderberg}, {Sollerman},
  {Stephens}, {Stratta}, {Ukwatta}, {Watson}, {Westra}, {Wold}, \&
  {Wolf}}]{Tanvir2009}
{Tanvir}, N.~R., {Fox}, D.~B., {Levan}, A.~J., {et~al.} 2009, \nat, 461, 1254

\bibitem[{{Tanvir} {et~al.}(2019){Tanvir}, {Fynbo}, {de Ugarte Postigo},
  {Japelj}, {Wiersema}, {Malesani}, {Perley}, {Levan}, {Selsing}, {Cenko},
  {Kann}, {Milvang-Jensen}, {Berger}, {Cano}, {Chornock}, {Covino},
  {Cucchiara}, {D'Elia}, {Gargiulo}, {Goldoni}, {Gomboc}, {Heintz}, {Hjorth},
  {Izzo}, {Jakobsson}, {Kaper}, {Kr{\"u}hler}, {Laskar}, {Myers},
  {Piranomonte}, {Pugliese}, {Rossi}, {S{\'a}nchez-Ram{\'\i}rez}, {Schulze},
  {Sparre}, {Stanway}, {Tagliaferri}, {Th{\"o}ne}, {Vergani}, {Vreeswijk},
  {Wijers}, {Watson}, \& {Xu}}]{Tanvir2019}
{Tanvir}, N.~R., {Fynbo}, J.~P.~U., {de Ugarte Postigo}, A., {et~al.} 2019,
  \mnras, 483, 5380

\bibitem[{{Tanvir} {et~al.}(2021{\natexlab{b}}){Tanvir}, {Le Floc'h},
  {Christensen}, {Caruana}, {Salvaterra}, {Ghirlanda}, {Ciardi}, {Maio},
  {D'Odorico}, {Piedipalumbo}, {Campana}, {Noterdaeme}, {Graziani}, {Amati},
  {Bagoly}, {Bal{\'a}zs}, {Basa}, {Behar}, {De Cia}, {Della Valle}, {De
  Pasquale}, {Frontera}, {Gomboc}, {G{\"o}tz}, {Horvath}, {Hudec},
  {Mereghetti}, {O'Brien}, {Osborne}, {Paltani}, {Rosati}, {Sergijenko},
  {Stanway}, {Sz{\'e}csi}, {Toth}, {Urata}, {Vergani}, \& {Zane}}]{Tanvir2021}
{Tanvir}, N.~R., {Le Floc'h}, E., {Christensen}, L., {et~al.}
  2021{\natexlab{b}}, Experimental Astronomy, 52, 219

\bibitem[{{Tanvir} {et~al.}(2012){Tanvir}, {Levan}, {Fruchter}, {Fynbo},
  {Hjorth}, {Wiersema}, {Bremer}, {Rhoads}, {Jakobsson}, {O'Brien}, {Stanway},
  {Bersier}, {Natarajan}, {Greiner}, {Watson}, {Castro-Tirado}, {Wijers},
  {Starling}, {Misra}, {Graham}, \& {Kouveliotou}}]{Tanvir2012}
{Tanvir}, N.~R., {Levan}, A.~J., {Fruchter}, A.~S., {et~al.} 2012, \apj, 754,
  46

\bibitem[{{Telikova} {et~al.}(2022){Telikova}, {Balashev}, {Noterdaeme},
  {Krogager}, \& {Ranjan}}]{Telikova2022}
{Telikova}, K.~N., {Balashev}, S.~A., {Noterdaeme}, P., {Krogager}, J.~K., \&
  {Ranjan}, A. 2022, \mnras, 510, 5974

\bibitem[{{Th{\"o}ne} {et~al.}(2013){Th{\"o}ne}, {Fynbo}, {Goldoni}, {de
  Ugarte}, {Campana}, {Vergani}, {Covino}, {Kr{\"u}hler}, {Kaper}, {Tanvir},
  {Zafar}, {D'Elia}, {Gorosabel}, {Greiner}, {Groot}, {Hammer}, {Jakobsson},
  {Klose}, {Levan}, {Milvang-Jensen}, {Nicuesa}, {Palazzi}, {Piranomonte},
  {Tagliaferri}, {Watson}, {Wiersema}, \& {Wijers}}]{Thone2013}
{Th{\"o}ne}, C.~C., {Fynbo}, J.~P.~U., {Goldoni}, P., {et~al.} 2013, \mnras,
  428, 3590

\bibitem[{{Tinsley}(1979)}]{Tinsley1979}
{Tinsley}, B.~M. 1979, \apj, 229, 1046

\bibitem[{{Tumlinson} {et~al.}(2017){Tumlinson}, {Peeples}, \&
  {Werk}}]{Tumlinson2017}
{Tumlinson}, J., {Peeples}, M.~S., \& {Werk}, J.~K. 2017, \araa, 55, 389

\bibitem[{{Vergani} {et~al.}(2009){Vergani}, {Petitjean}, {Ledoux},
  {Vreeswijk}, {Smette}, \& {Meurs}}]{Vergani2009}
{Vergani}, S.~D., {Petitjean}, P., {Ledoux}, C., {et~al.} 2009, \aap, 503, 771

\bibitem[{{Vergani} {et~al.}(2015){Vergani}, {Salvaterra}, {Japelj}, {Le
  Floc'h}, {D'Avanzo}, {Fernandez-Soto}, {Kr{\"u}hler}, {Melandri}, {Boissier},
  {Covino}, {Puech}, {Greiner}, {Hunt}, {Perley}, {Petitjean}, {Vinci},
  {Hammer}, {Levan}, {Mannucci}, {Campana}, {Flores}, {Gomboc}, \&
  {Tagliaferri}}]{Vergani2015}
{Vergani}, S.~D., {Salvaterra}, R., {Japelj}, J., {et~al.} 2015, \aap, 581,
  A102

\bibitem[{{Verhamme} {et~al.}(2008){Verhamme}, {Schaerer}, {Atek}, \&
  {Tapken}}]{Verhamme2008}
{Verhamme}, A., {Schaerer}, D., {Atek}, H., \& {Tapken}, C. 2008, \aap, 491, 89

\bibitem[{{Verhamme} {et~al.}(2006){Verhamme}, {Schaerer}, \&
  {Maselli}}]{Verhamme2006}
{Verhamme}, A., {Schaerer}, D., \& {Maselli}, A. 2006, \aap, 460, 397

\bibitem[{{Vernet} {et~al.}(2011){Vernet}, {Dekker}, {D'Odorico}, {Kaper},
  {Kjaergaard}, {Hammer}, {Randich}, {Zerbi}, {Groot}, {Hjorth}, {Guinouard},
  {Navarro}, {Adolfse}, {Albers}, {Amans}, {Andersen}, {Andersen}, {Binetruy},
  {Bristow}, {Castillo}, {Chemla}, {Christensen}, {Conconi}, {Conzelmann},
  {Dam}, {de Caprio}, {de Ugarte Postigo}, {Delabre}, {di Marcantonio},
  {Downing}, {Elswijk}, {Finger}, {Fischer}, {Flores}, {Fran{\c c}ois},
  {Goldoni}, {Guglielmi}, {Haigron}, {Hanenburg}, {Hendriks}, {Horrobin},
  {Horville}, {Jessen}, {Kerber}, {Kern}, {Kiekebusch}, {Kleszcz}, {Klougart},
  {Kragt}, {Larsen}, {Lizon}, {Lucuix}, {Mainieri}, {Manuputy}, {Martayan},
  {Mason}, {Mazzoleni}, {Michaelsen}, {Modigliani}, {Moehler}, {M{\o}ller},
  {Norup S{\o}rensen}, {N{\o}rregaard}, {P{\'e}roux}, {Patat}, {Pena}, {Pragt},
  {Reinero}, {Rigal}, {Riva}, {Roelfsema}, {Royer}, {Sacco}, {Santin},
  {Schoenmaker}, {Spano}, {Sweers}, {Ter Horst}, {Tintori}, {Tromp}, {van
  Dael}, {van der Vliet}, {Venema}, {Vidali}, {Vinther}, {Vola}, {Winters},
  {Wistisen}, {Wulterkens}, \& {Zacchei}}]{Vernet2011}
{Vernet}, J., {Dekker}, H., {D'Odorico}, S., {et~al.} 2011, \aap, 536, A105

\bibitem[{{Vielfaure} {et~al.}(2021){Vielfaure}, {Vergani}, {Gronke}, {Japelj},
  {Palmerio}, {Fynbo}, {Malesani}, {Milvang-Jensen}, {Salvaterra}, \&
  {Tanvir}}]{Vielfaure2021}
{Vielfaure}, J.~B., {Vergani}, S.~D., {Gronke}, M., {et~al.} 2021, \aap, 653,
  A83

\bibitem[{{Vielfaure} {et~al.}(2020){Vielfaure}, {Vergani}, {Japelj}, {Fynbo},
  {Gronke}, {Heintz}, {Malesani}, {Petitjean}, {Tanvir}, {D'Elia}, {Kann},
  {Palmerio}, {Salvaterra}, {Wiersema}, {Arabsalmani}, {Campana}, {Covino}, {De
  Pasquale}, {de Ugarte Postigo}, {Hammer}, {Hartmann}, {Jakobsson},
  {Kouveliotou}, {Laskar}, {Levan}, \& {Rossi}}]{Vielfaure2020}
{Vielfaure}, J.~B., {Vergani}, S.~D., {Japelj}, J., {et~al.} 2020, \aap, 641,
  A30

\bibitem[{{Vreeswijk} {et~al.}(2012){Vreeswijk}, {Ledoux}, {De Cia}, \&
  {Smette}}]{Vreeswijk2012}
{Vreeswijk}, P.~M., {Ledoux}, C., {De Cia}, A., \& {Smette}, A. 2012, Memorie
  della Societa Astronomica Italiana Supplementi, 21, 14

\bibitem[{{Vreeswijk} {et~al.}(2013){Vreeswijk}, {Ledoux}, {Raassen}, {Smette},
  {De Cia}, {Wo{\'z}niak}, {Fox}, {Vestrand}, \& {Jakobsson}}]{Vreeswijk2013}
{Vreeswijk}, P.~M., {Ledoux}, C., {Raassen}, A.~J.~J., {et~al.} 2013, \aap,
  549, A22

\bibitem[{{Vreeswijk} {et~al.}(2007){Vreeswijk}, {Ledoux}, {Smette}, {Ellison},
  {Jaunsen}, {Andersen}, {Fruchter}, {Fynbo}, {Hjorth}, {Kaufer}, {M{\o}ller},
  {Petitjean}, {Savaglio}, \& {Wijers}}]{Vreeswijk2007}
{Vreeswijk}, P.~M., {Ledoux}, C., {Smette}, A., {et~al.} 2007, \aap, 468, 83

\bibitem[{{Watson} {et~al.}(2007){Watson}, {Hjorth}, {Fynbo}, {Jakobsson},
  {Foley}, {Sollerman}, \& {Wijers}}]{Watson2007}
{Watson}, D., {Hjorth}, J., {Fynbo}, J.~P.~U., {et~al.} 2007, \apjl, 660, L101

\bibitem[{{Watson} {et~al.}(2013){Watson}, {Zafar}, {Andersen}, {Fynbo},
  {Gorosabel}, {Hjorth}, {Jakobsson}, {Kr{\"u}hler}, {Laursen}, {Leloudas}, \&
  {Malesani}}]{Watson2013}
{Watson}, D., {Zafar}, T., {Andersen}, A.~C., {et~al.} 2013, \apj, 768, 23

\bibitem[{{Whalen} {et~al.}(2008){Whalen}, {Prochaska}, {Heger}, \&
  {Tumlinson}}]{Whalen2008}
{Whalen}, D., {Prochaska}, J.~X., {Heger}, A., \& {Tumlinson}, J. 2008, \apj,
  682, 1114

\bibitem[{{White} {et~al.}(2021){White}, {Bauer}, {Baumgartner}, {Bautz},
  {Berger}, {Cenko}, {Chang}, {Falcone}, {Fausey}, {Feldman}, {Fox}, {Fox},
  {Fruchter}, {Fryer}, {Ghirlanda}, {Gorski}, {Grant}, {Guiriec}, {Hart},
  {Hartmann}, {Hennawi}, {Kann}, {Kaplan}, {Kennea}, {Kocevski}, {Kouveliotou},
  {Lawrence}, {Levan}, {Lidz}, {Lien}, {Littenberg}, {Mas-Ribas}, {Moss},
  {O'Brien}, {O'Meara}, {Palmer}, {Pasham}, {Racusin}, {Remillard}, {Roberts},
  {Roming}, {Rud}, {Salvaterra}, {Sambruna}, {Seiffert}, {Sun}, {Tanvir},
  {Terrile}, {Thomas}, {van der Horst}, {Verstrand}, {Willems}, {Wilson-Hodge},
  {Young}, {Amati}, {Bozzo}, {Karczewski}, {Hernandez-Monteagudo}, {Rebolo
  Lopez}, {Genova-Santos}, {Martin}, {Granot}, {Bemiamini}, {Gil}, \&
  {Burns}}]{White2021}
{White}, N.~E., {Bauer}, F.~E., {Baumgartner}, W., {et~al.} 2021, in Society of
  Photo-Optical Instrumentation Engineers (SPIE) Conference Series, Vol. 11821,
  Society of Photo-Optical Instrumentation Engineers (SPIE) Conference Series,
  1182109

\bibitem[{{Wiseman} {et~al.}(2017){Wiseman}, {Perley}, {Schady}, {Prochaska},
  {de Ugarte Postigo}, {Kr{\"u}hler}, {Yates}, \& {Greiner}}]{Wiseman2017}
{Wiseman}, P., {Perley}, D.~A., {Schady}, P., {et~al.} 2017, \aap, 607, A107

\bibitem[{{Wolfe} {et~al.}(2003){Wolfe}, {Prochaska}, \&
  {Gawiser}}]{Wolfe_SFRmedium2003}
{Wolfe}, A.~M., {Prochaska}, J.~X., \& {Gawiser}, E. 2003, \apj, 593, 215

\bibitem[{{Wolfe} {et~al.}(2008){Wolfe}, {Prochaska}, {Jorgenson}, \&
  {Rafelski}}]{Wolfe2008}
{Wolfe}, A.~M., {Prochaska}, J.~X., {Jorgenson}, R.~A., \& {Rafelski}, M. 2008,
  \apj, 681, 881

\bibitem[{{Yang} {et~al.}(2022){Yang}, {Leethochawalit}, {Treu},
  {Roberts-Borsani}, {Brada{\v{c}}}, {Birrer}, {Castellano}, {Merlin},
  {Fontana}, {Amorin}, \& {Trenti}}]{Yang2022}
{Yang}, L., {Leethochawalit}, N., {Treu}, T., {et~al.} 2022, \mnras, 514, 1148

\bibitem[{{Zafar} \& {M{\o}ller}(2019)}]{zafar19}
{Zafar}, T. \& {M{\o}ller}, P. 2019, \mnras, 482, 2731

\bibitem[{{Zafar} {et~al.}(2013){Zafar}, {P{\'e}roux}, {Popping}, {Milliard},
  {Deharveng}, \& {Frank}}]{zafar13}
{Zafar}, T., {P{\'e}roux}, C., {Popping}, A., {et~al.} 2013, \aap, 556, A141

\end{thebibliography}

\begin{appendix}
\section{Additional Figures and Tables}

\begin{figure*}
   \centering
   \includegraphics[width=1\textwidth]{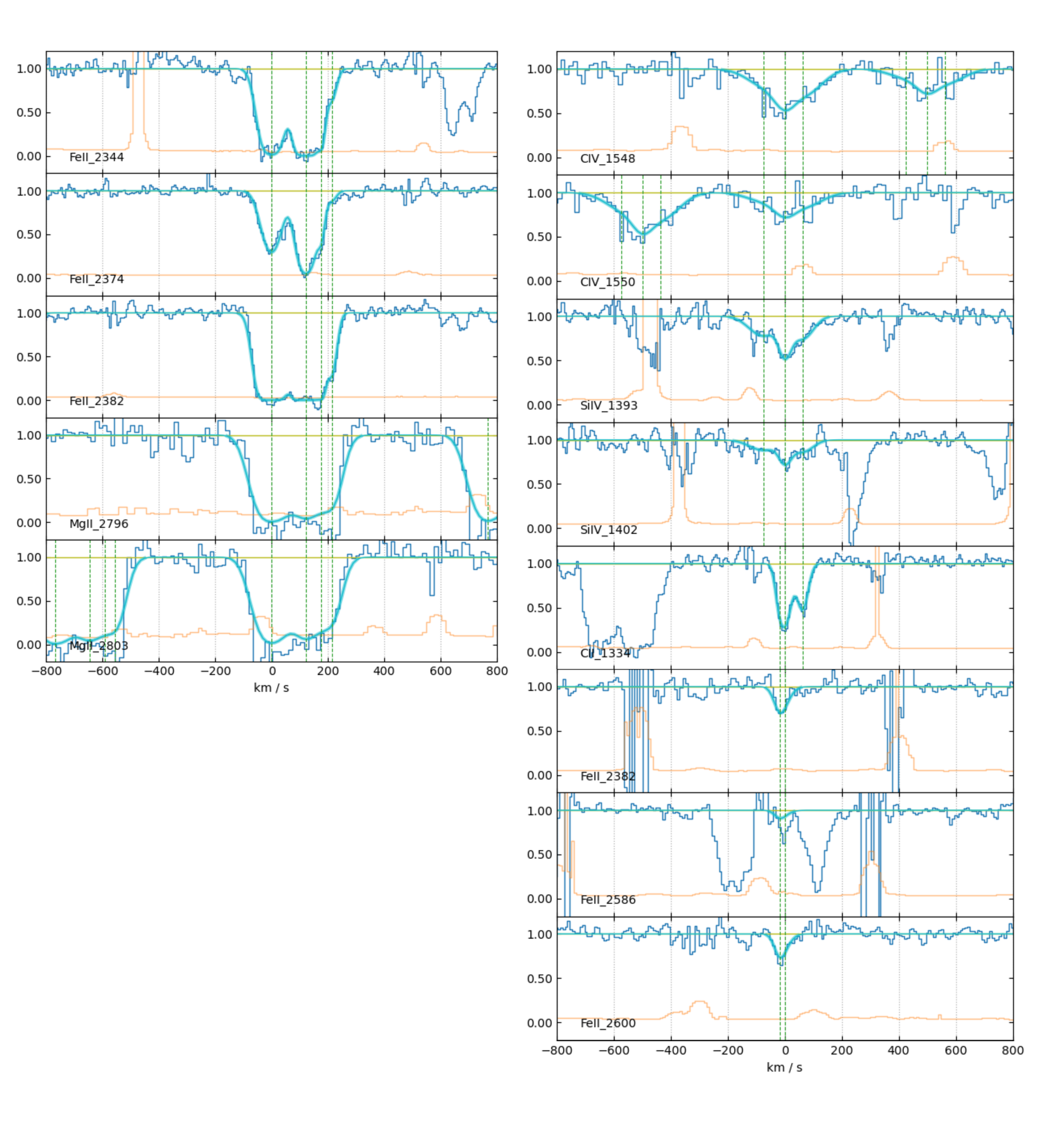}
    \caption{VLT/X-shooter optical and NIR spectrum of the intervening systems. {\it Left panel}: \feii{} and \mgii{} absorption of the $z=2.8296$ foreground absorber. {\it Right panel}: Low-and high-ionization absorption of the $z=5.7390$ foreground absorber. For both panels the zero velocity is fixed to the above-mentioned redshift for each absorber.}
         \label{intervening}
   \end{figure*}
   
\clearpage

   \begin{figure*}
  \centering
  \includegraphics[width=0.95\textwidth]{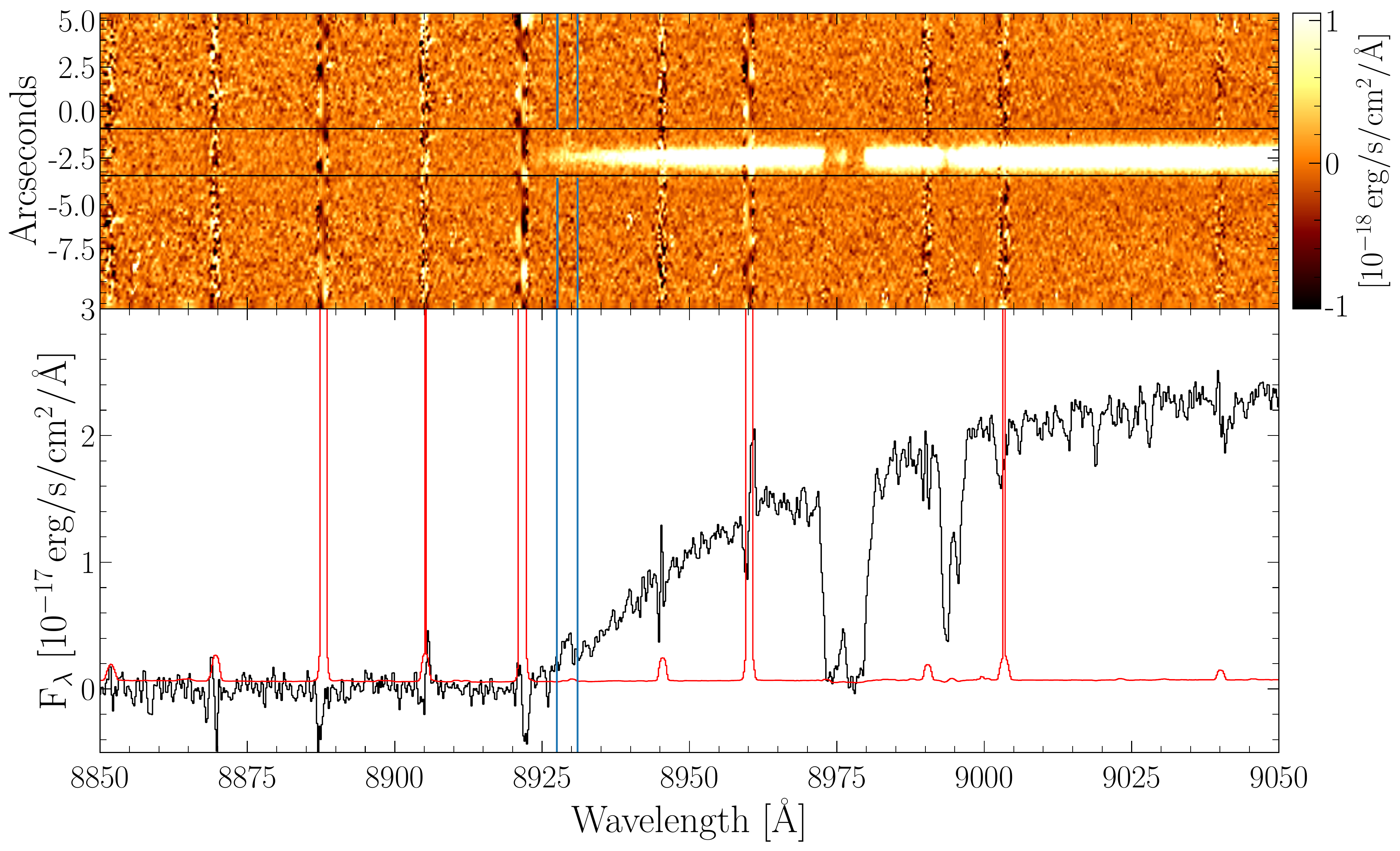}
      \caption{Tentative detection of Lyman-$\alpha$ emission at $z=6.3449$. {\it Upper panel}: part of the X-shooter 2D spectrum. Wavelengths are reported in the observer frame.
      {\it Lower panel}: 1D spectrum corresponding to the extraction between the black horizontal lines of the 2D spectrum (from $-0.8\arcsec$ to $-3.3\arcsec$). In both panels the tentative Lyman-$\alpha$ emission detection is indicated by the blue lines surrounding it. The red line corresponds to the noise spectrum.}
          \label{figlya}
  \end{figure*}
  
   \begin{figure*}
\centering
\subfloat{\label{L_distance_correlation}
\centering
\includegraphics[width=0.45\linewidth]{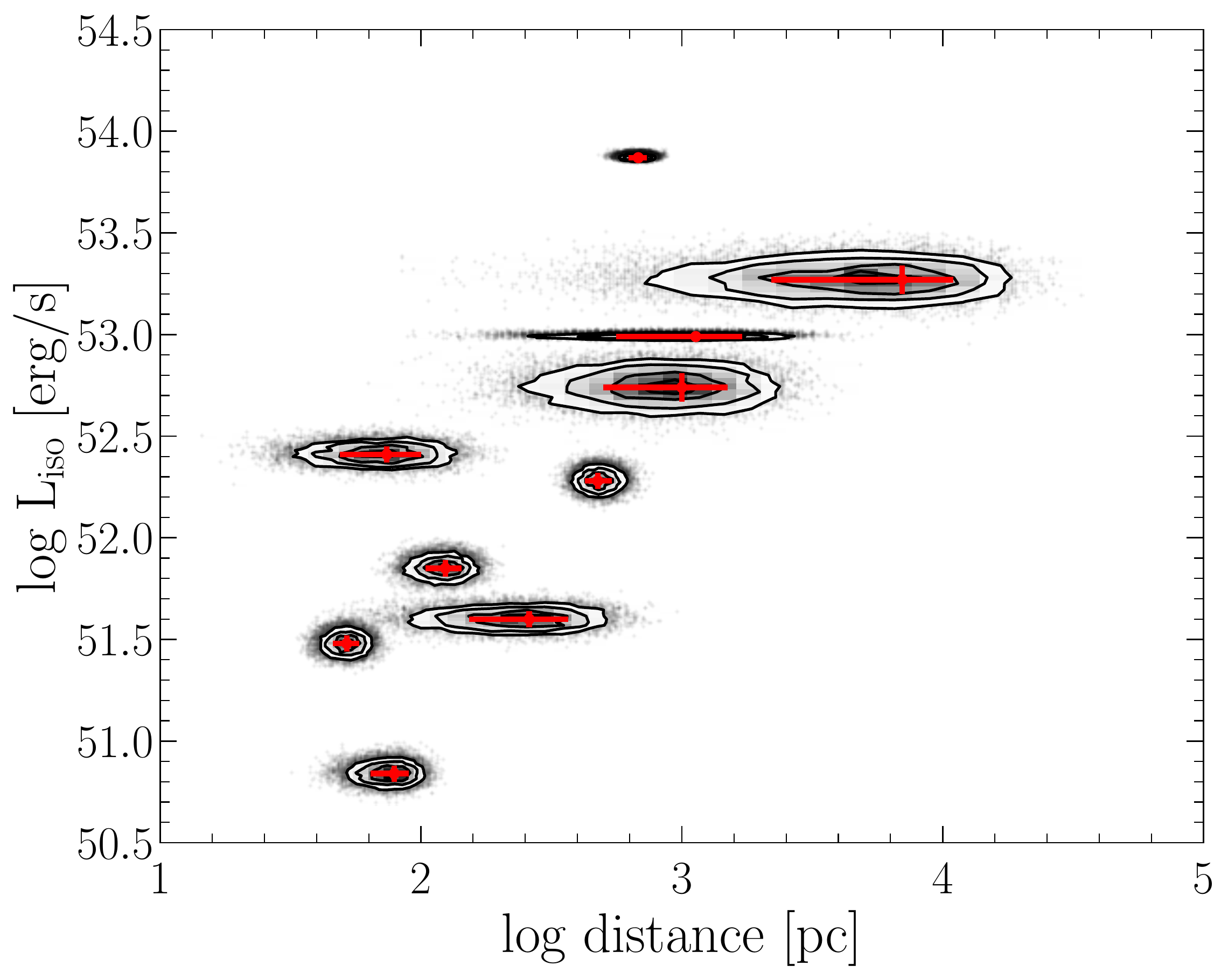}
}
\hfill
\subfloat{\label{rho_spearman}
\centering
\includegraphics[width=0.47\linewidth]{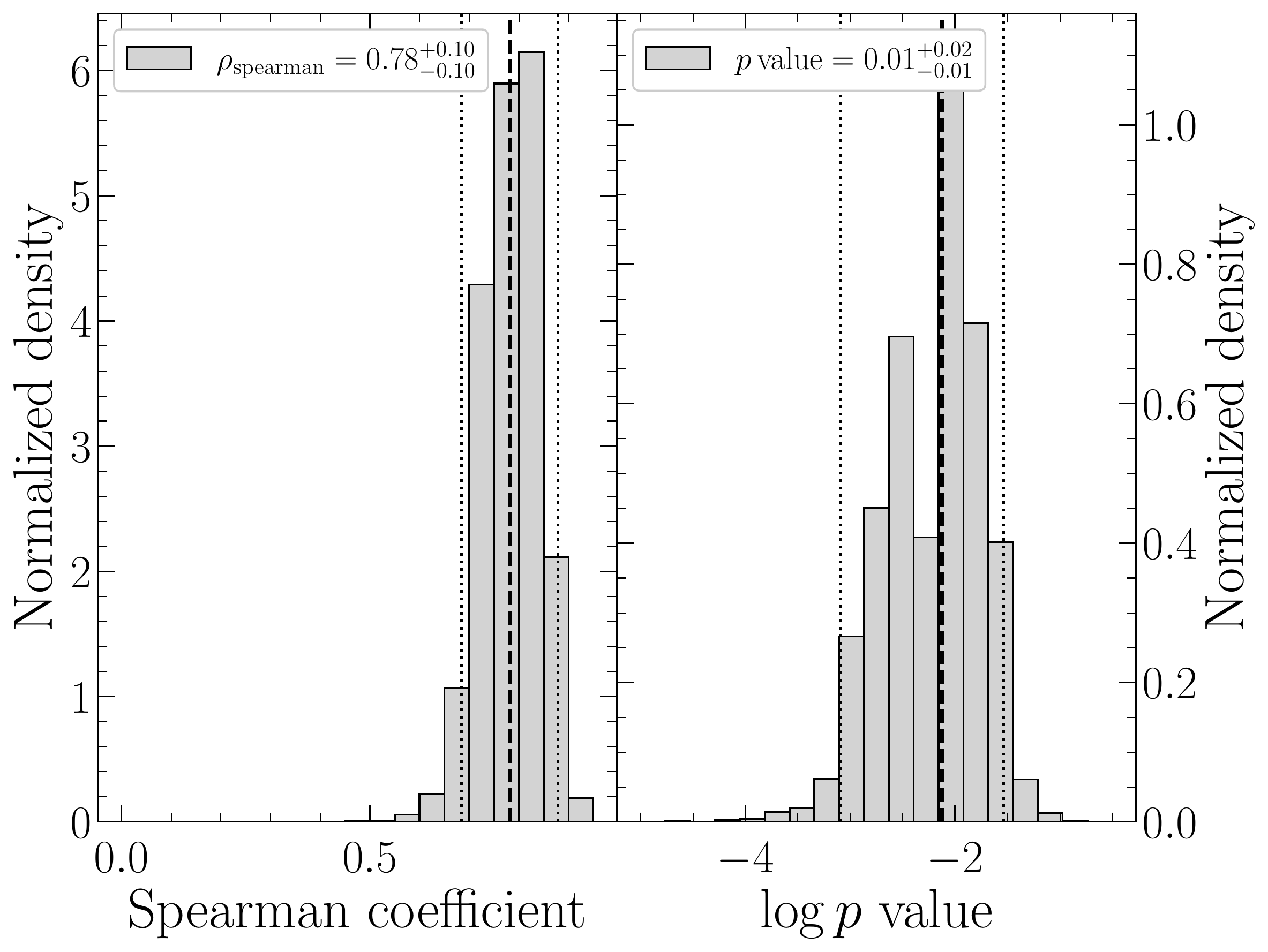}
}
\caption{Correlation between the distance of the closest low-ionization gas clouds and the GRB luminosity. {\it Left panel}: Monte Carlo simulation of the $L_{\rm iso}$-distance relation. In red the $L_{\rm iso}$ (erg\,s$^{-1}$) and distance (parsecs) values with their error bars for the GRBs with available measurements, while the contours represent the result of the simulation of N=10000 realizations. {\it Right panel}: Probability density function of the Spearman coefficient and the associated $p$-value for each realization. The black dashed line represents the median value while the dotted ones represent the 2\,$\sigma$ intervals.}
\label{Correlation}
\end{figure*}

 \begin{figure*}
   \centering
      \includegraphics[width=0.8\textwidth]{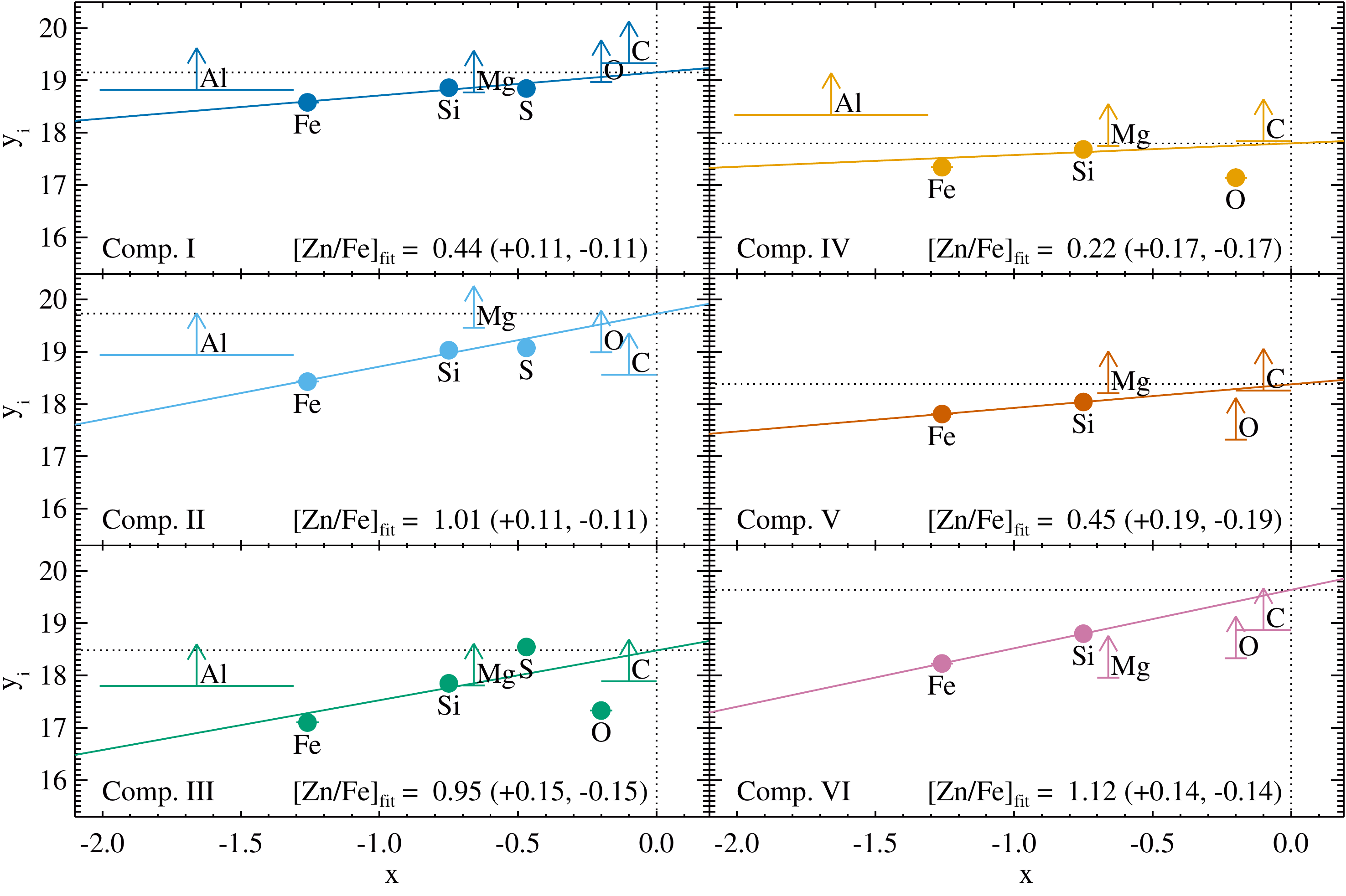}
      \caption{Similar to Fig. \ref{depletion tot}, but for the individual components of the absorption-line profile. In this case, the information on the \hi{} is not available, and therefore we use a slightly different notation, where $y_i$ can be interpreted as an equivalent metal column and it is defined in Ramburuth-Hurt et al. in prep. Again, the slope of the linear relations that can be fit through the data represent the overall strength of dust depletion, [Zn/Fe]$_{\rm fit}$, but this time the $y$-intercept at $x=0$ gives an estimate of the combination between the total metallicity and the \hi{} column. Here $y_i$ represents an equivalent metal column. The slope of the linear fit to the data (solid line) determines the overall strength of depletion [Zn/Fe]$_{\rm fit}$, as labeled.}
         \label{depletion comp}
   \end{figure*}

\begin{figure*}
   \centering
   \includegraphics[width=0.8\textwidth]{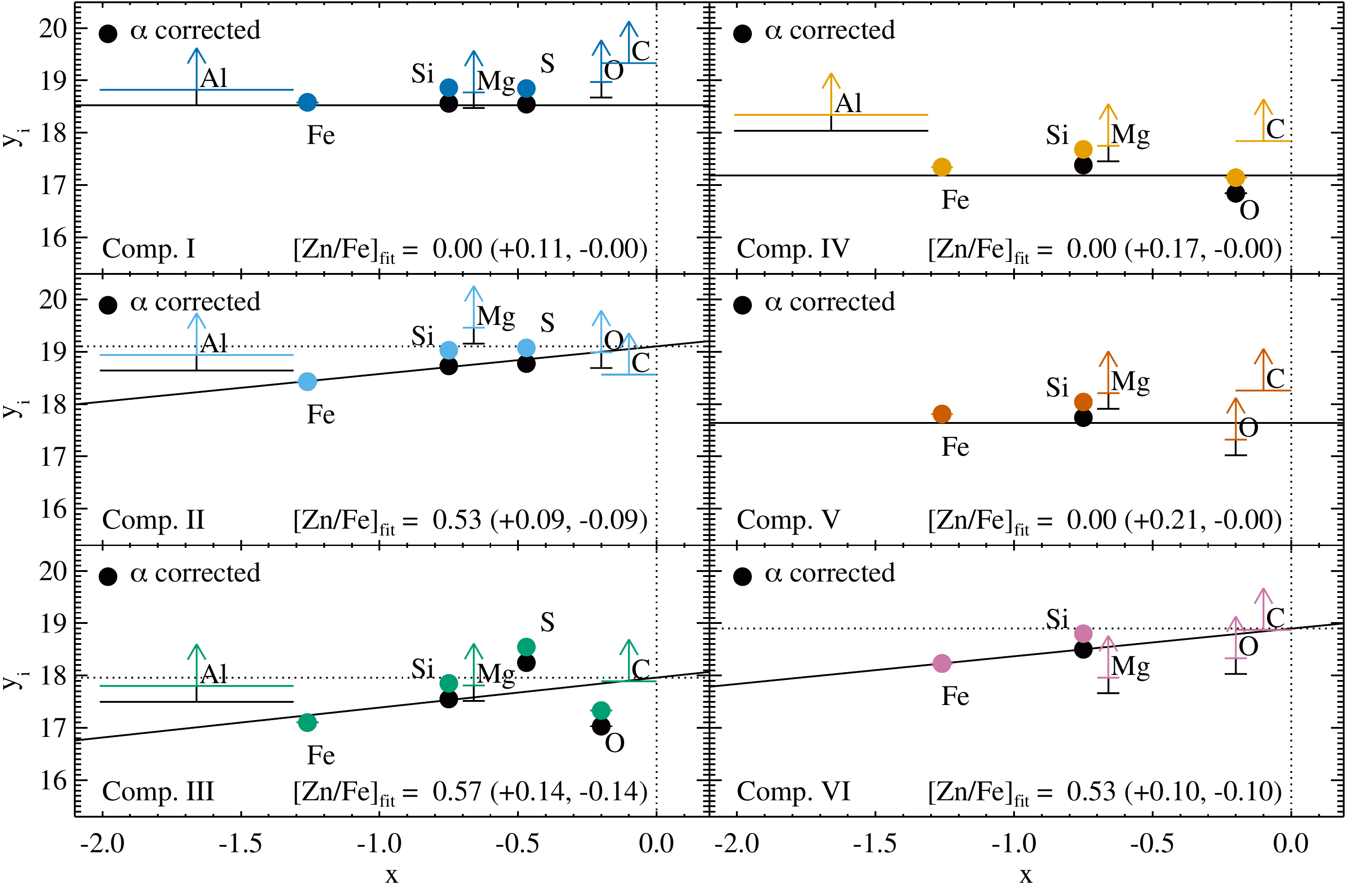}
      \caption{Same as Fig.\,\ref{depletion comp}, but with $\alpha$-element enhancement correction for the relevant elements (0.3\,dex; black points and limits).}
         \label{depletion comp alpha}
   \end{figure*}

\begin{table*}
\centering
\caption{Properties derived from the total metal abundances and component by component. The total metallicity ([M/H]$_{\rm tot}$), dust depletion [Zn/Fe]$_{\rm fit}$, dust-to-metal ratio ($DTM$) and dust extinction ($A_{\rm V}\,(mag)$) are reported for the analysis performed not taking $\alpha$-element enhancement into account.}
\label{table_results_noalpha}
\begin{tabular}{c c c c c c c c}       
\hline\hline
Without $\alpha$-element corr.  & $I$ & $II$ & $III$ & $IV$ & $V$ & $VI$ & Tot\\
\hline
\rule{0pt}{3ex}
[M/H]$_{\rm tot}$ & & & & & & & $-1.01\pm0.14$\\[2ex]

[Zn/Fe]$_{\rm fit}$&
$0.44^{+0.11}_{-0.11}$&
$1.01^{+0.11}_{-0.11}$&
$0.95^{+0.15}_{-0.15}$&
$0.22^{+0.17}_{-0.17}$&
$0.45^{+0.19}_{-0.19}$&
$1.12^{+0.14}_{-0.14}$&
$0.89\pm0.12$\\[2ex]

$DTM$ & 
$0.23^{+0.03}_{-0.03}$&
$0.38^{+0.04}_{-0.04}$&
$0.37^{+0.04}_{-0.04}$&
$0.13^{+0.04}_{-0.04}$&
$0.23^{+0.04}_{-0.04}$&
$0.41^{+0.05}_{-0.05}$&
$0.36\pm0.04$\\[2ex]

$A_{\rm V}\,(mag)$ & & & & & & & $0.04\pm0.02$\\[0.5ex]
\hline
\hline
\end{tabular}
\end{table*}

\begin{figure*}
\centering
\raisebox{2cm}{\includegraphics[width=0.3\textwidth]{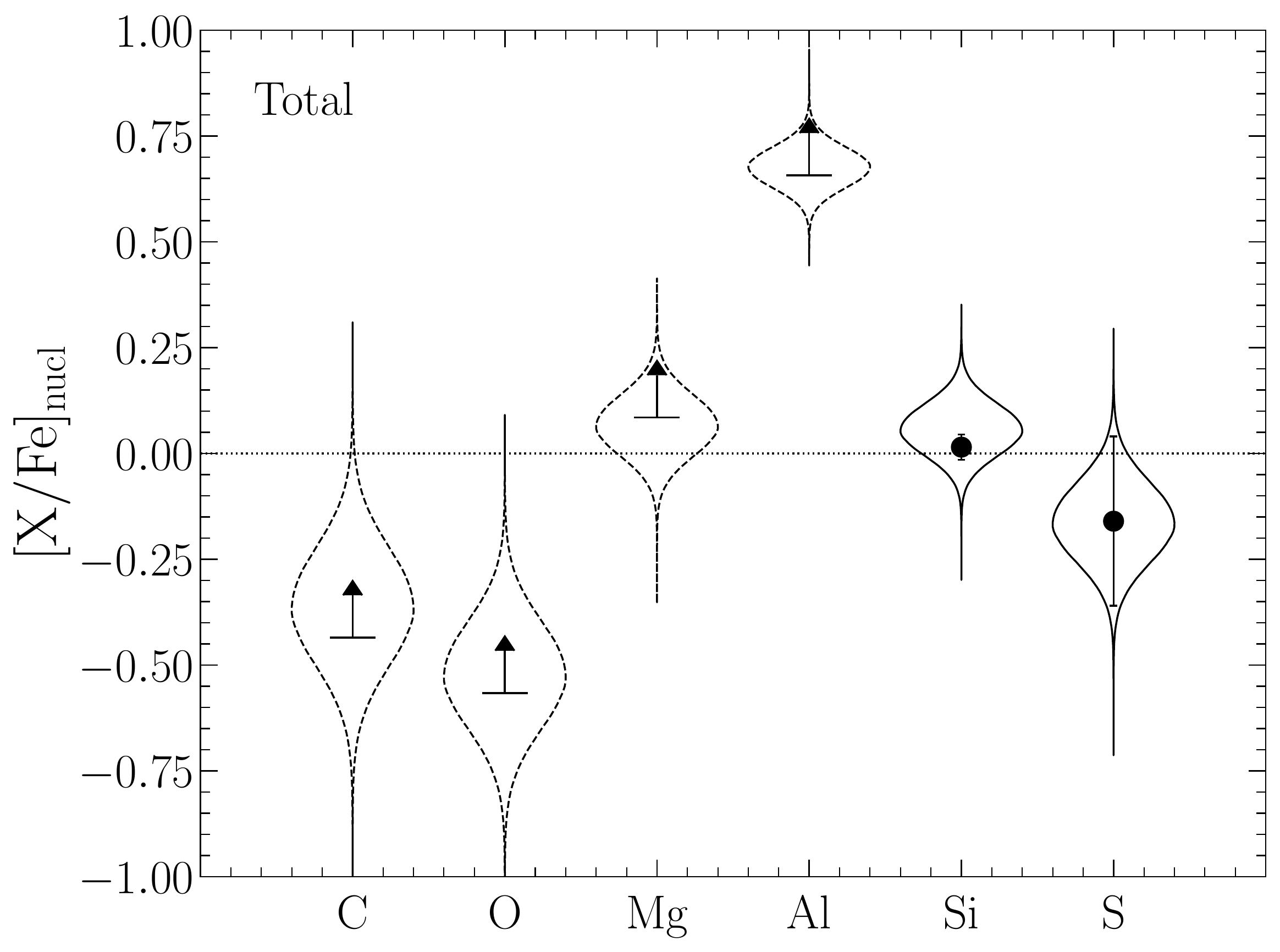}}
\hspace{0.5cm}
\includegraphics[width=0.65\textwidth]{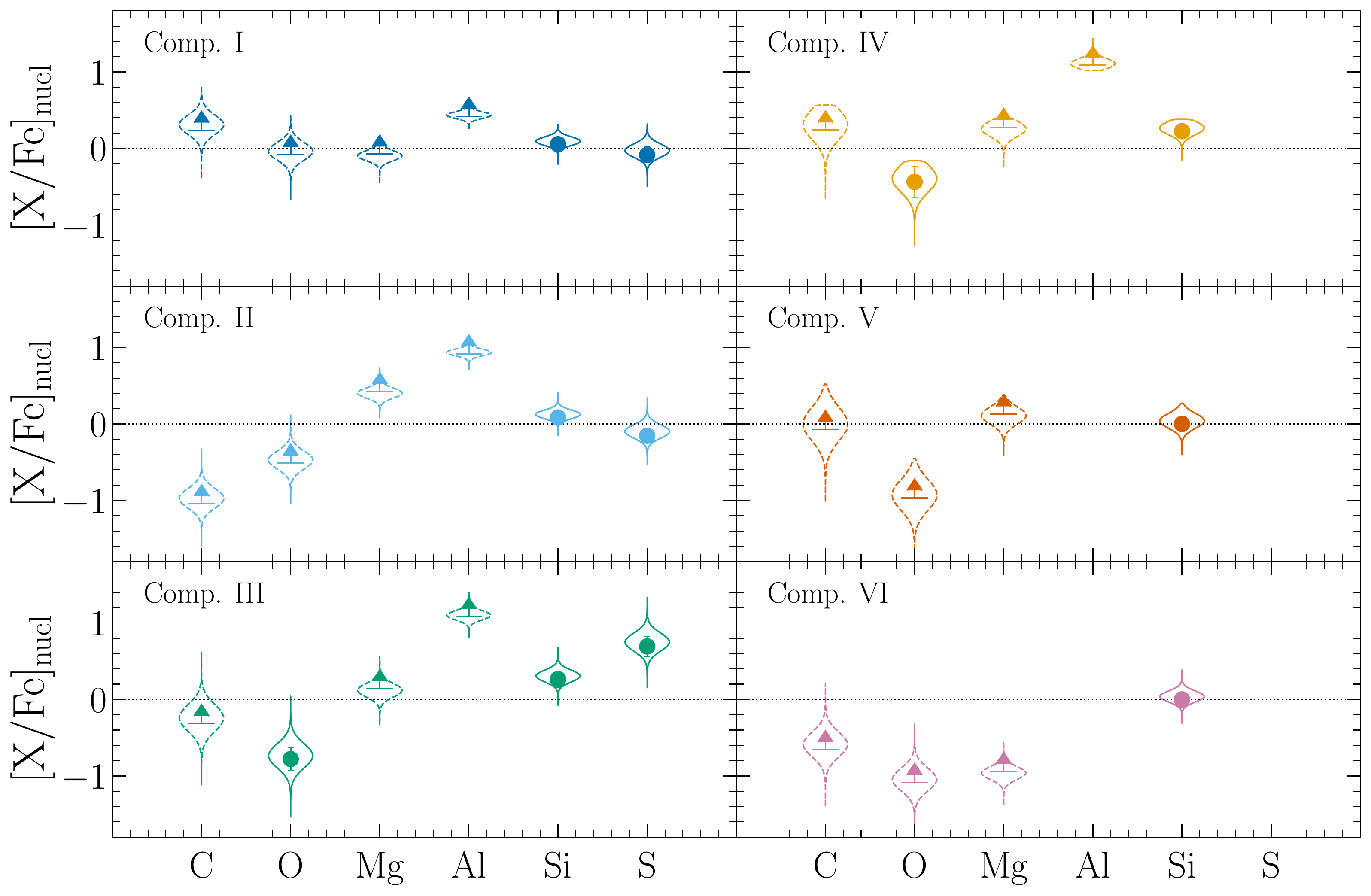}
\caption{Same as Fig. \ref{multicomp_alpha}, but without taking into account $\alpha$-element enhancement correction.}
\label{multicomp_noalpha}
\end{figure*}

\begin{table*}
\caption{[$X$/Fe] residuals of the depletion pattern fitting (see Figs. \ref{depletion tot}, \ref{depletion comp} and \ref{depletion comp alpha}). The 
values and errors are calculated only using the observational errors of the measured column densities without taking into account the uncertainties due to the dust depletion correction.}  
\label{table_abb} 
\centering
\begin{tabular}{c c c c c c c c}       
\hline\hline
With $\alpha$-element corr.  & $I$ & $II$ & $III$ & $IV$ & $V$ & $VI$ & Tot\\
\hline
\rule{0pt}{3ex}
[C/Fe] & $>0.75$ & $>-0.48$ & $>0.12$ & $>0.50$ & $>0.45$ & $>0.03$ & $>0.22$\\

[O/Fe] & $>0.38$ & $>0.00$ & $-0.37\pm0.15$ & $-0.2\pm0.2$ & $>-0.49$ & $>-0.46$ & $>0.03$\\

[Mg/Fe] & $>0.19$ & $>0.71$ & $>0.36$ & $>0.41$ & $>0.40$ & $>-0.59$ & $>0.42$\\

[Al/Fe] & $>0.24$ & $>0.72$ & $>0.93$ & $>1.00$ & & & $>0.43$\\

[Si/Fe] & $0.28\pm0.06$ & $0.33\pm0.04$ & $0.46\pm0.10$ & $0.34\pm0.09$ & $0.23\pm0.08$ & $0.30\pm0.03$ & $0.30\pm0.03$\\

[S/Fe] & $0.26\pm0.10$ & $0.23\pm0.09$ & $0.99\pm0.13$ & & & & $0.3\pm0.2$\\[1ex]
\hline
\hline
Without $\alpha$-element corr.  & $I$ & $II$ & $III$ & $IV$ & $V$ & $VI$ & Tot\\
\hline
\rule{0pt}{3ex}
[C/Fe] & $>0.24$ & $>-1.04$ & $>-0.31$ & $>0.24$ & $>-0.07$ & $>-0.66$ & $>-0.43$\\

[O/Fe] & $>-0.08$ & $>-0.51$ & $-0.78\pm0.15$ & $-0.4\pm0.2$ & $>-0.97$ & $>-1.08$ & $>-0.57$\\

[Mg/Fe] & $>-0.07$ & $>0.43$ & $>0.14$ & $>0.28$ & $>0.13$ & $>-0.94$ & $>0.08$\\

[Al/Fe] & $>0.42$ & $>0.91$ & $>1.08$ & $>1.09$ & & & $>0.66$\\

[Si/Fe] & $0.06\pm0.06$ & $0.08\pm0.04$ & $0.26\pm0.10$ & $0.23\pm0.09$ & $0.00\pm0.08$ & $0.00\pm0.03$ & $0.02\pm0.03$\\

[S/Fe] & $-0.08\pm0.10$ & $-0.15\pm0.09$ & $0.69\pm0.13$ &  &  & & $-0.2\pm0.2$\\[1ex]
\hline
\hline
\end{tabular}
\end{table*}

\vspace{1cm}

\begin{table*}
\caption{[$X$/Fe] residuals of the depletion pattern fitting (see Figs. \ref{depletion tot}, \ref{depletion comp} and \ref{depletion comp alpha}). The values and uncertainties are obtained including MC simulations to take into account the dust depletion errors. We stress that the impact of dust depletion on the nuclear abundances is correlated between elements; i.e., a higher dust depletion correction lowers all [$X$/Fe] values except for [Al/Fe] which it raises, and vice-versa.}
\label{table_abb_depl} 
\centering
\begin{tabular}{c c c c c c c c}       
\hline\hline
With $\alpha$-element corr.  & $I$ & $II$ & $III$ & $IV$ & $V$ & $VI$ & Tot\\
\hline
\rule{0pt}{3ex}
[C/Fe]  &   $0.73^{+0.06}_{-0.09}$  & $-0.41^{+0.10}_{-0.10}$  &  $0.20^{+0.16}_{-0.16}$ &  $0.44^{+0.09}_{-0.14}$  &  $0.36^{+0.12}_{-0.18} $ &  $0.10^{+0.12}_{-0.12} $ &   $0.29^{+0.10}_{-0.10}$\\[2ex]

[O/Fe]  &   $0.35^{+0.06}_{-0.09}$  & $0.04^{+0.09}_{-0.10}$  &  $-0.33^{+0.15}_{-0.15}$ & $-0.28^{+0.09}_{-0.13}$  & $-0.60^{+0.11}_{-0.16}$  & $-0.42^{+0.11}_{-0.11}$  &  $0.07^{+0.09}_{-0.09}$\\[2ex]

[Mg/Fe] &   $0.13^{+0.03}_{-0.05}$  & $ 0.69^{+0.05}_{-0.05}$  &  $0.35^{+0.08}_{-0.08}$ &  $0.32^{+0.05}_{-0.07}$  &  $0.30^{+0.06}_{-0.09} $ &  $-0.61^{+0.06}_{-0.06}$  &  $0.40^{+0.05}_{-0.05}$\\[2ex]

[Al/Fe] &   $0.29^{+0.03}_{-0.02}$  & $ 0.74^{+0.04}_{-0.04}$  &  $0.95^{+0.06}_{-0.06}$ &  $1.07^{+0.05}_{-0.03}$  &             &          &                    $0.45^{+0.04}_{-0.04}$\\[2ex]

[Si/Fe] &   $0.28^{+0.03}_{-0.04}$  & $ 0.37^{+0.05}_{-0.05}$  &  $0.50^{+0.07}_{-0.07}$ &  $0.32^{+0.04}_{-0.06}$  &  $0.20^{+0.05}_{-0.08} $ &  $0.34^{+0.05}_{-0.05} $ &   $0.34^{+0.05}_{-0.05}$\\[2ex]

[S/Fe]  &   $0.26^{+0.04}_{-0.06}$  & $ 0.28^{+0.07}_{-0.07}$  &  $1.05^{+0.11}_{-0.11}$ &  &  &  &   $0.28^{+0.07}_{-0.07}$\\[1ex]
\hline
\hline
Without $\alpha$-element corr.  & $I$ & $II$ & $III$ & $IV$ & $V$ & $VI$ & Tot\\
\hline
\rule{0pt}{3ex}
[C/Fe] &   $0.31^{+0.13}_{-0.13}$ &   $-0.97^{+0.13}_{-0.13}$ & $-0.24^{+0.17}_{-0.17}$  &  
$0.29^{+0.16}_{-0.18}$&   $0.00^{+0.21}_{-0.22}$ &   $-0.59^{+0.16}_{-0.16}$ & $-0.36^{+0.14}_{-0.14}$\\[2ex]

[O/Fe] &  $-0.04^{+0.12}_{-0.12}  $  & $ -0.47^{+0.12}_{-0.12}  $ & $ -0.74^{+0.16}_{-0.16}    $ &   $ -0.42^{+0.15}_{-0.17}   $ &  $ -0.93^{+0.20}_{-0.20}$   &  $-1.05^{+0.15}_{-0.15}$  & $  -0.52^{+0.13}_{-0.13}$\\[2ex]

[Mg/Fe]&   $-0.09^{+0.07}_{-0.07}  $ &  $ 0.40^{+0.07}_{-0.07}   $ & $0.12^{+0.09}_{-0.09}  $ &   $0.25^{+0.08}_{-0.10}     $ &  $  0.11^{+0.11}_{-0.11}$  &   $-0.96^{+0.08}_{-0.08}$ &  $  0.07^{+0.07}_{-0.07}$\\[2ex]

[Al/Fe]&   $0.44^{+0.04}_{-0.04}$ &  $ 0.93^{+0.04}_{-0.04}$ & $1.10^{+0.06}_{-0.06}   $  &  $       1.12^{+0.06}_{-0.06}$ &               &                                &    $0.68^{+0.05}_{-0.05}$\\[2ex]

[Si/Fe]&   $0.10^{+0.06}_{-0.06}   $&   $0.12^{+0.06}_{-0.06}    $ & $0.31^{+0.08}_{-0.08} $  &  $ 0.26^{+0.07}_{-0.08}    $ &  $  0.04^{+0.09}_{-0.10}$  &   $0.04^{+0.07}_{-0.07}$ &   $ 0.06^{+0.06}_{-0.06}$\\[2ex]

[S/Fe] &   $-0.03^{+0.09}_{-0.09}  $  & $ -0.10^{+0.09}_{-0.09}  $ & $  0.75^{+0.12}_{-0.12}$ & & & & $  -0.16^{+0.09}_{-0.09}$\\[1ex]

\hline
\hline
\end{tabular}
\end{table*}

\end{appendix}

\end{document}